\documentclass[floats,floatfix,showpacs,amssymb,prd,twocolumn,superscriptaddress,nofootinbib,nolongbibliography,reprint]{revtex4-1}

\usepackage{amssymb,amsmath,verbatim,mathtools,needspace,enumitem,etoolbox,graphicx,physics,microtype,afterpage,xspace,tabularx,lmodern,multirow,bm}
\usepackage{textcomp, gensymb}

\usepackage[normalem]{ulem}
\usepackage{placeins}
\usepackage{comment}
\usepackage[dvipsnames, usenames]{xcolor}
\definecolor{linkcolor}{rgb}{0.0,0.3,0.5}
\usepackage[unicode, colorlinks=true, linkcolor=linkcolor, citecolor=linkcolor, filecolor=linkcolor, urlcolor=linkcolor, linktocpage, breaklinks]{hyperref}
\usepackage[all]{hypcap}
\usepackage[T1]{fontenc}
\usepackage[utf8]{inputenc}
\usepackage{mathrsfs}
\usepackage[usenames,dvipsnames]{xcolor}
\hypersetup{colorlinks=true,citecolor=romared,linkcolor=romared,urlcolor=romared}

\definecolor{romared}{RGB}{142,0,28}

\newcommand{\be}{\begin{equation}}
\newcommand{\ee}{\end{equation}}

\def\be{\begin{equation}}
\def\ee{\end{equation}}
\newcommand{\beq}{\begin{eqnarray}}
\newcommand{\eeq}{\end{eqnarray}}

\usepackage{aas_macros}
\usepackage{makecell}
\usepackage{soul}
\usepackage[nolist,nohyperlinks]{acronym}
\usepackage{orcidlink}
\usepackage{lipsum}
\usepackage[ruled,lined]{algorithm2e}

\acrodef{LSC}[LSC]{LIGO Scientific Collaboration}
\acrodef{BH}{black hole}
\acrodef{NS}{neutron star}
\acrodef{PN}{Post-Newtonian}
\acrodef{BBH}{binary black-hole}
\acrodef{BNS}{binary neutron-star}
\acrodef{NSBH}{neutron-star black-hole}
\acrodef{NR}{numerical relativity}
\acrodef{GW}{gravitational wave}
\acrodef{PSD}{power spectral density}
\acrodef{aLIGO}{Advanced Laser interferometer Gravitational-Wave Observatory}
\acrodef{AZDHP}{aLIGO zero detuned high power density}
\acrodef{GR}{general relativity}
\acrodef{PE}{parameter estimation}
\acrodef{LAL}{LIGO algorithm library}
\acrodef{TPI}{tensor-product interpolant}
\acrodef{SVD}{singular value decomposition}
\acrodef{SNR}{signal-to-noise ratio}
\acrodef{ODE}{ordinary differential equation}
\acrodef{PDE}{partial differential equation}
\acrodef{ROM}{reduced order model}
\acrodef{QNM}{quasi-normal mode}
\acrodef{IMR}{inspiral-merger-ringdown}
\acrodef{LVK}{LIGO-Virgo-KAGRA}
\acrodef{SXS}{Simulating eXtreme Spacetimes}

\newcommand{\jhu}{\affiliation{William H. Miller III Department of Physics and Astronomy, Johns Hopkins University, 3400 North Charles Street, Baltimore, Maryland, 21218, USA}}
\newcommand{\AEI}{\affiliation{Max Planck Institute for Gravitational Physics (Albert Einstein Institute), Am M\"uhlenberg 1, Potsdam 14476, Germany}}

\graphicspath{{../Plots/}}

\newcommand{\orcid}[1]{\href{https://orcid.org/#1}{\includegraphics[width=10pt]{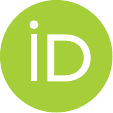}}}
\renewcommand{\vec}[1]{\boldsymbol{#1}}

\newcommand{\ben}{\begin{enumerate}}
\newcommand{\een}{\end{enumerate}}

\def\be{\begin{equation}}
\def\ee{\end{equation}}
\def\beq{\begin{eqnarray}}
\def\eeq{\end{eqnarray}}

\def\apj{\rm{ApJ}}
\def\apjl{\rm{ApJ}}
\def\apjs{\rm{ApJS}}
\def\aap{\rm{A\&A}}

\def\nat{\rm{Nature}}

\graphicspath{{./Plots/}}

\allowdisplaybreaks

\begin{document}

\pagenumbering{arabic}

\title{Probing minihalo lenses with diffracted gravitational waves}

\author{Mark Ho-Yeuk Cheung \orcid{0000-0002-7767-3428}}
\jhu
\author{Ken K. Y. Ng \orcid{0000-0003-3896-2259}}
\jhu
\author{Miguel Zumalac\'{a}rregui \orcid{0000-0002-9943-6490}}
\AEI
\author{Emanuele Berti \orcid{0000-0003-0751-5130}}
\jhu

\date{\today}

\begin{abstract}
    When gravitational waves pass near a gravitating object, they are deflected, or lensed.
    If the object is massive, such that the wavelength of the waves is small compared to its gravitational size, lensed gravitational wave events can be identified when multiple signals are detected at different times.
    However, when the wavelength is long, wave-optics diffraction effects will be important, and a lensed event can be identified by looking for frequency-dependent modulations to the gravitational waveform, without having to associate multiple signals.
    For current ground-based gravitational wave detectors observing stellar-mass binary compact object mergers, wave-optics effects are important for lenses with masses $\lesssim 1000 M_{\odot}$.
    Therefore, minihalos below this mass range could potentially be identified by lensing diffraction.
    The challenge with analyzing these events is that the frequency-dependent lensing modulation, or the amplification factor, is prohibitively expensive to compute for Bayesian parameter inference.
    In this work, we use a novel time-domain method to construct interpolators of the amplification factor for the Navarro-Frenk-White (NFW), generalized singular isothermal sphere (gSIS) and cored isothermal sphere (CIS) lens models.
    Using these interpolators, we perform Bayesian inference on gravitational-wave signals lensed by minihalos injected in mock detector noise, assuming current sensitivity of ground-based detectors.
    We find that we could potentially identify an event when it is lensed by minihalos and extract the values of all lens parameters in addition to the parameters of the GW source.
    All of the methods are implemented in \texttt{glworia}~\cite{glworia_github}, the accompanying open-source \texttt{python} package, and can be generalized to study lensed signals detected by current and next-generation detectors.

\end{abstract}

\maketitle

\section{Introduction}

Gravitational lensing is a pivotal concept in our understanding of the Universe~\cite{1992grle.book.....S,Bartelmann:2010fz}.
Lensing occurs when a massive celestial object, such as a galaxy or star,
affects the path of light, or (more comprehensively) any form of radiation, such as electromagnetic (EM) waves or gravitational waves (GWs).
According to Einstein's theory of general relativity~\cite{Einstein:1916vd}, the curvature of spacetime around the massive object curves the path of the waves, leading to a distorted perception of the source, or even to the observation of multiple images of the same source.

The lensing of EM waves has been an integral aspect of astrophysical and cosmological research.
Through the study of EM lensing, scientists have successfully discovered exoplanets orbiting distant stars~\cite{Bond:2004qd}, observed distant objects that would otherwise be too faint to detect~\cite{Coe:2012wj,2022Natur.603..815W}, probed the nature of dark matter~\cite{Clowe:2003tk,Markevitch:2003at}, and measured cosmological parameters~\cite{Planck:2018vyg}.

The first GW event, GW150914, was detected by the LIGO gravitational-wave detectors in 2015~\cite{LIGOScientific:2016aoc}.  The radiation was produced by the merger of two distant black holes (BHs), and it opened up an entirely new way of observing and understanding the universe~\cite{Maggiore:2007ulw}.  Since then, the LIGO and Virgo detectors have detected around a hundred GW events, sourced by merging BHs and neutron stars~\cite{LIGOScientific:2018mvr,LIGOScientific:2020ibl,KAGRA:2021vkt}.

Similar to EM waves, GWs can also be lensed~\cite{Ohanian:1974ys}.
The detection of lensed GWs could lead to significant advancements in fundamental physics, astrophysics, and cosmology.
Most typical GW events are transients, and the lensing of these events can be used to perform cosmography~\cite{Oguri:2019fix,Liao:2022gde,Sereno:2011ty,Liao:2017ioi,Cao:2019kgn,Jana:2022shb}.
GW lensing can potentially be used to break the mass sheet degeneracy, which affects EM lensing~\cite{Cremonese:2021puh}.
Tests of GR can be performed using GW birefringence~\cite{Ezquiaga:2020dao,Wang:2021gqm,Goyal:2023uvm} or the GW propagation speed~\cite{Baker:2016reh,Collett:2016dey}.
If multiple images were observed, we could potentially localize the host galaxy of the GW source~\cite{Hannuksela:2020xor,Yu:2020agu}.

This paper is focused on exploring lensed GWs, with special attention to wave optics effects.
This refers to the consideration of the wave-like behavior of the lensed GWs, including diffraction and interference.
These effects become significant when the wavelength of the waves is comparable to the gravitational length-scale $GM / c^2$ of the lensing substructure~\cite{Takahashi:2003ix}, where $M$ is the (appropriately defined) mass of the lens, $G$ is Newton's constant, and $c$ is the speed of light.
Wave optics effects can lead to frequency-dependent modifications in the GW amplitude, which can reveal critical information about the lens.

The science case of GW lensing is even richer when we consider wave-optics effects.
While in the geometrical optics limit a lensed waveform is degenerate with an unlensed one with a trivial magnification and phase shift (unless higher harmonics are included~\cite{Ezquiaga:2020gdt}, or when the BBH merger is eccentric or precessing~\cite{Dai:2017huk}), wave-optics effects are frequency-dependent, meaning that only a single signal is required for a conclusive detection.
Intermediate-mass BHs could diffract GWs and leave an observable imprint on detected signals~\cite{Lai:2018rto,Gais:2022xir}, while
primordial BHs, massive compact halo objects (MACHOs) and other substructures could also be identified in a similar fashion~\cite{Liu:2023ikc,Basak:2021ten,Meena:2019ate,Jung:2017flg,Diego:2019rzc,Urrutia:2021qak,Zhou:2022yeo,Tambalo:2022wlm}.
The structure of low-mass halo lenses could be probed when lensed events are detected by ground-based~\cite{Dai:2018enj,Oguri:2020ldf,Guo:2022dre} and space-based detectors~\cite{Caliskan:2022hbu,Savastano:2023spl,Caliskan:2023zqm,Sereno:2010dr,Gao:2021sxw,Choi:2021bkx,Cremonese:2021ahz,Guo:2022dre,Fairbairn:2022xln}.
Individual stars residing in a galaxy could also introduce measurable GW lensing effects~\cite{Christian:2018vsi,Diego:2019lcd,Cheung:2020okf,Mishra:2021xzz,Suvorov:2021uvd,Meena:2022unp,Yeung:2023mbs,Shan:2023ngi,Shan:2023qvd}.

The diffraction of GWs can be used to probe the abundance of compact dark matter halos and their morphology. Many theories of dark matter predict such objects:
ultralight bosonic field can produce solitonic cores and compact axion structures~\cite{Hui:2016ltb,Arvanitaki:2019rax}, self-interacting dark matter theories predict the gravitothermal collapse of subhalos~\cite{Gilman:2021sdr,Adhikari:2022sbh}, and warm dark matter can form massive prompt cusps~\cite{Delos:2022yhn,Delos:2023exh}.
While primordial BHs with masses $\gtrsim 20M_\odot$ are severely constrained by observations of the cosmic microwave background~\cite{Serpico:2020ehh}, extended dark matter objects in that range are subject to looser limits, driven by lensing of EM sources~\cite{Esteban-Gutierrez:2023qcz,Blaineau:2022nhy,Oguri:2017ock,Zumalacarregui:2017qqd}.
GW lensing is not yet competitive in this range~\cite{Basak:2021ten}, but future limits from LIGO-Virgo-KAGRA~\cite{Jung:2017flg} will likely be more stringent than the limits from EM sources such as fast radio bursts~\cite{Munoz:2016tmg}. In the future, third-generation detectors will improve current bounds by $\sim 4$ orders of magnitude~\cite{GilChoi:2023qrz}.

Lensed GWs have been actively searched~\cite{Hannuksela:2019kle, LIGOScientific:2021izm, LIGOScientific:2023bwz,Liu:2020par,Dai:2020tpj} with no conclusive discoveries.
These searches work mostly in the geometrical optics regime, where different GW event triggers are compared against each other to uncover lensed events with multiple images arriving at different times~\cite{Haris:2018vmn,Lo:2021nae,Janquart:2022wxc,Ali:2022guz,Ezquiaga:2023xfe}.
Given a catalog of detected events, potential subthreshold counterpart images are also being looked for~\cite{Li:2019osa,Li:2023zdl,McIsaac:2019use}.
The rate of detecting strongly lensed pairs of GW signals in the LIGO-Virgo-KAGRA detector network running at design sensitivity has been forecast to range from once to a few tens per year~\cite{Ng:2017yiu,Li:2018prc,Xu:2021bfn,Wierda:2021upe}, with a lower limit of $10^{-5} {\rm yr}^{-1}$~\cite{Smith:2017mqu}.
Lensing effects will also affect parameter estimation of both individual events and GW populations~\cite{Oguri:2018muv,Cao:2014oaa,Dai:2016igl,Cusin:2020ezb,Mukherjee:2021qam}.

While most lensing studies have focused on the geometrical optics regime, there has been active research on wave optics effects in the past decades.
Femtolensing of gamma ray bursts with diffraction has been considered~\cite{Gould:1991td,Barnacka:2012bm}, but the finite size of the source~\cite{Matsunaga:2006uc} leads to poor constraints on the MACHO population~\cite{Katz:2018zrn}.
Similar studies have been undertaken for fast radio burst sources in the geometrical optics limit~\cite{Munoz:2016tmg,Dai:2017twh,Laha:2018zav} and with wave-optics effects included~\cite{Zheng:2014rpa,Eichler:2017eid,Katz:2019qug,Leung:2022vcx,Jow:2020rcy}, with the caveat that these will be affected by scintillation.
Pioneering work on GW lensing with diffraction was also performed two decades ago~\cite{Takahashi:2003ix,Takahashi:2004mc}.

Other than searches for multiple-image lensing in the geometrical optics regime, searches for diffracted GW events have also been undertaken, where a single signal could lead to a conclusive detection of lensing~\cite{Lai:2018rto,Kim:2022lex}.
However, these studies assume point-mass lens profiles, meaning that they are adequate for searching for GWs lensed by PBHs or stars, but not for more diffuse lenses like minihalos.
The detection rate of GW signals diffracted by small-mass halos depends significantly on the halo profile, and could range from being close to zero to thousands per year~\cite{Guo:2022dre}.
Nonetheless, analysis methods assuming diffuse lens profiles will be needed in order to detect or constrain the rates of these minihalo lensing events.

On the technical side, the field has recently made significant progress in tackling the computational challenges of computing the wave-optics effects of lensing for general lens profiles.
The computation of wave-optics effects requires solving a highly-oscillatory Fresnel-Kirchhoff integral, with analytic solutions only for simple cases.
A number of numerical approaches have been used to solve the integral, including a direct quadrature-type integration, transforming the problem into a series of contour integrals~\cite{Ulmer:1994ij,Mishra:2021xzz}, and analytic continuation, i.e. Picard-Lefschetz type methods~\cite{Feldbrugge:2019fjs,Jow:2022pux,Tambalo:2022plm}.
Approximate methods were also used to speed up calculations~\cite{Takahashi:2004mc,Savastano:2023spl}.

While we can compute the wave-optics effects with the methods mentioned above, efficiency is also important.
When a GW signal is detected, one has to perform Bayesian parameter estimation (PE) to extract the value of the lensing-related parameters from the data.
For this purpose, the lensed waveform model has to be evaluated $\sim O(10^7)$ times.
For the point-mass lens case an analytic solution to the amplification factor is known and can be efficiently called, so PE can be performed~\cite{Lai:2018rto,Seo:2021psp,Yeung:2023mbs}.
However, for other lens models the numerical evaluation of the amplification factor is prohibitively expensive for PE.
To the best of our knowledge, the only publicly available software package capable of performing PE for lens models other than the point mass and SIS lenses is the \texttt{Gravelamps} package~\cite{Wright:2021cbn}, where the amplification factor is computed by direct integration and interpolated over the frequency and the impact parameter of the source.

In this work, we will focus on the detection of wave optics effects in GWs lensed by minihalos and measured by ground-based detectors.
In the frequency band of ground-based detectors, these effects are measurable for lens masses of $\sim 1 - 1000 M_\odot$.
In this paper we introduce \texttt{glworia}~\cite{glworia_github} (gravitational lensing in the wave optics regime: interpolator for the amplification), a \texttt{python} package for computing the amplification factor for arbitrary spherically-symmetric lens models with one lens parameter $l$ other than the (redshifted) mass of the lens $M_{Lz}$ and the (one-dimensional) impact parameter $y$ of the source.%
\footnote{In this work, we will call this a ``one-parameter'' lens model, because $M_{Lz}$ only affects the frequency scale of the computations and $y$ does not depend on the profile of the lens.
    We will also refer to $l$ (but not $M_{Lz}$) as the ``lens parameter''.}
The amplification factors are then interpolated in the time domain, and PE can be performed for the lens mass, impact parameter, and the lens parameter.

In Sec.~\ref{sec:preliminaries} we review the mathematics behind computing the lensing amplification factor.
In Sec.~\ref{sec:integral} we explain our numerical implementation of the contour integration method and how we construct the interpolation table for the amplification factor.
In Sec.~\ref{sec:PE} we perform Bayesian parameter estimation on mock lensing signals injected into LIGO detector noise.
In Sec.~\ref{sec:discussion} we discuss the implications of our results and potential directions of future work.

\section{Preliminaries}\label{sec:preliminaries}

\begin{figure}
    \centering
    \includegraphics[width=0.49\textwidth]{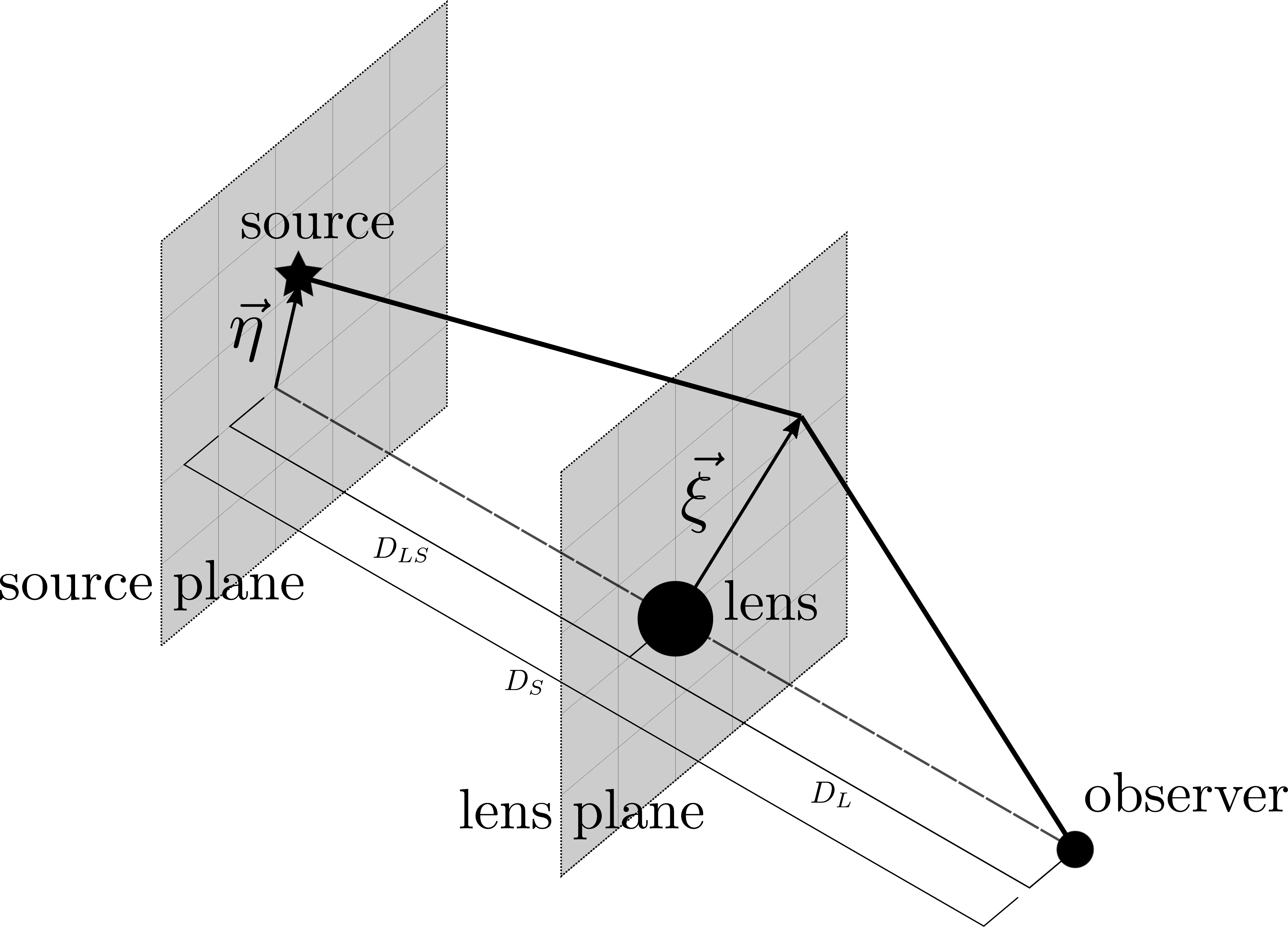}
    \caption{
        A sketch of the source-lens-observer setup.
        The source plane and lens planes are the planes perpendicular to the optical axis at the source and at the lens, respectively.
        The relevant angular diameter distances are $D_L$ (from the observer to the lens), $D_S$ (from the observer to the source), and $D_{LS}$ (from the lens to the source).
        The displacement vector from the optical axis to the source in the source plane is $\vec{\eta}$; $\vec{\xi}$ is the coordinate vector in the lens plane.
        Figure adapted from~\cite{Cheung:2020okf}.
    }
    \label{fig:lensing}
\end{figure}

\subsection{The lensing integral}

The following derivation follows closely Ref.~\cite{1992grle.book.....S}.
In this work, we will use geometric units such that $c = G = 1$.

While we will study gravitational-wave lensing with wave-optics diffractive effects, it will still be useful to inherit some terminology that is used in the geometrical-optics regime.
For example, while in the wave-optics limit we might not see discrete images that are well resolved in the sky or in the arrival time, we will still call the critical points of the time delay function (arrival time) the ``images''.
To further simplify notation, we will define the ``strong-lensing'' regime to be the case when there are multiple such images, i.e., there are additional critical points other than the trivial minimum of the time delay, and ``weak-lensing'' the case when there is only one image, i.e., only the global minimum exists.
Such a definition might be different from other works in the literature, especially in EM lensing.

The spacetime metric of an isolated, slowly moving, noncompact matter distribution is given by (neglecting the expansion of the Universe for now)
\begin{equation}
    ds^2 = \left(1 + 2U \right) c^2 dt^2 - \left( 1 - 2U \right) d\vec{x}^2,
\end{equation}
where $U$ is the gravitational potential of the matter.
On a (future-directed) null curve, $ds^2 = 0$.
The quantity $n = 1 - 2 U$ can be viewed as an effective refractive index of the gravitational field.
If a light pulse is emitted at $t = 0$ at the source, its arrival time at a stationary observer at fixed spatial coordinates will be
\begin{align}
    t & = \int \left( 1 - 2U \right) dl. \nonumber \\
      & = l - 2 \int U dl, \label{eq:time_delay}
\end{align}
to first order in $U$, where $l$ is the overall Euclidean length of the light travel path.
Now consider light pulses that travel from the source to the observer in Fig.~\ref{fig:lensing}, defining $\vec{\xi}$ to be the location of the lens in the lens plane and $\vec{\eta}$ to be that of the source in the source plane, where both are defined with respect to the optical axis, with both planes perpendicular to the axis.
We further define the angular diameter distances $D_L$ between the lens and the observer, $D_S$ between the source and the observer, and $D_{LS}$ between the lens and the source.
Assuming that $\xi = |\vec{\xi}|$ and $\eta = |\vec{\eta}|$ are much smaller than $D_L$ and $D_S$, the path length is
\begin{equation}
    l \approx D_{LS} + D_L + \dfrac{1}{2 D_{LS}}(\vec{\xi} - \vec{\eta})^2 + \dfrac{1}{2 D_{L}} \xi^2.
\end{equation}
The potential $U$ is linear in the mass distribution, so we can derive a Green's function to solve the integral in the second term of Eq.~\eqref{eq:time_delay}.
For a point mass, $U = 4 M G / r c^2$, and the integral from the source $S$ to the image position $I$ on the lens plane is
\begin{equation}
    \int^I_S U dl = G M \ln \dfrac{\xi}{2 D_{LS}},
\end{equation}
where we neglect higher-order terms.
The integral from the lens plane to the observer is analogous, and the total integral is
\begin{equation} \label{eq:PM_integral}
    \int U dl = G M \ln \dfrac{\xi}{\xi_0} + {\rm const.}\, ,
\end{equation}
where $\xi_0$ is an arbitrary constant which can be chosen as convenient.

In fact, it is often convenient to use a length-scale $\xi_0$ that is related to an appropriately defined mass of the lens.
For example, for a point mass lens with a redshifted mass $M_{Lz}$, $\xi_0$ can be chosen to be the radius of the critical curve on the lens plane (to be discussed later), or the Einstein radius $\theta_E$:
\begin{equation}\label{eq:Einstein_radius}
    \xi_0 = \theta_E \equiv \sqrt{\dfrac{4 D_L D_{LS}}{(1 + z_L) D_S} M_{Lz}}.
\end{equation}
For a point mass lens, the Einstein radius $\theta_E$ is the radius of the critical curve, where the magnification of the image will be formally infinite if the image lies on it.
In fact, we can use the same definition, Eq.~\eqref{eq:Einstein_radius}, for any lens model.
In this way, defining either one of $M_{Lz}$ or $\xi_0$ fixes the other quantity.
This implies that the definition of $M_{Lz}$ could be as arbitrary as that of $\xi_0$, and it often is not the same as other definitions of the mass of halos, e.g. the virial mass.

We will consider lenses with an extent $L$ much smaller than $D_L$ and $D_{LS}$.
In that case, the deflection of the ray occurs mainly when it is close to the lens, so we can assume that the light travel path consists of two straight line segments with an abrupt change in direction on the lens plane.
Then, for a deflection angle $\hat{\alpha}$, if $\hat{\alpha} L$ (the perpendicular displacement of the ray when it is close to the lens) is small compared to the length scale on which $U$ changes, the integral $\int U dl$ can be performed over the unperturbed ray.
As the integral for the point-mass case is a function of $\xi$, which can be seen as an impact parameter, the integral only depends on the perpendicular distance of the mass elements to the ray, but not on their distribution along the direction parallel to the ray.
Therefore, we can project the mass distribution onto the lens plane before integrating.
This is called the thin-lens approximation.

Exploiting the linearity of $U$ with respect to mass elements and employing the thin-lens approximation, we can use Eq.~\eqref{eq:PM_integral} to write
\begin{align}
    \hat{\psi}(\vec{\xi}) & \equiv 2 \int U dl \nonumber                                                                                                    \\
                          & = 4 \int d^2 \xi^\prime \Sigma(\vec{\xi^\prime}) \ln \left(\dfrac{|\vec{\xi} - \vec{\xi^\prime}|}{\xi_0} \right) + {\rm const.}
\end{align}
for a general lens distribution, where $\Sigma(\vec{\xi^\prime})$ is the density profile projected onto the 2D lens plane.
A ray originating from $\vec{\eta}$ on the source plane, passing through $\vec{\xi}$ on the lens plane and arriving at the observer would correspond to a light travel time
\begin{equation}
    t = (1 + z_L)\phi(\vec{\xi}, \vec{\eta}) + {\rm const.} \, ,
\end{equation}
where $\phi(\vec{\xi}, \vec{\eta})$ is the Fermat potential
\begin{equation}
    \phi(\vec{\xi}, \vec{\eta}) = \dfrac{D_L D_S}{2 D_{LS}}\left(\dfrac{\vec{\xi}}{D_L} - \dfrac{\vec{\eta}}{D_S}\right)^2 - \hat{\psi}(\vec{\xi}),
\end{equation}
and we have included a factor of $(1 + z_L)$ to account for cosmological redshift: see Ref.~\cite{1992grle.book.....S} for a detailed derivation.
It is convenient to define $\vec{x} = \vec{\xi}/\xi_0$ and $\vec{y} = \vec{\eta}D_L/\xi_0 D_S$, which are the angular positions of $\vec{\xi}$ and $\vec{\eta}$ with respect to the optical axis in the sky according to the observer, normalized by $\xi_0$.
Then, the arrival time of a ray at the observer can be rewritten as
\begin{multline}
    T(\vec{x}, \vec{y}) \\ \equiv \dfrac{D_S \xi_0^2}{D_L D_{LS}} (1 + z_L)\left(\dfrac{1}{2}|\vec{x} - \vec{y}|^2 - \psi(\vec{x}) + \phi_m(\vec{y})\right),
\end{multline}
where $\psi(\vec{x}) \equiv \hat{\psi}(\xi_0 \vec{x})$, and $\phi_m(\vec{y})$ is chosen such that the minimum value of $T(\vec{x}, \vec{y})$ is zero.

In the eikonal approximation, the metric perturbation corresponding to GWs can be written as
\begin{equation}
    h_{\mu \nu} = \Phi e_{\mu \nu},
\end{equation}
where $e_{\mu \nu}$ is the polarization tensor of the GW, whose change is of the order $U$ in our case and can be assumed to be constant.
Then, the problem reduces to finding the scalar function $\Phi$, which satisfies the Helmholz equation in the frequency domain
\begin{equation}
    (\nabla^2 + \omega^2) \tilde{\Phi} = 4 \omega^2 U \tilde{\Phi}.
\end{equation}

Now, the field at the location of the observer could be solved by the Fresnel-Kirchhoff diffraction integral (Eq.~\eqref{eq:Kirchhoff} below).
Because the integral only works if the wave travels freely beyond the integration surface, we need to define a surface $E^\prime$ parallel to the lens plane but closer to the observer, and assume that the rays do not interact with the lens anymore after passing through $E^\prime$.
Let the perpendicular distance between $E^\prime$ and the observer be $D_L^\prime$.
Then, applying the eikonal approximation, the phase of the rays at $E^\prime$ is
\begin{equation}
    S = \omega (\tilde{\phi} - D_L^\prime) + \alpha(\vec{\eta}),
\end{equation}
where we used $S = \omega \int n dl$, and $\alpha(\vec{\eta})$ is a constant that does not depend on the location of the ray on the lens plane.
If $ D_L - D_L^\prime \ll 1$, the amplitude of the wave at $E^\prime$ is approximately the same as that at $E$.
This is because the alteration of amplitude (i.e., magnification) of the observed wave is attributed to the focusing of the waves due to small-angle scattering.
The effect of focusing is only significant over a long distance (i.e., when the waves are traveling from $E^\prime$ to the observer), but it has a negligible effect for the short journey from $E$ to $E^\prime$.
Thus, if the complex amplitude of the wave without the lens is $A e^{-i \alpha(\eta)}$ (where $\alpha$ has been included in the definition to simplify the calculations, without loss of generality), the amplitude of the wave at $E^\prime$ will be
\begin{equation}
    \Phi_{E^\prime} (\vec{\xi^\prime}) = \dfrac{D_S A}{D_{LS}} e^{i \omega (\tilde{\phi} - D_L^\prime)}.
\end{equation}
Then, plugging this into the Kirchhoff integral,
\begin{equation}\label{eq:Kirchhoff}
    \tilde{\Phi}_O = \dfrac{1}{4\pi} \int_{E^\prime} d^2 \xi^\prime \left[\Phi_{E^\prime} \dfrac{\partial}{\partial n}\left( \dfrac{e^{i \omega D_L^\prime}}{D_L^\prime}\right) - \dfrac{e^{i \omega D_L^\prime}}{D_L^\prime} \dfrac{\partial}{\partial n} \Phi_{E^\prime} \right],
\end{equation}
the lensing amplification factor $\tilde{F} = \tilde{\Phi}_O / A$ at the observer is (including the effects of cosmological redshift now, see Ref.~\cite{1992grle.book.....S} for details)
\begin{align}
    \tilde{F} & = (1 + z_L)\dfrac{\omega}{2 \pi i} \dfrac{D_S}{D_L D_{LS}} \int_E d^2 \xi e^{i \omega \tilde{\phi}(\vec{\xi, \eta})} \label{eq:lensing_integral_unnorm} \\
              & = \dfrac{w} {2\pi i} \int_E d^2 x e^{i w T}, \label{eq:lensing_integral}
\end{align}
where we have defined the dimensionless angular frequency
\begin{equation}
    w = (1 + z_L)\dfrac{D_S \xi_0^2}{D_L D_{LS}} \omega.
\end{equation}
In deriving Eq.~\eqref{eq:lensing_integral_unnorm} we have used the thin lens approximation, neglected the derivatives of slowly varying functions (i.e., we only keep the derivatives of the oscillatory components with factors of $e^{i \omega D_L}$), and taken the limit $E^\prime \to E$.

Hence, we have derived the effects of lensing on a gravitational waveform.
If the unlensed frequency-domain waveform strain in the detector is $\tilde{h}$, the lensed waveform $\tilde{h}_L$ is
\begin{equation} \label{eq:lens_wf}
    \tilde{h}_L(f) = \tilde{F}(f) \tilde{h}(f),
\end{equation}
with $2 \pi f = \omega$ and $\tilde{F}$ given by Eq.~\eqref{eq:lensing_integral}.

\subsection{Time-domain integral}
\label{subsec:time_domain_integral}

Assuming that the lens is spherically symmetric, the integral in Eq.~\eqref{eq:lensing_integral} can be transformed into a 1D integral and evaluated numerically.
However, the integral is highly oscillatory, and it is nontrivial to compute when $w$ is large, e.g. $w \gtrsim 10^3$.
An alternative strategy is to solve the integral in the time domain.
We start by Fourier transforming the quantity $\tilde{F}/(i \omega)$,
\begin{align}
    \tilde{I}(\tau) & = \dfrac{1}{2 \pi} \int d^2 x \int dw e^{i w (\phi(\vec{x}, \vec{y}) - \tau)} \nonumber                      \\
                    & = \int d^2 x \delta(T(\vec{x}, \vec{y}) - \tau)                                    \label{eq:contour_delta}  \\
                    & = \sum_k \oint_{\gamma_k} \dfrac{ds}{|\nabla_{\vec{x}}T(\vec{x}(\tau, s), \vec{y})|}, \label{eq:contour_int}
\end{align}
where $\delta$ is the Dirac delta function, $\nabla_{\vec{x}}$ is the gradient operator with respect to the lens plane coordinates $\vec{x}$, and the summation in Eq.~\eqref{eq:contour_int} is over all of the contours $\gamma_k$ of constant $\tau$ parameterized by $s$.
Going from Eq.~\eqref{eq:contour_delta} to Eq.~\eqref{eq:contour_int} we have made use of the properties of delta functions.

In this form, Eq.~\eqref{eq:contour_int} has a conceptually simple interpretation.
For a ``contour ribbon'' bounded by two contour lines at fixed $T = \tau$ and $T=\tau + d\tau$, the width of such a ribbon at the parameterized point $s$ is proportional to $1 / |\nabla T|$.
Thus, $\tilde{I}(\tau)$ can be viewed as the lens-plane cross-sectional area of the rays that will arrive at the time $\tau$.
Because each point on the lens plane should receive approximately the same flux per unit area, if the source were a pulse, $\tilde{I}(\tau)$ would then be proportional to the power received at time $\tau$.
For a signal that is continuous in time, e.g. $h(\tau)$, the response is the convolution of $h(\tau)$ with $\tilde{I}(\tau)$, and the Fourier transform of the convolution is the multiplication of the two components in the frequency domain, as in Eq.~\eqref{eq:lens_wf}.
Note that $\tilde{I}$ is the Fourier transform of $\tilde{F}/(i \omega)$, and the factor $i \omega$ can be understood as the same factor that appears when applying the Huygens-Fresnel principle.

In this work, we will solve the lensing integral in terms of Eq.~\eqref{eq:contour_int}.
We will discuss its numerical implementation in Sec.~\ref{sec:integral}.

\subsection{Geometrical optics limit}

When $w$ is large, the integral in Eq.~\eqref{eq:lensing_integral} is dominated by the stationary points of $T(\vec{x}, \vec{y})$, so we can use the stationary phase approximation to obtain
\begin{equation} \label{eq:F_geom}
    F_{\rm geom}(f) = \sum_j \sqrt{|\mu_j|} e^{i w T(\vec{x}_j, \vec{y}) - i \pi n_j},
\end{equation}
where the summation is over all of the stationary points (image positions) of $T(\vec{x}, \vec{y}$).
The Morse indices $n_j$ take the values $0, \frac{1}{2}, 1$ for minima, maxima, and saddle points of $T$, respectively.
The magnification $\mu_j$ of an image is given by
\begin{equation}
    \mu_j^{-1} = {\rm det}({\rm Hess}_{\vec{x}} T(\vec{x}, \vec{y})|_{\vec{x} = \vec{x}_j}).
\end{equation}
For spherically symmetric lenses, this is explicitly given by
\begin{equation}\label{eq:spherical_magnification}
    \mu_j^{-1} = \left( 1 - \dfrac{\alpha(x)}{x} \right) \left( 1 - \dfrac{d \alpha(x)}{d x}\right),
\end{equation}
where $x = |\vec{x}|$, $\vec{\alpha}(\vec{x}) = \nabla \psi(\vec{x})$, and $\alpha(x) = |\vec{\alpha}|$.

It is evident from Eq.~\eqref{eq:spherical_magnification} that $\mu_j$ blows up when $\alpha(x) / x = 1$ or $d\alpha(x)/dx = 1$.
These are locations of fold catastrophes.
For spherically symmetric lenses, they are circles in the lens plane and they are called ``critical curves''.
If an image lies on a critical curve, the image has an infinite magnification.
The source position $\vec{y}$ and image positions $\vec{x}_j$ are related by the lens equation
\begin{equation}
    \nabla_{\vec{x}} T(\vec{x}_j, \vec{y}) = \vec{x}_j - \vec{y} - \vec{\alpha}(\vec{x}_j) = 0.
\end{equation}
Therefore, we can map the critical curves from the lens plane back into the source plane to obtain the caustics of the lens.
For spherically symmetric lenses, these are circles in the source plane with radius $y_{\rm crit}$.
For the lens profiles considered in this work, when $y < y_{\rm crit}$, there will be either two or three images, depending on whether the profile is cuspy, and when $y > y_{\rm crit}$ there will be only one image.
The case $y = y_{\rm crit}$ corresponds to a fold catastrophe where two of the images merge into one at infinite magnification.

\subsection{Lens models}

In this work, we will be using three classes of lens models.
For simplicity, we will only consider models with a single parameter $l$ (in addition to the redshifted lens mass $M_{Lz}$ and impact parameter $y$).
These will include the Navarro–Frenk–White (NFW) profile, the generalized singular isothermal sphere (gSIS) profile, and the cored isothermal sphere (CIS) profile.
Wave-optics lensing by the gSIS and CIS profiles have been studied in detail in Refs.~\cite{Tambalo:2022plm,Tambalo:2022wlm}.

\subsubsection{NFW profile}

The density profile of an NFW lens is given by~\cite{Navarro:1995iw}
\begin{equation}
    \rho(r) = \dfrac{\rho_0}{r/r_s ( 1 + r/r_s)^2},
\end{equation}
where $r_s$ is a scale radius that determines where the lens transitions from an $r^{-1}$ dependence to an $r^{-3}$ dependence, and $\rho_0$ is a normalization constant.
We define the characteristic convergence
\begin{equation}
    \kappa = \dfrac{\rho_s r_s}{\Sigma_{\rm crit}},
\end{equation}
where $\Sigma_{\rm crit} = D_S / 4 \pi D_L D_{LS} $.
If we choose $\xi_0 = \kappa$, the lensing potential can be calculated to be~\cite{Bartelmann:1996hq}
\begin{equation}
    \psi(x) = \begin{cases}
        \dfrac{\kappa}{2} \left[\left(\ln \dfrac{x}{2}\right)^2 - \left( {\rm arctanh} \sqrt{1 - x^2} \right)^2 \right] & \mbox{if} \ x \leq 1, \\
                                                                                                                        &                       \\
        \dfrac{\kappa}{2} \left[\left(\ln \dfrac{x}{2}\right)^2 + \left( {\rm arctan} \sqrt{x^2 - 1} \right)^2 \right]  & \mbox{if} \ x > 1.
    \end{cases}
\end{equation}

\subsubsection{gSIS profile}

The gSIS model is a generalization of the singular isothermal sphere model (SIS), where the slope of the profile is allowed to vary.
The density profile for a gSIS lens is given by
\begin{equation}
    \rho(r) = \rho_0 \left( \dfrac{r_s}{r} \right)^{k + 1},
\end{equation}
and the SIS profile is recovered by setting $k = 1$ and fixing the constants appropriately.
By choosing
\begin{equation}
    \xi_0 = \left(\dfrac{2 \beta_k}{2 - k} \dfrac{\rho_0 r_s}{\Sigma_{\rm crit}} \right)^{1/k} r_s,
\end{equation}
where $\beta_k = \sqrt{\pi} \Gamma(k/2) / \Gamma((k+1)/2)$ and $\Gamma$ is the gamma function,
the lensing potential is
\begin{equation}
    \psi(x) = \dfrac{x^{2 - k}}{2 - k}.
\end{equation}

When $k > 1$, the slope of the profile is steeper than the SIS profile, i.e., matter is more concentrated.
Analogous to the point-mass lens model, there will always be two images in this regime (a minimum and a saddle point in the time delay function $T$), with a cusp in the function $T$ at the center of the lens (which is different from the point mass lens case, where there will be a pole).
When $k < 1$, the profile is broader than the SIS profile. There are three images when the source is within the caustic and one image when it is outside, similar to the NFW and CIS cases.

\subsubsection{CIS profile}

The CIS model (also called the NIS, nonsingular isothermal sphere) is another modification of the SIS model, where a core replaces the singular cusp at its center.
The density profile is given by \cite{1987ApJ...320..468H}
\begin{equation}
    \rho(r) = \rho_0 \dfrac{r_c^2}{r^2 + r_c^2},
\end{equation}
where $r_c$ is the characteristic radius of the core and $\rho_0$ is a normalization constant.
Choosing the scale
\begin{equation}
    \xi_0 = \dfrac{2 \pi \rho_0 r_c^2}{\Sigma_{\rm crit}},
\end{equation}
the lensing potential is
\begin{equation}
    \psi(x) = \sqrt{x_c^2 + x^2} + x_c \log \dfrac{2 x_c}{ \sqrt{x_c^2 + x^2} + x_c},
\end{equation}
where $x_c = r_c / \xi_0$.

\section{Computing the lensing integral}\label{sec:integral}

In Sec.~\ref{subsec:time_domain_integral} we showed that the problem of computing the effect of lensing on a GW signal is reduced to computing contour integrals on a set of contour lines of the time delay function $T(\vec{x}, \vec{y})$, as in Eq.~\eqref{eq:contour_int}.
For a fine enough array in time $\tau[n]$, if the contour integral $\tilde{I}[n] = \tilde{I}(\tau[n])$ is computed, the frequency domain amplification $F(f)$ can be obtained by an inverse discrete Fourier Transform (IDFT),
\begin{equation}
    F[n] = 2 \pi i f[n] \, {\rm IDFT} \! \left[ \tilde{I}[n] \right],
\end{equation}
where $f[n]$ is the corresponding discrete Fourier Transform (DFT) frequency vector, and we have assumed that the elements of $\tau[n]$ are uniformly spaced in time.

\begin{figure*}
    \centering
    \includegraphics[width=0.99\textwidth]{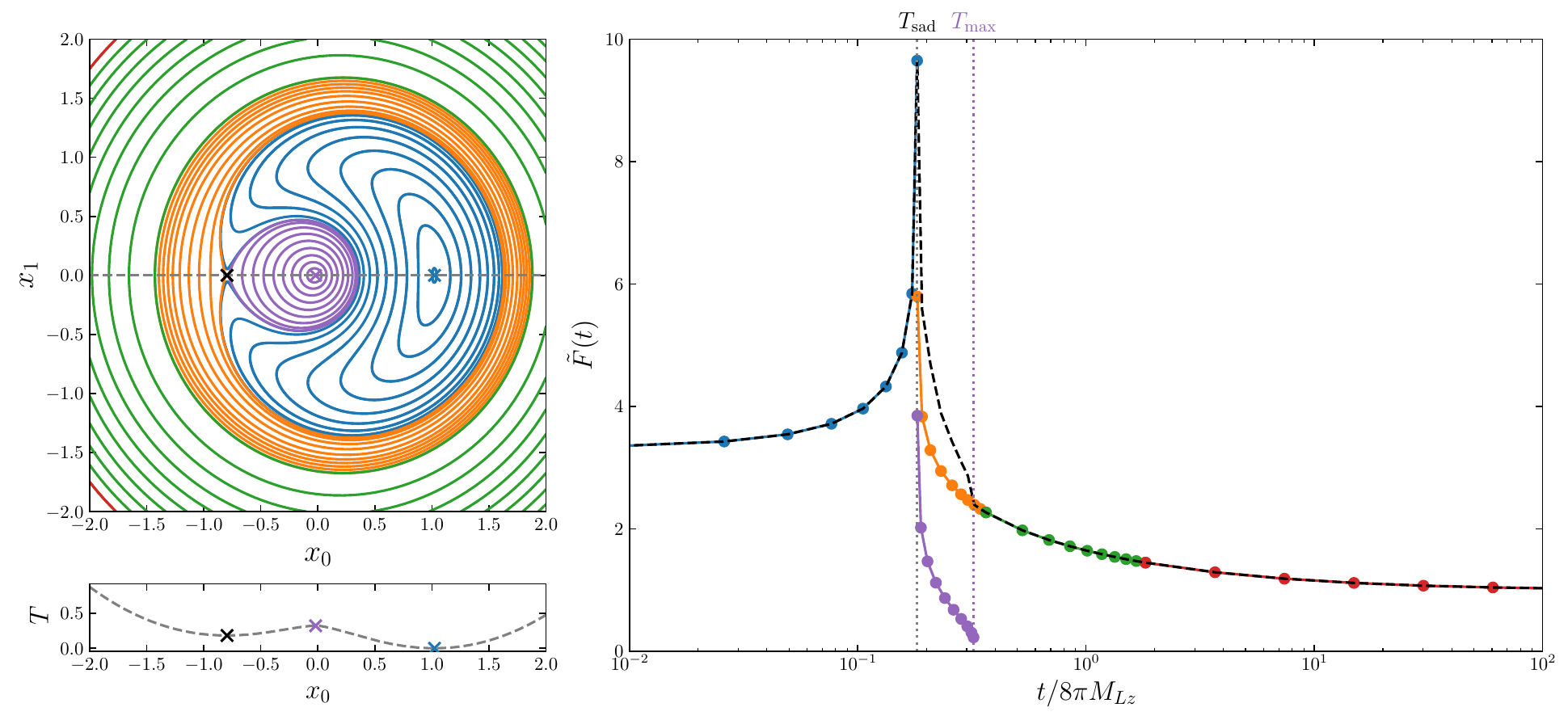}
    \caption{
        Contours of the time delay function $T(\vec{x}, \vec{y})$ and the time domain amplification factor $\tilde{I}(t)$.
        Top left: the contours levels of $T(\vec{x}, \vec{y})$. The blue, black and purple crosses label the minimum, saddle point and maximum image positions, respectively.
        Contour lines with different colors correspond to the circles of the same color in the right panel, and each color corresponds to a different segment of $\tilde{I}(t)$ where a different interpolation table will be constructed: see Sec.~\ref{subsec:interpolation}.
        Bottom left: a cross-sectional view of the top-left panel at $x_1 = 0$ (the grey dashed line), with the cross-sectional value of $T$ plotted and image positions labeled.
        Right: The time domain amplification $T(\vec{x}, \vec{y})$, obtained by integrating over the contours shown in the top left panel.
        Each circle corresponds to an integration over a contour of the same color shown in the top left panel.
        The black dashed line is the summed contribution over all contours of that particular value of $\tau$.
        The vertical dashed lines label $T_{\rm sad}$ and $T_{\rm max}$, the time delays of the saddle point and maximum images.
    }
    \label{fig:time_domain}
\end{figure*}

\subsection{Time-domain integral}

The computation of $\tilde{I}[n]$ requires summing over the contribution from disjoint contours corresponding to the same time $\tau$.
For the lens models considered, depending on the topology of the time delay function $T(\vec{x}, \vec{y})$ (i.e., the number of critical points) and the value of $\tau$, we have to either consider only one contour loop, or sum over two of them.
For the case of weak lensing (where there is only one critical point image, corresponding a minimum) there will always be only one contour loop for any $\tau$.

For all cases considered, there will be at most a minimum, a saddle point, and a maximum of $T(\vec{x}, \vec{y})$.
For noncuspy profiles (those with a finite $\rho(r)$ at $r = 0$), when the impact parameter $y < y_{\rm crit}$, there will be three images (a minimum, a saddle point, and a maximum), while when $y > y_{\rm crit}$ there will only be one image (a minimum).
However, when the profile is cuspy (e.g. for the gSIS lens, or the SIS limit of the CIS lens) the maximum is replaced by a cusp, so there will only be a minimum and saddle point image for $y < y_{\rm crit}$ in this case.

In Fig.~\ref{fig:time_domain} we show the contour line topology for an NFW lens in the strong lensing regime (three images).
Due to the spherical symmetry of the lens profile, we can assume that the source position lies on the horizontal axis, i.e., $\vec{y}
    = (y_0, y_1) = (y_0, 0)$.
Then, the images will also lie on the horizontal axis, and we can solve for their positions with Newton's method, or any 1D root-finding algorithm.
The contour lines for any constant $\tau$ will pass through the horizontal axis twice.
As shown in the top left panel of Fig.~\ref{fig:time_domain}, all of the blue and purple contour lines will pass through the line between the minimum and maximum images once, and the other contours will pass through the line extending from the saddle point image to negative spatial infinity once, meaning that we will be able to locate a point on any contour line by using a 1D root-finding algorithm.
Once a point is located, we can trace out the corresponding contour line by using the Runge-Kutta method while performing the contour integral.
The specifics of our implementation are explained in Appendix~\ref{app:contour_step}.

When $T_{\rm min} < \tau < T_{\rm sad}$ or $\tau > T_{\rm max}$, there is only one contour loop (the blue loops for the former case, and the green or red in the latter).
However, if we follow the evolution of the contour lines from low to high $\tau$, at $\tau = T_{\rm sad}$ the contour lines break into two, and they merge back together at $\tau = T_{\rm max}$.
Therefore, for $T_{\rm sad} < \tau < T_{\rm max}$, we will have to sum over the two contour loops when computing $\tilde{I}(\tau)$.
As shown in the right panel of Fig.~\ref{fig:time_domain}, $\tilde{I}$ is discontinuous at $T_{\rm sad}$ and $T_{\rm max}$.
Between $T_{\rm sad}$ and $T_{\rm max}$, we have to sum over the contributions of the orange and purple contours to obtain the full $\tilde{I}(\tau)$, the black dashed line.
For the case of single-image weak lensing, there will always be only one contour for every $\tau$.

In Fig.~\ref{fig:time_domain}, we do not use an array of $\tau$s that are uniformly spaced.
This is because $\tilde{I}(\tau)$ varies faster in the vicinity of an image, so it is better to use a finer resolution in $\tau$ close to the arrival time of the images.
Before performing the IDFT we will interpolate the results back onto a uniform grid in $\tau$.

\subsection{Transforming to the frequency domain}

To obtain the frequency-domain amplification $F(w)$, we Fourier transform the derivative of $\tilde{I}(\tau)$.
As $\tilde{I}(\tau)$ is computed on a discrete set of $\tau$'s, we will need to resort to a numerical differentiation scheme if we want to work with the derivative directly.
However, as only the Fourier transform of the derivative is required, we can instead multiply by a factor of $i w$ after transforming $\tilde{I}(\tau)$ to compensate for the derivative.

From Fig.~\ref{fig:time_domain} it is apparent that $\tilde{I}(\tau)$ does not tend to the same limit as $\tau \to \pm \infty$.
In fact, $\tilde{I} = 0$ for $\tau < 0$, i.e. before the signal has arrived (not shown in the figure), and $\tilde{I} \to 1$ for $\tau \to \infty$.
The difference between the two limits would cause spectral leakage in the Fourier transform.
A way to mitigate this effect is by using a windowing function~\cite{Diego:2019lcd,Cheung:2020okf}, but we found that subtracting a constant before Fourier transforming as implemented in Ref.~\cite{Shan:2022xfx} gives more accurate results.
Explicitly, we perform the transformation
\begin{equation}
    F(w) = i w \, {\rm IFT} [\tilde{I}(\tau)] + \sqrt{\mu_{\rm min}},
\end{equation}
where $\mu_{\rm min}$ is the magnification of the minimum image, and $\rm IFT$ denotes the inverse Fourier transform, which is performed numerically as a discrete inverse Fourier transform.
In other implementations of the contour integration method, e.g. in Refs.~\cite{Ulmer:1994ij,Mishra:2021xzz,Tambalo:2022plm,Tambalo:2022wlm}, the discontinuities attributed to the saddle point and maximum images are also removed in a similar manner, but we find that we do not need to do that to achieve satisfactory accuracy for our purposes.

In practice, we would like to target a wide range of $M_{Lz}$ spanning orders of magnitude, and the unlensed GWs also span orders of magnitude in frequency.
Therefore, we will also have to compute $F(w)$ over orders of magnitude in $w$.
As the frequency bins of DFT results are in linear scale, we will have to patch together multiple results at different orders of magnitude to efficiently compute $F(w)$ over the required range.
We will divide our frequency space into four domains, in ascending order in $w$.
We first define $T_{\rm high}$ to be the latest time of occurrence of a significant feature (peaks or kinks) in $\tilde{I}(\tau)$.
Explicitly,
\begin{equation}
    T_{\rm high} = \begin{cases}
        \max (T_{\rm sad}, T_{\rm max}) \quad & \text{for strong lensing}, \\
        \max_\tau \tilde{I}(\tau) \quad       & \text{for weak lensing}.
    \end{cases}
\end{equation}
In the first domain, we take $F(w) = 1$, because $F(w) \to 1$ and $w \to 0$.
In the second domain, we compute $F(w)$ by Fourier transforming $\tilde{I}(\tau)$ over $N$ equally-spaced (interpolated) sampling points with $\tau$ between $0$ and $20 T_{\rm high}$.
The third domain is computed similarly to the second, but with the higher bound of $\tau$ extended to $\min (2000 T_{\rm high}, 1000)$.
In the fourth domain, we use the geometrical optics approximation to compute $F(w)$, i.e., Eq.~\eqref{eq:F_geom}.
Three transition frequencies $w_1, w_2, w_3$ are required to define the boundaries between the four domains in frequency space.
We use the lowest positive frequency within the DFT frequency bins as $w_1$, while we use $w_2 = 2.5 / T_{\rm high}$ and $ w_3 = 250 / T_{\rm high}$ or $50 / T_{\rm high}$ for strong lensing and weak lensing, respectively.
We need to use a lower value of $w_3$ (i.e., transition to the geometrical optics limit earlier) for weak lensing because the phase is not recovered as accurately at high frequencies with our full wave-optics calculations.

The above procedure works for any spherically symmetric lens model with a single image (minimum) or three images (minimum, saddle, maximum).
However, if the center of the profile is a cusp, e.g. for the gSIS lens, the central image will be replaced by a nonsmooth kink.
This only requires a minor modification of the procedure above, because the topology of the contour lines still follows that shown in the top left panel of Fig.~\ref{fig:time_domain}.
While the root-finding algorithm will not be able to identify the kink at the center, we can simply specify by hand that there is effectively an ``image'' at the origin and the contour lines have morphologies similar to the case when the image was smooth.
On the other hand, we keep in mind that we should not include the contribution of such an ``image'' when computing the geometrical optics amplification, because it is not a true image after all.

\begin{figure*}
    \centering
    \includegraphics[width=0.99\textwidth]{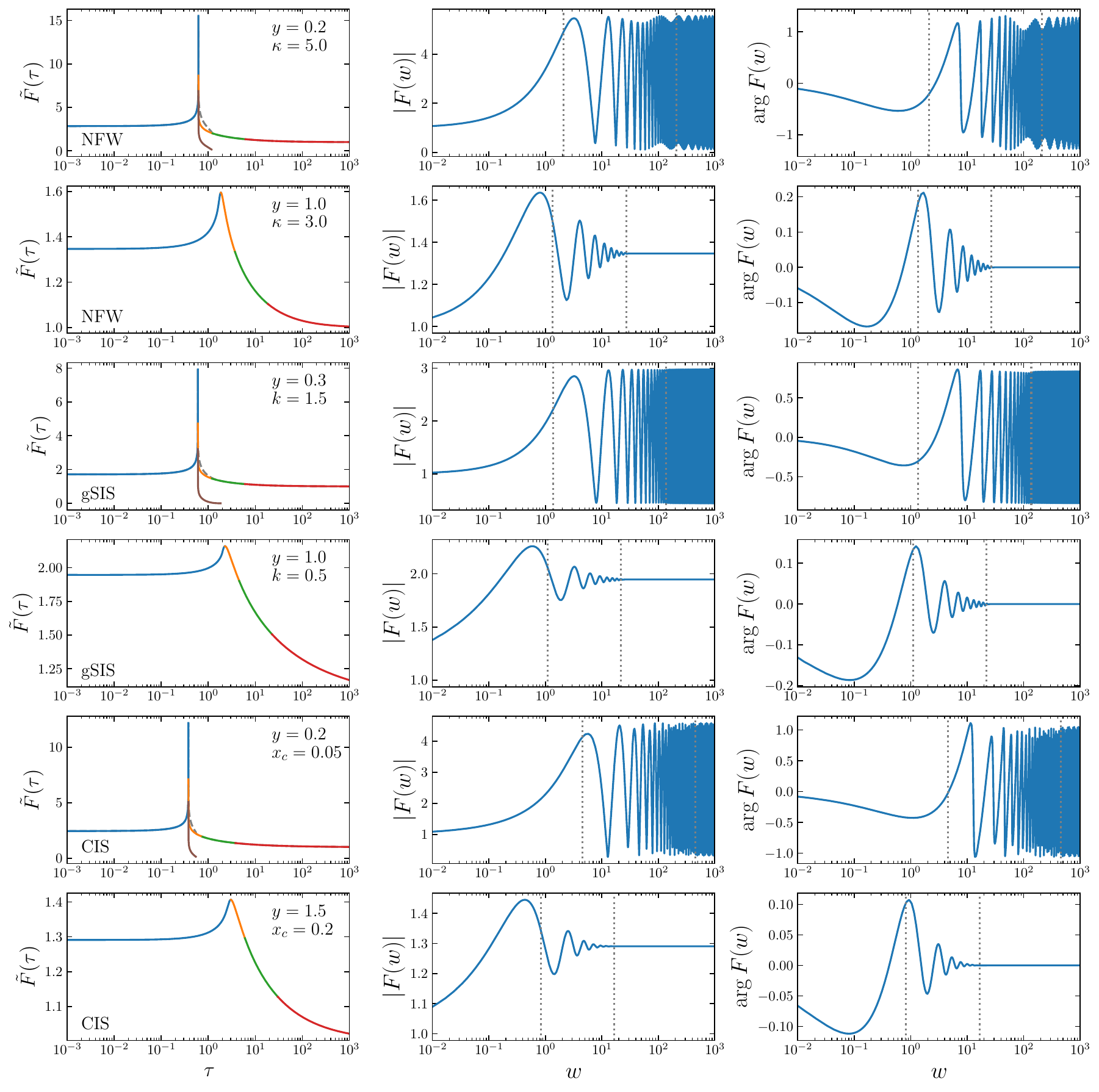}
    \caption{
        Examples of amplification factors for the NFW (first two rows), gSIS (middle two rows) and CIS (bottom two rows) lens models.
        The first row for each lens model corresponds to the strong lensing case with multiple images, while the second row shows the weak lensing case with a single image.
        We show the time domain amplification $\tilde{I}(\tau)$ (left column) as well as the magnitude (middle column) and phase (right column) of the frequency domain amplification $F(w)$.
        Curves with different colors in the left column correspond to different segments with a separate interpolation table (see Sec.~\ref{subsec:interpolation}), while the black dashed lines for the strong lensing cases are the values of $\tilde{I}(\tau)$ summed over all contours at the same $\tau$.
        The horizontal grey dashed lines in the middle and right columns correspond to the transition frequencies where we patch together results in different frequency domains, with the highest frequency domain using the geometrical optics results.
    }
    \label{fig:freq_domain}
\end{figure*}

Using the above procedure, we can produce the frequency domain amplification $F(w)$ for the NFW, gSIS and CIS lenses, both in the case of single-image weak lensing and multiple-image strong lensing, and the resulting time domain and frequency domain amplifications are plotted in Fig.~\ref{fig:freq_domain}.
For the strong lensing case, we always have to sum over two contour lines when we are in the regime $T_{\rm sad} < \tau < T_{\rm max}$, where $T_{\rm max}$ is either the arrival time of the maximum image or the corresponding time delay of the kink at the origin, if it exists.
The frequency domain amplification is oscillatory in the high-frequency limit, because it corresponds to interference of multiple images, which can also be seen in Eq.~\eqref{eq:F_geom}.
The amplification oscillates at intermediate frequencies but approaches a constant at high $w$, corresponding to $\sqrt{|\mu_{\rm min}|}$.
It is the intermediate oscillations due to the wave-optics effects that encode the information about the lens.
If the wavelength is short (hence the frequency is high) and wave-optics effects are unimportant, the amplification is constant, so the lensing effects will be completely degenerate with the luminosity distance.
Hence, wave-optics effects help us identify lensed events, even when there is only a single image.

\subsection{Time-domain interpolation}\label{subsec:interpolation}

The computation of the frequency-domain integral takes $O(1)$ seconds on a GPU.
While further optimizations could improve the speed, this is still inadequate for performing full Bayesian parameter estimation on detected GW events.
For the spherically symmetric lens models considered in this work, we only have to consider three parameters that are related to the lensing set up: the redshifted lens mass $M_{Lz}$, the source position $y = |\vec{y}|$, and the additional parameter that characterizes the lens profile, i.e., $\kappa$ for NFW, $k$ for gSIS, and $x_c$ for CIS (recall that we call such a lens parameter collectively as $l$, where $l \in \{\kappa, k, x_c\}$ depending on the lens model of concern).
The lens mass $M_{Lz}$ only affects which frequency range $w = 8 \pi M_{Lz} f$ to use, given a frequency $f$.
Therefore, the problem of computing amplification factors is essentially a 2D problem in parameter space with an additional dimension in frequency, so the dimension is low enough that we can construct an interpolation table for rapidly calling the results.

A challenge of building an interpolation table is that the amplification factor is oscillatory in the frequency domain.
When performing parameter estimation, we might want our prior range to span orders of magnitude in $M_{Lz}$, and when $M_{Lz}$ is large we will reach the high-$w$ oscillatory regime.
This could require a lot of points in $w$ when constructing the interpolation table, which could become a bottleneck in terms of speed and memory.
However, as shown in Figs.~\ref{fig:time_domain} and \ref{fig:freq_domain}, the amplification in the time domain is a piece-wise smooth and nonoscillatory function, while the transition boundaries between the pieces are the image time delays, which are easy to compute.
Therefore, we can build an interpolation table for $\tilde{I}(\tau)$, and we can perform a (relatively computationally cheap) Fourier transform to obtain $F(w)$ every time we want to call the amplification factor.

To avoid the nonsmooth features in $\tilde{I}(\tau)$, we divide the function into different pieces with boundaries related to the image time delays.
These different segments are plotted with different colors in Figs.~\ref{fig:time_domain} and \ref{fig:freq_domain}.
For the multiple image strong-lensing case, they correspond to sets of contour lines with the same topology around the images, with additional segments divided at different orders of magnitude for $\tau$ to adjust the resolution in time.
For example, the blue segment corresponds to contour lines with $T_{\rm min} < \tau < T_{\rm sad}$ enclosing the minimum image, the orange and purple segments correspond to those with $T_{\rm sad} < \tau < T_{\rm max}$, with the purple one corresponding to the contours immediately surrounding the maximum image, and the green and red contours correspond to the contours with $\tau > T_{\rm max}$ surrounding all of the images, approximately larger and larger circles as $\tau \to \infty$.
We transition from the green to the red segment to adjust the resolution of the interpolation nodes in time as $\tau$ increases.
For the weak lensing case, while all of the contours have the same morphology (they surround the minimum image and become approximately circular as $\tau \to \infty$ without ever breaking into multiple contours), we still divide $\tilde{I}(\tau)$ into segments because there could be a rather sharp peak at an intermediate time (see, e.g. the second panel on the left column of Fig.~\ref{fig:freq_domain}), and we also want to change the resolution of the interpolation nodes when $\tau$ increases by orders of magnitude.
To avoid the potentially sharp peak, we divide between the blue and orange segments at the location where $\tilde{I}(\tau)$ attains a maximum value.
This location in time is nontrivial and has to be located numerically.
We then divide between the orange, green and red segments approximately by the order of magnitude of $\tau$.

An interpolation scheme works well because each segment and their boundaries (if defined appropriately) vary smoothly with the lensing parameters $l$ and $y$.
The interpolation will then proceed in two steps: interpolating the values of the boundaries, and then interpolating $\tilde{I}(\tau)$ itself for each segment.

\begin{figure*}
    \centering
    \includegraphics[width=0.99\textwidth]{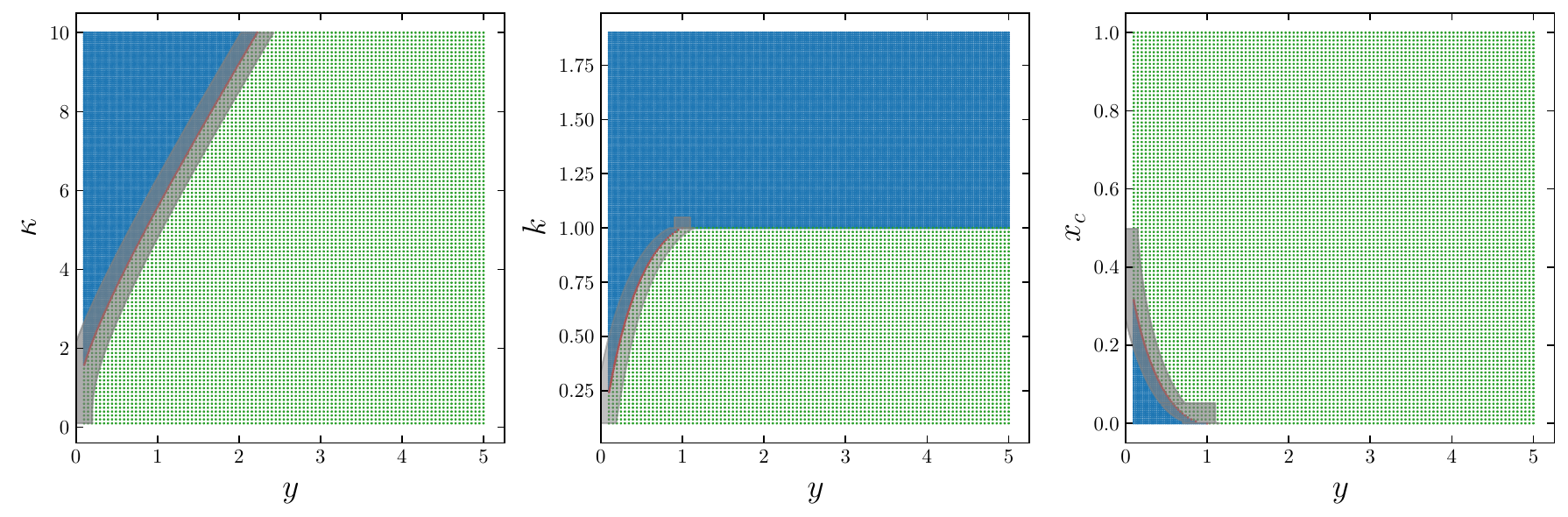}
    \caption{
        The interpolation domains used for parameter estimation of the NFW (left), gSIS (middle) and CIS (right) lenses.
        Each circle is a node for interpolation, where the time domain amplification and image time delays are computed.
        The blue, red and green points correspond to strong lensing, critical, and weak lensing points.
        We use double the resolution in each dimension within strong lensing regions.
        The grey regions are excision regions where we set the prior for parameter estimation to zero, because the results in those regions are pathological due to diverging behaviors near the caustic.
    }
    \label{fig:interp_domain}
\end{figure*}

The interpolation procedure is slightly different between the strong-lensing and weak-lensing cases because of the difference in contour line topology.
Therefore, we have to divide the 2D lensing parameter space into these two regimes when constructing the interpolation table.
Such a division can naturally be imposed by considering the caustic curves $y_{\rm crit}(l)$, which depend on the lens parameter $l$.
For $y < y_{\rm crit}$ we are in the strong lensing regime, and for $y > y_{\rm crit}$ we are in the weak lensing one.
The critical regime $y \approx y_{\rm crit}$ is pathological because the magnification of the image(s) blows up to infinity.
While this does not happen in physical scenarios due to the spatially finite nature of the source and diffraction effects, the geometrical optics computations will be affected, so we cannot transition to Eq.~\eqref{eq:F_geom} in that limit.
In principle, we can include diffraction corrections to Eq.~\eqref{eq:F_geom} in the critical limit.
However, given that the $y \approx y_{\rm crit}$ region does not take up a significant portion of parameter space, we choose to defer the treatment of this region to future work.
In the current work, we will excise this region from the parameter space by setting the prior probability to zero in the region when performing parameter estimation.

In Fig.~\ref{fig:interp_domain} we show the interpolation nodes in the 2D parameter space of $l$ and $y$ for all three lens models considered.
We excised the shaded region close to the caustic $y_{\rm crit}(l)$, and divided the whole parameter space into a strong lensing regime and a weak lensing regime.
We use double the resolution in each dimension for the strong lensing case because the quantities to be interpolated tend to vary more rapidly in that regime.

\section{Bayesian Inference of the lensing parameters}\label{sec:PE}

Given an interpolation table for $F(f)$ and a waveform model for $\tilde{h}(f)$, we can rapidly call the lensed waveform $\tilde{h}_L(f)$, and we can use standard sampling techniques to infer the properties of detected lensed GW events.
In this work we perform an injection-recovery analysis with the \texttt{Bilby} Bayesian inference package.

To test the observability of lensing effects, we perform injection runs for both the strong and weak lensing cases for all three lens models.
For the weak lensing cases, we perform injections with different source positions $y$ to determine the critical value of $y$ for which lensing is observable.
For the gSIS lens we also perform runs at the SIS limit ($k = 1$) between the weak and strong lensing regime.

\begin{figure*}
    \raggedright
    \includegraphics[width=0.32\textwidth]{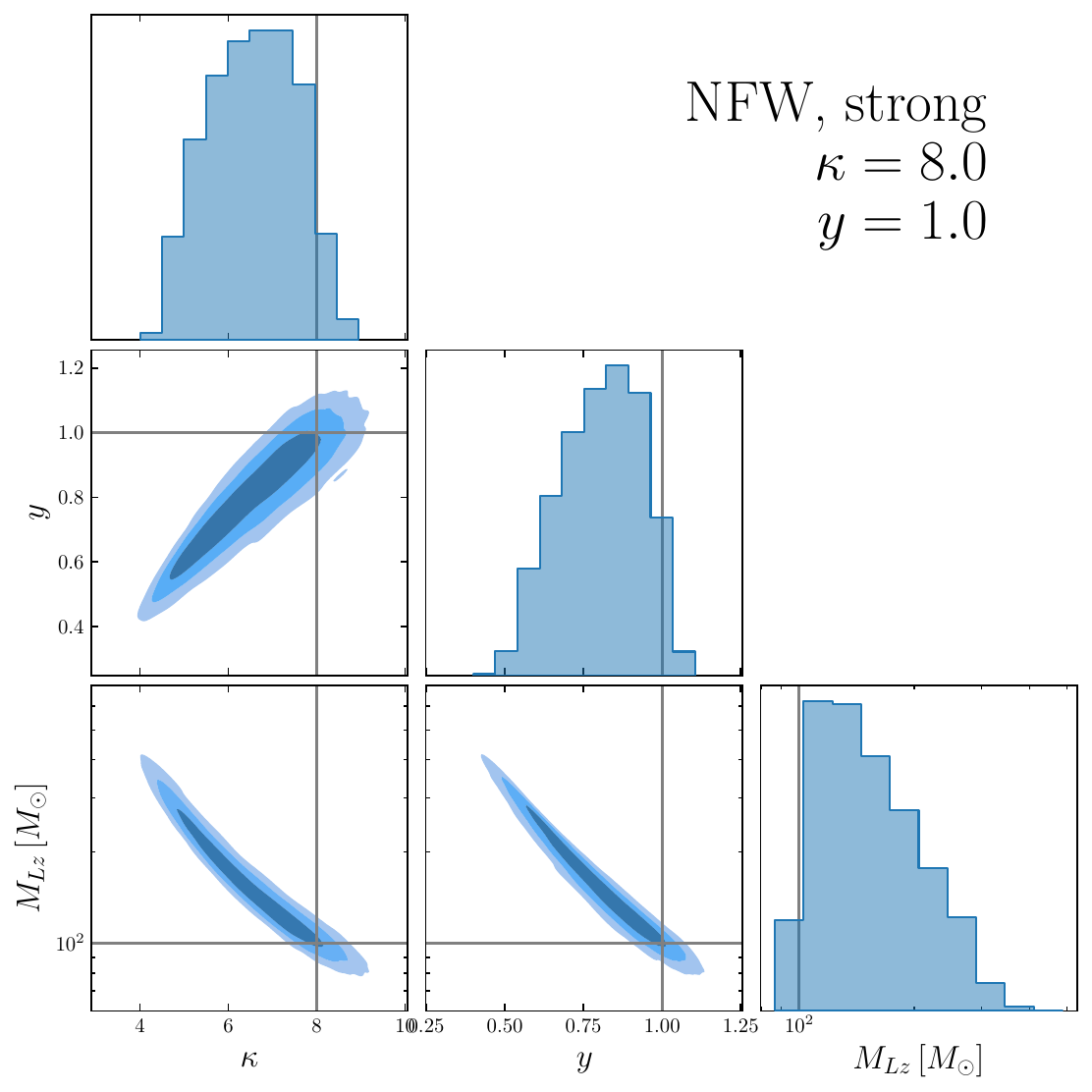} \\
    \includegraphics[width=0.32\textwidth]{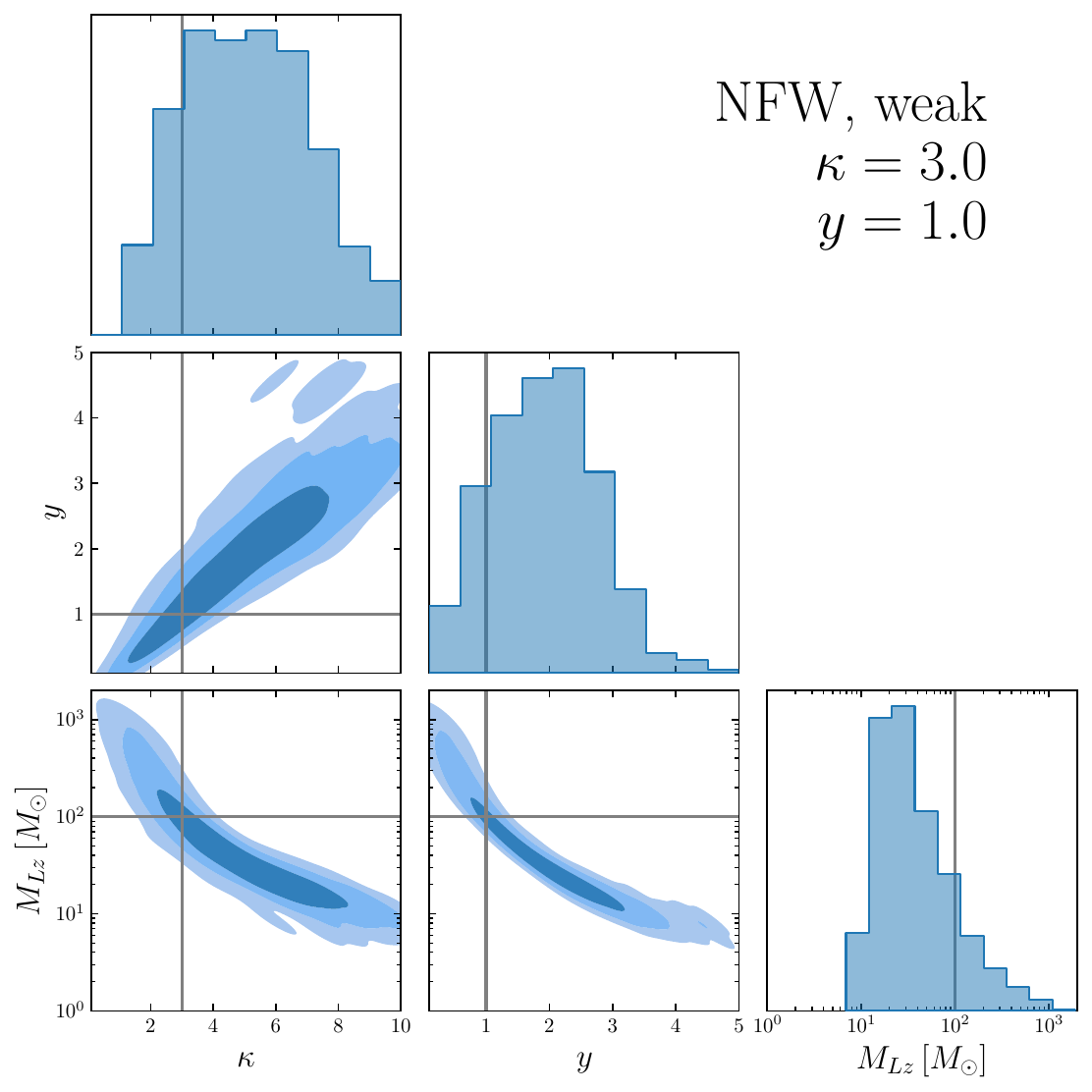}
    \includegraphics[width=0.32\textwidth]{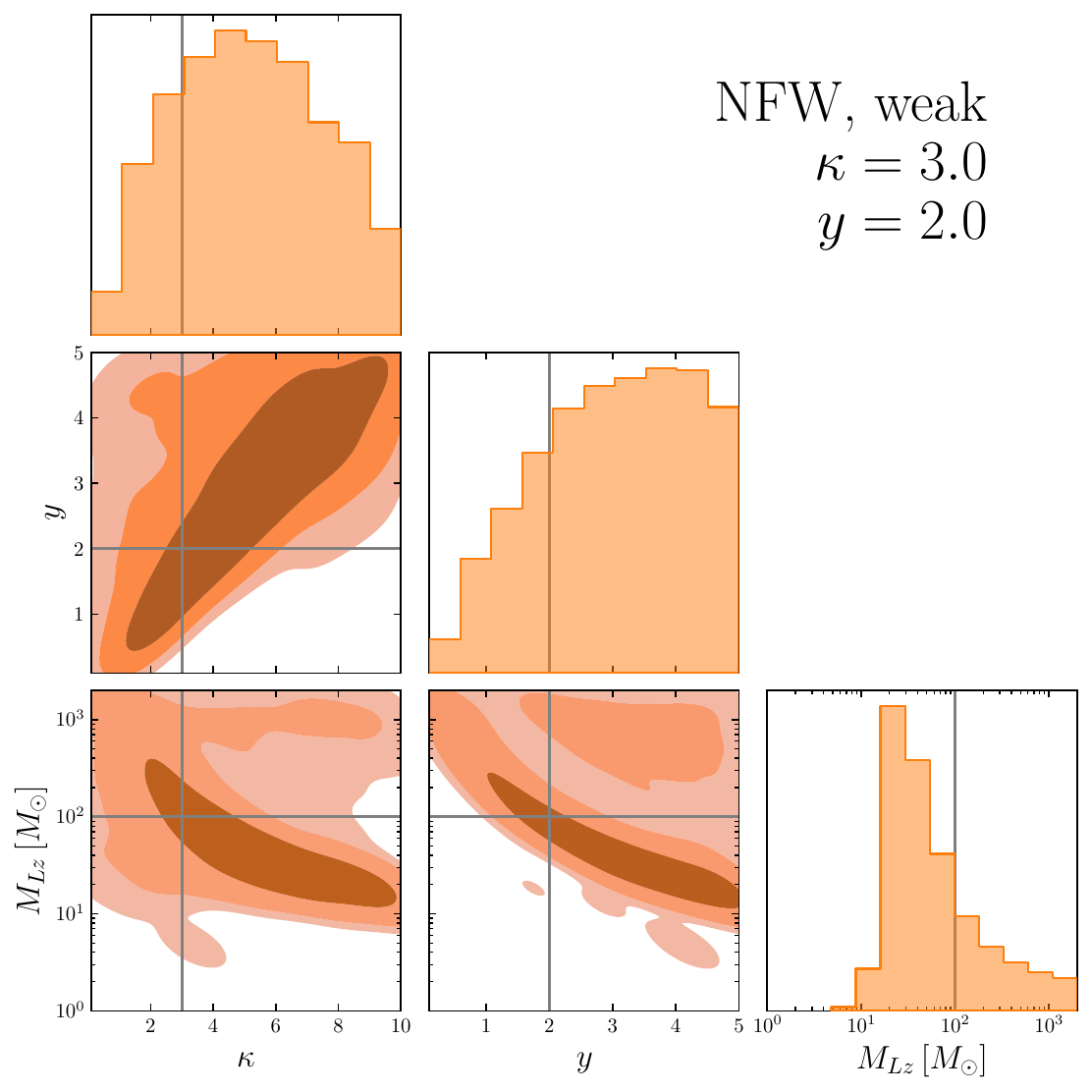}
    \includegraphics[width=0.32\textwidth]{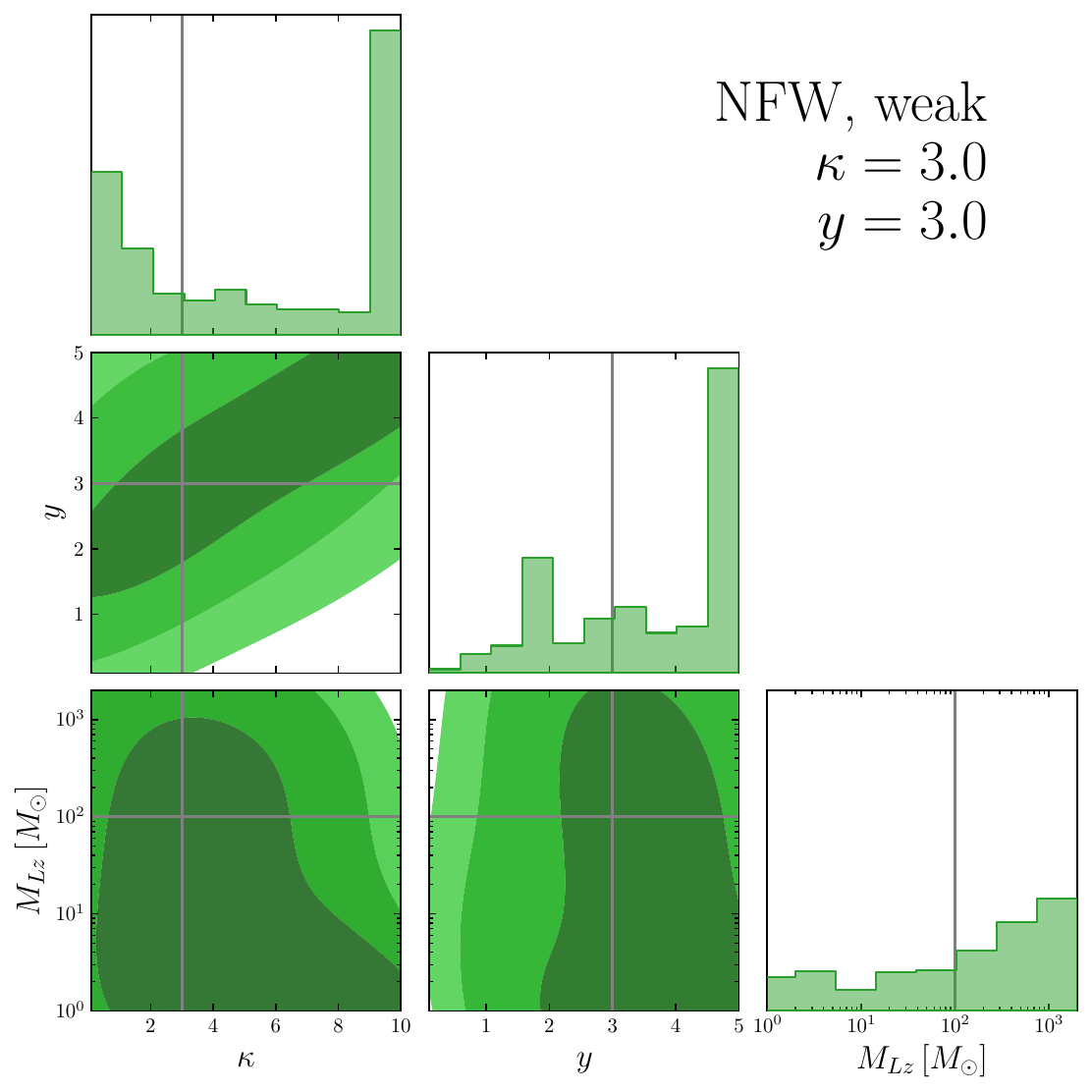}
    \caption{
        Parameter estimation results for the NFW lens in the strong (top row, one panel) and weak (bottom row, three panels) limit.
        For $\kappa = 3.0$ and $y = 3.0$ (bottom right panel), the strong spike in the posterior at the upper bounds of $\kappa$ and $y$ is an artifact of prior reweighting.
        The injected values are marked by the grey lines.
    }
    \label{fig:PE_NFW}
\end{figure*}

\begin{figure*}
    \raggedleft
    \includegraphics[width=0.32\textwidth]{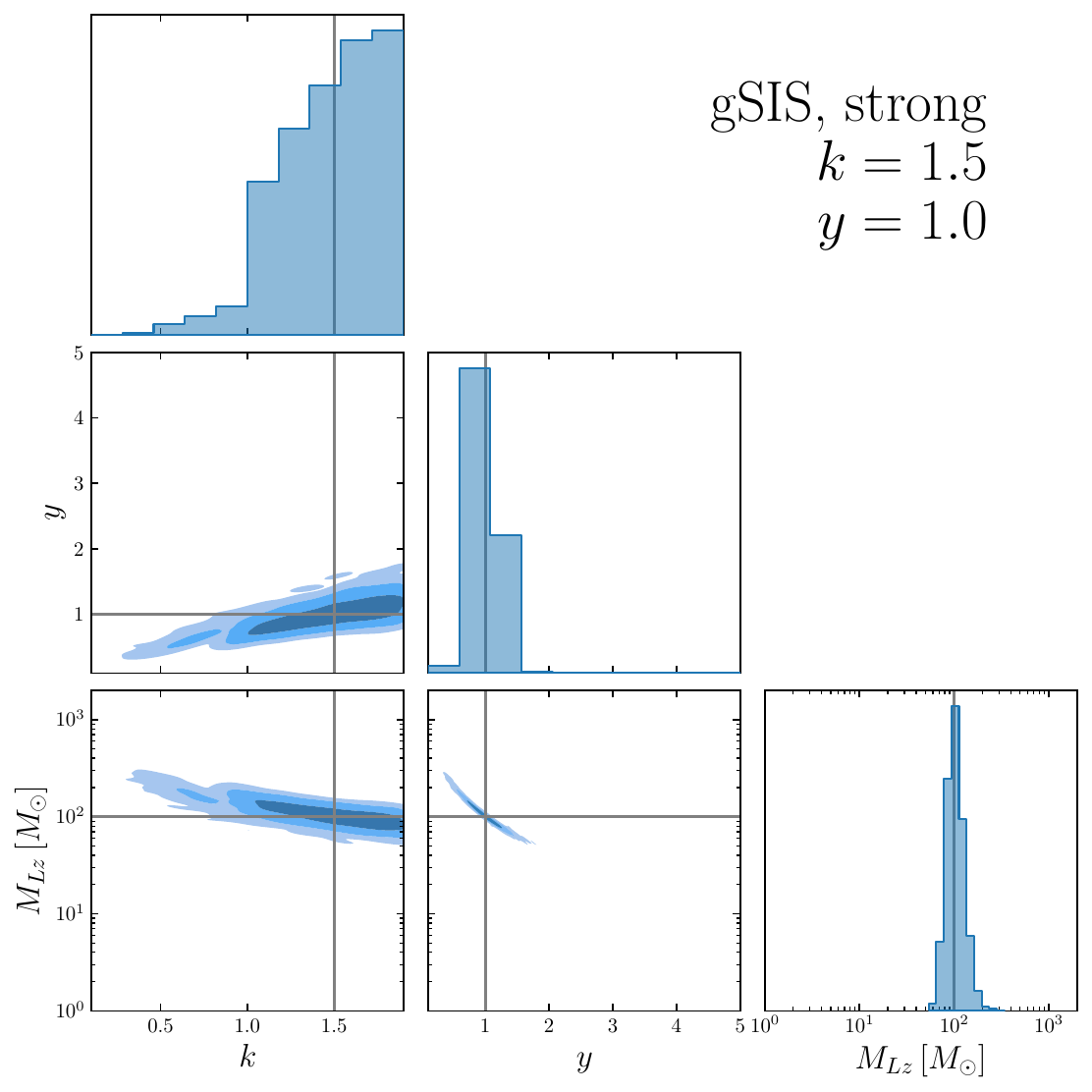}
    \includegraphics[width=0.32\textwidth]{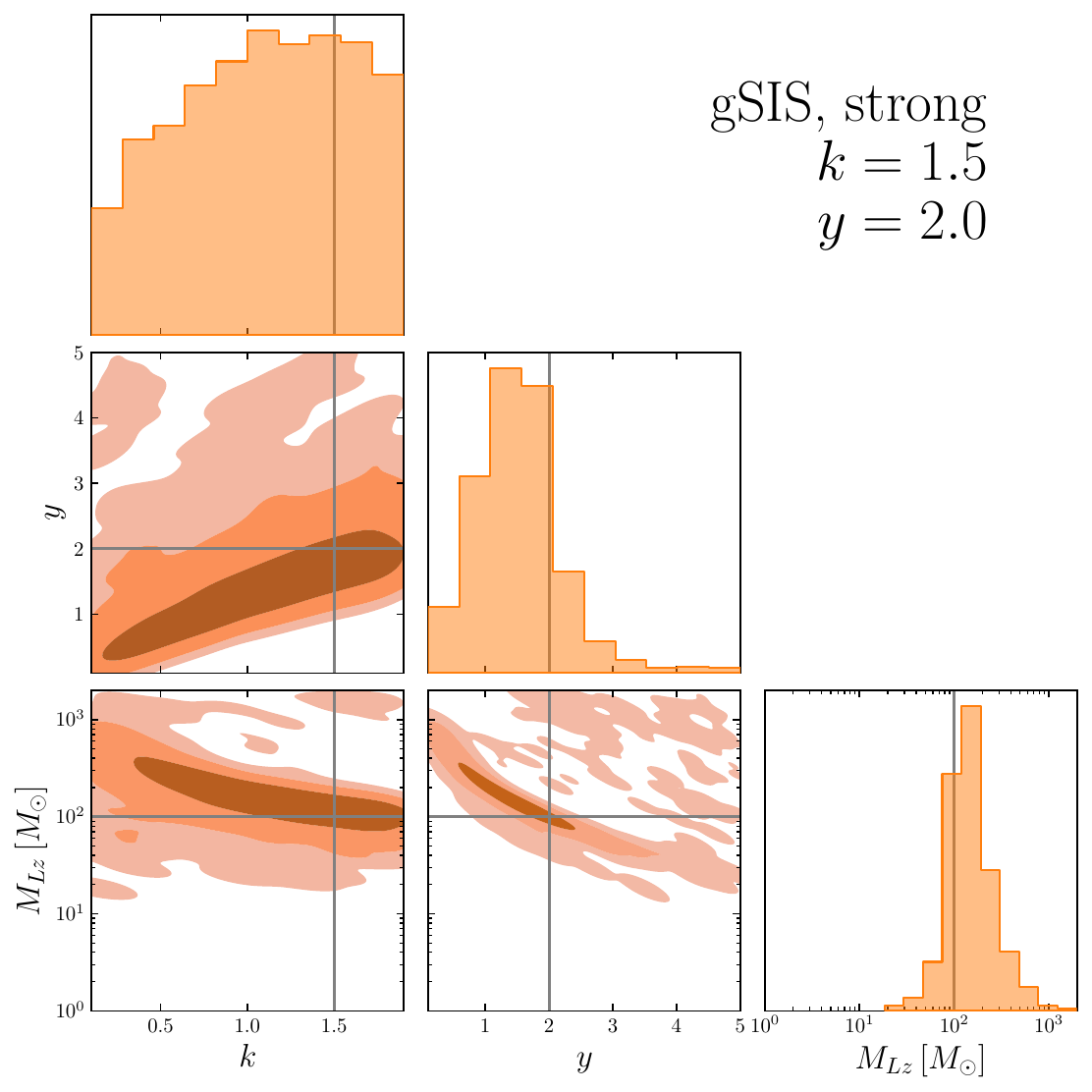}
    \includegraphics[width=0.32\textwidth]{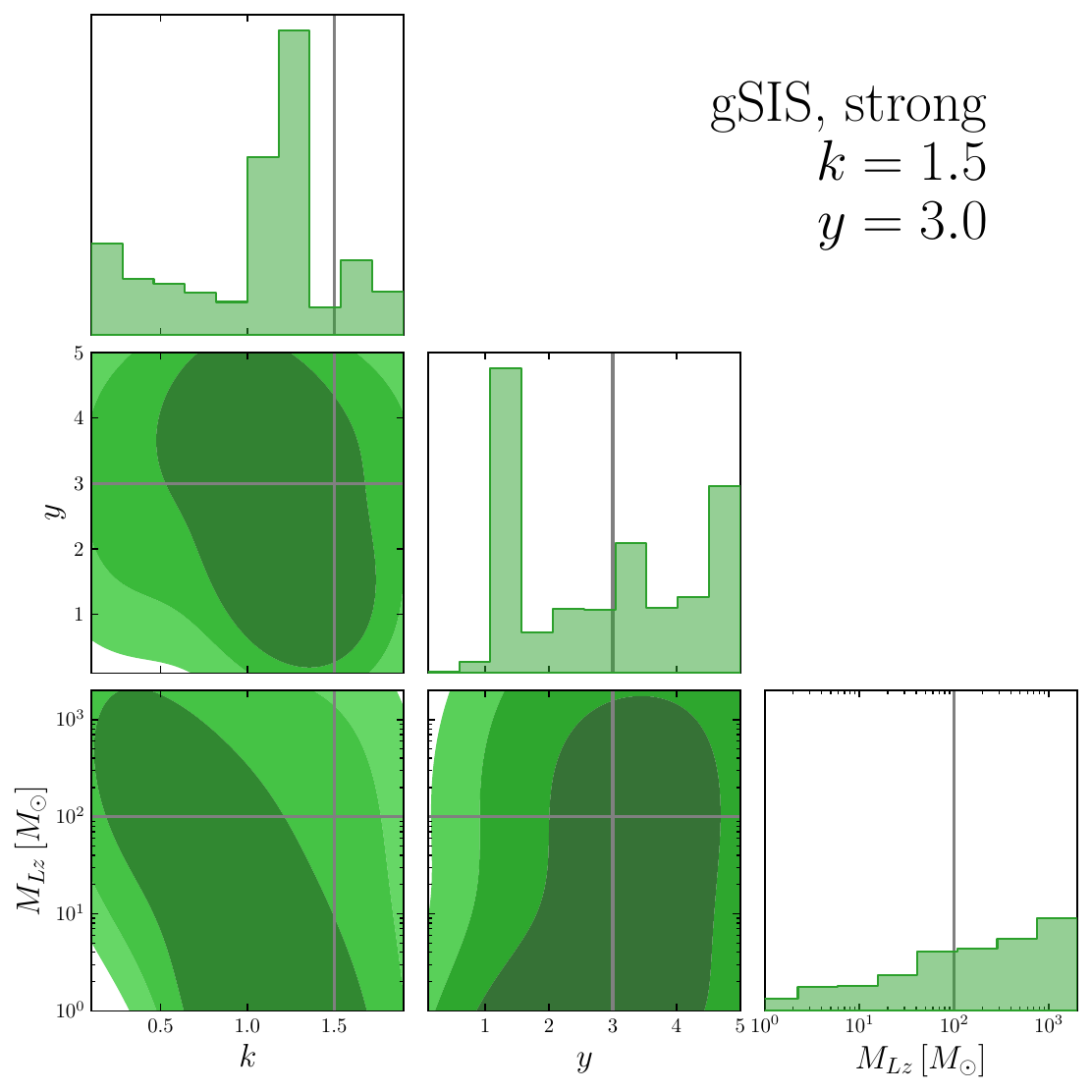} \\
    \includegraphics[width=0.32\textwidth]{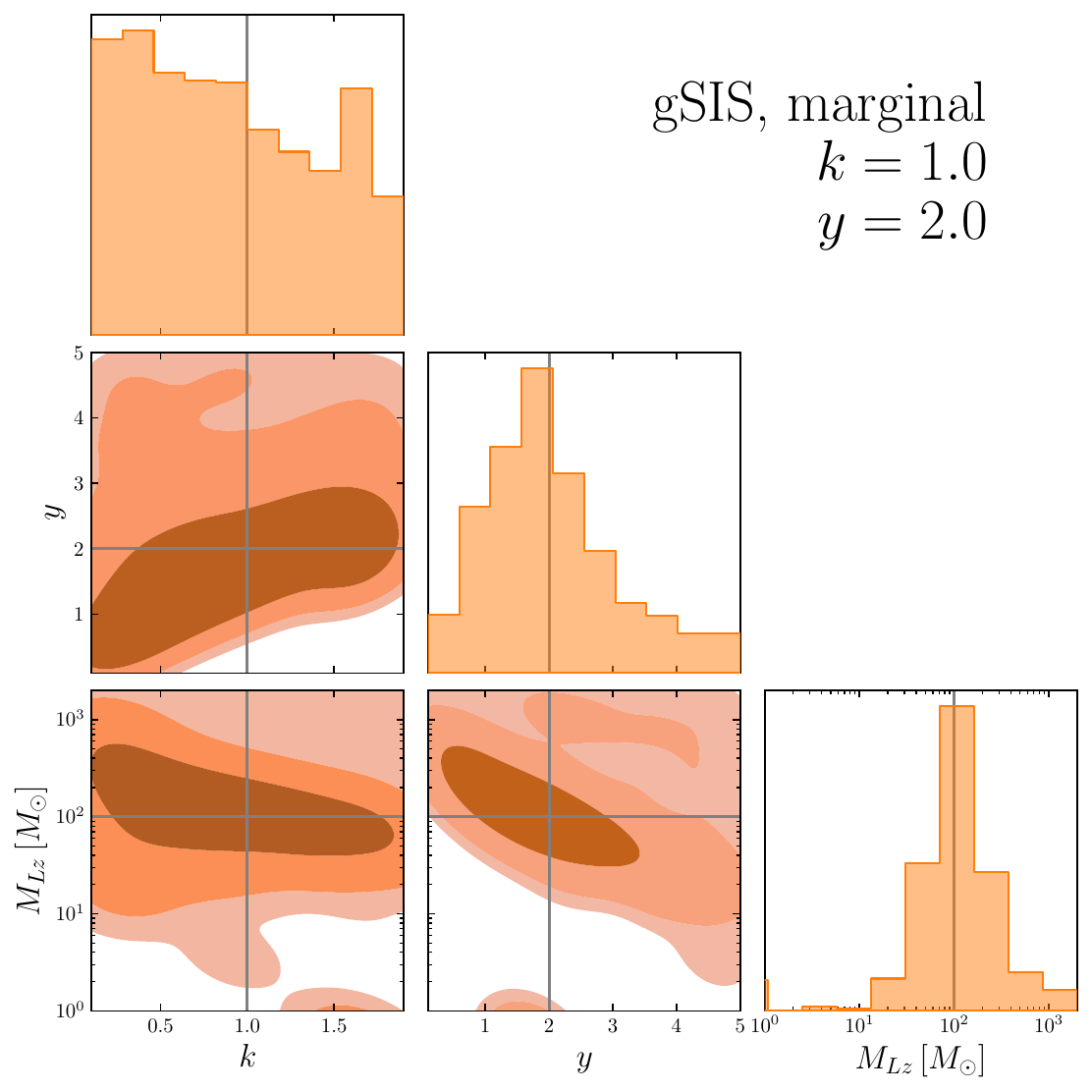}
    \includegraphics[width=0.32\textwidth]{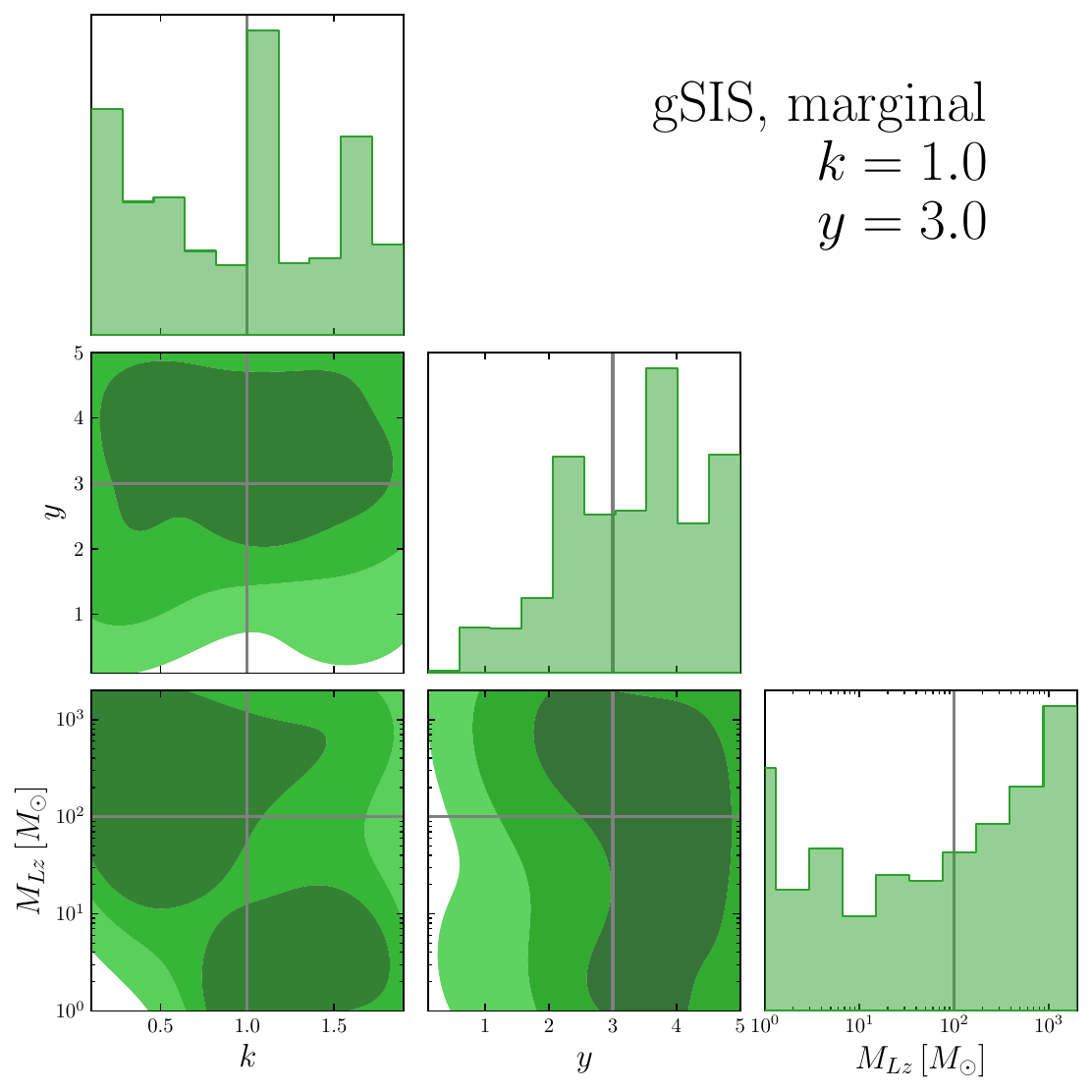}
    \\
    \includegraphics[width=0.32\textwidth]{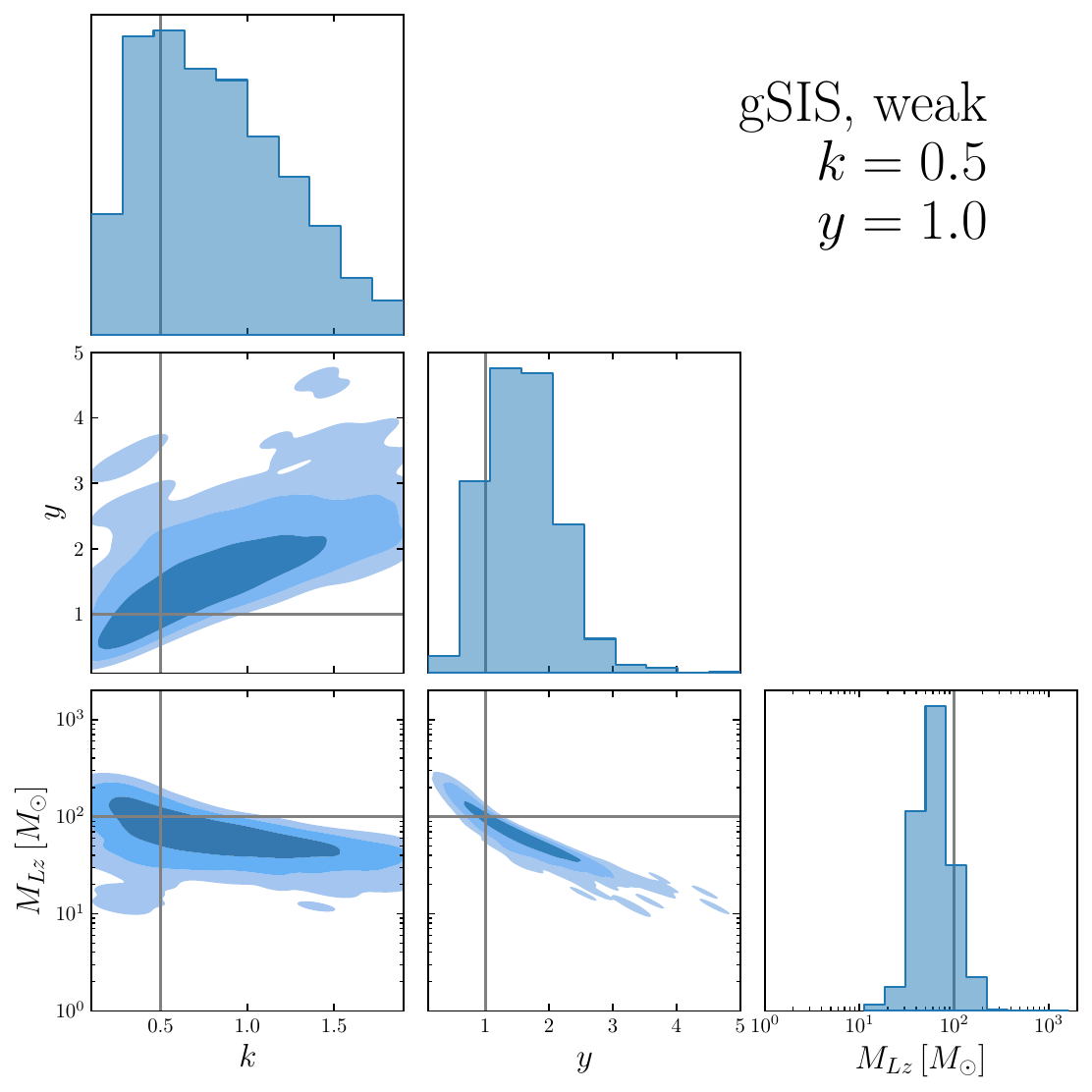}
    \includegraphics[width=0.32\textwidth]{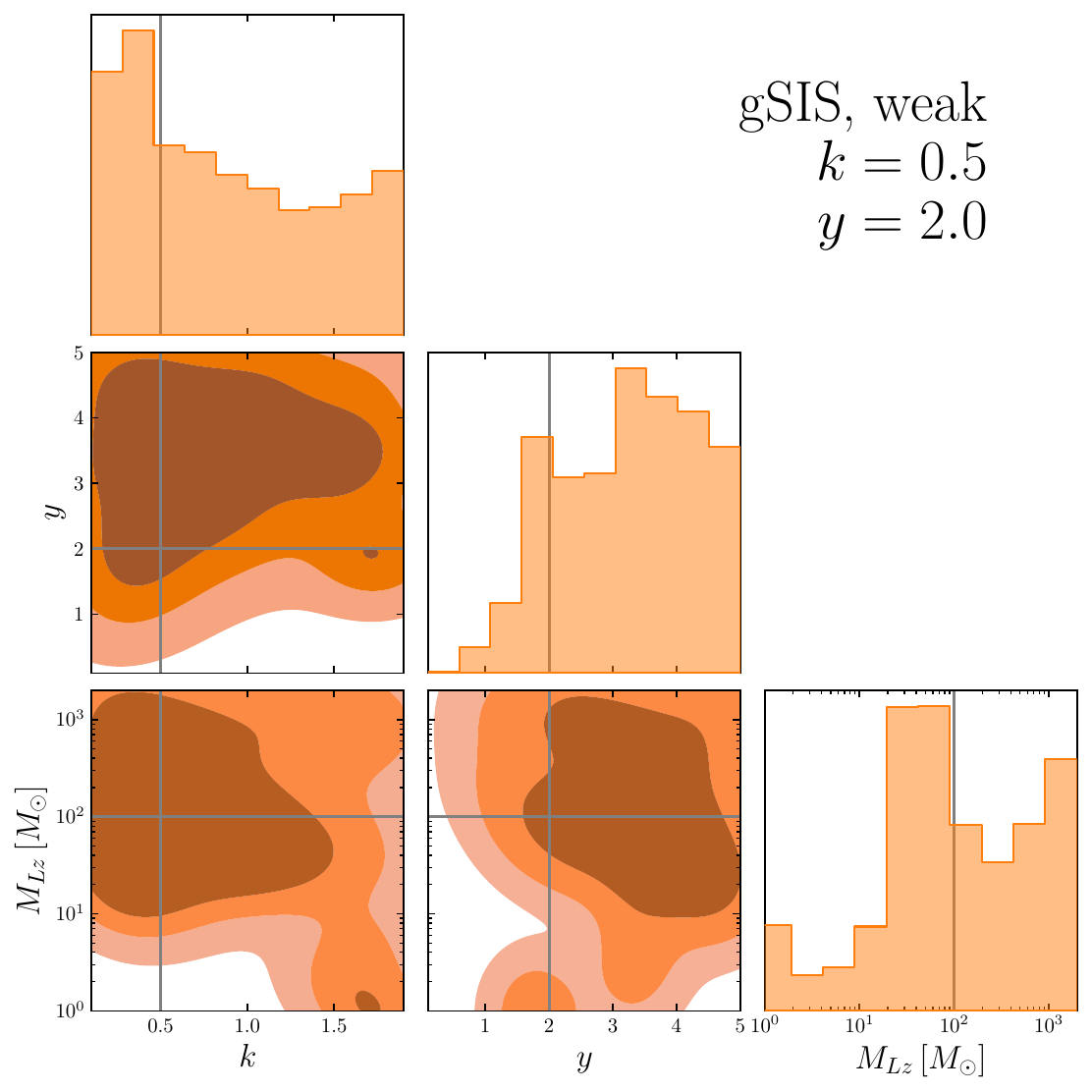}
    \includegraphics[width=0.32\textwidth]{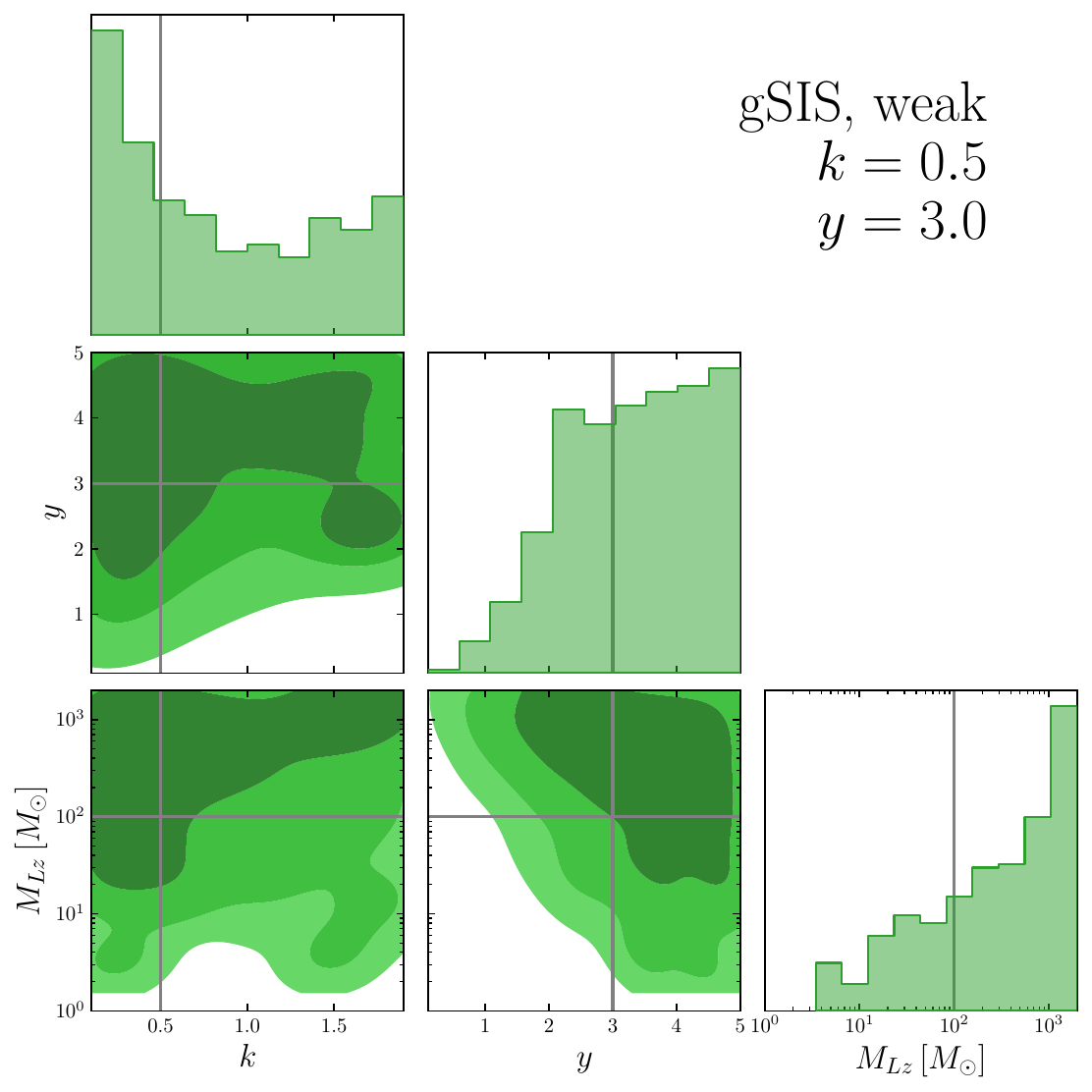}
    \caption{
        Parameter estimation results for the gSIS lens in the strong (top row, three panels), marginal (middle row, two panels) and weak (bottom row, three panels) lensing limit.
        The $k = 1.0$ marginal cases (middle row) correspond to the SIS lens, where there are two images and a central cusp on the lens plane.
        We skipped the $k = 1.0, y = 1.0$ case because the source lies on the caustic of the SIS lens.
        The injected values are marked by the grey lines.
    }
    \label{fig:PE_gSIS}
\end{figure*}

\begin{figure*}
    \includegraphics[width=0.32\textwidth]{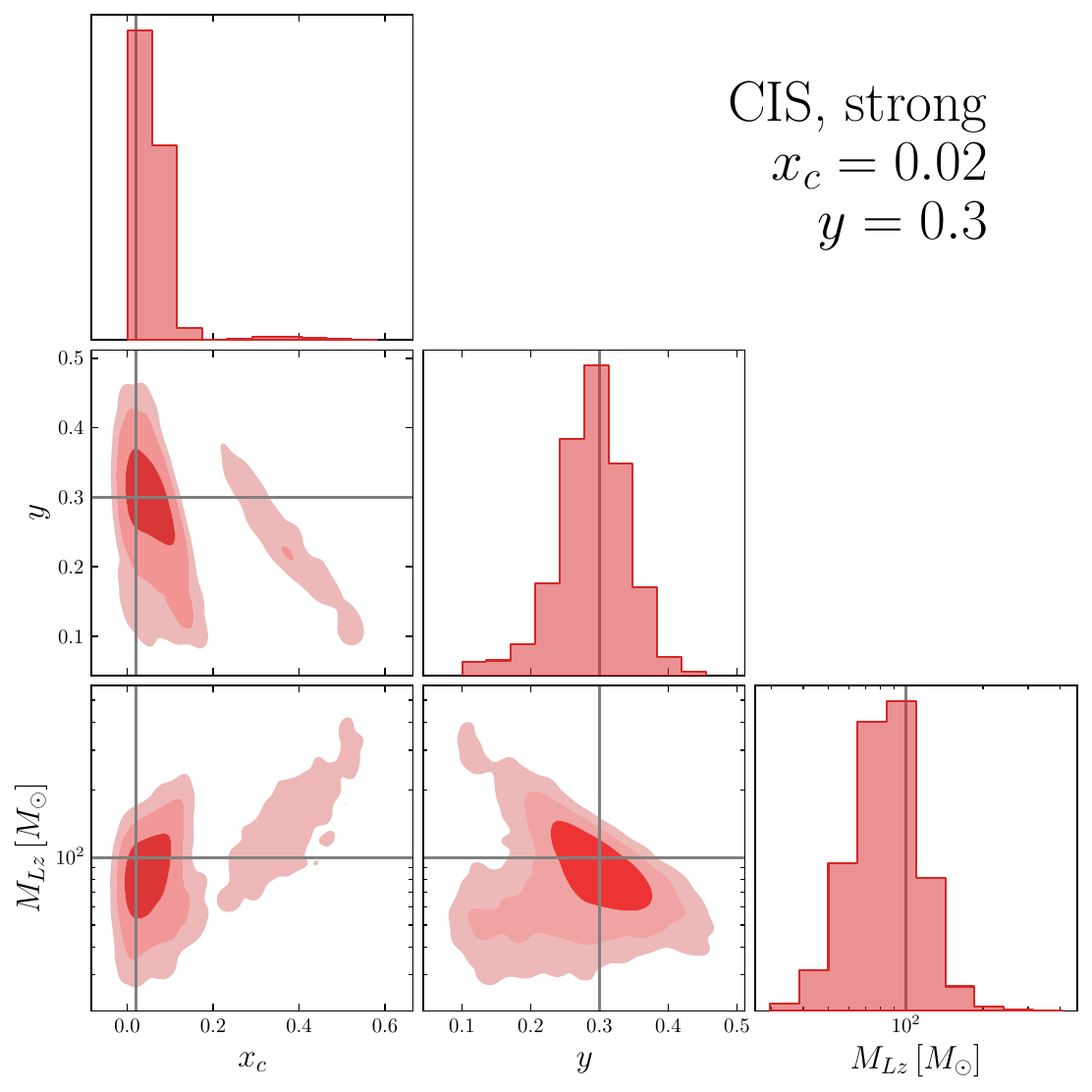} \\
    \includegraphics[width=0.32\textwidth]{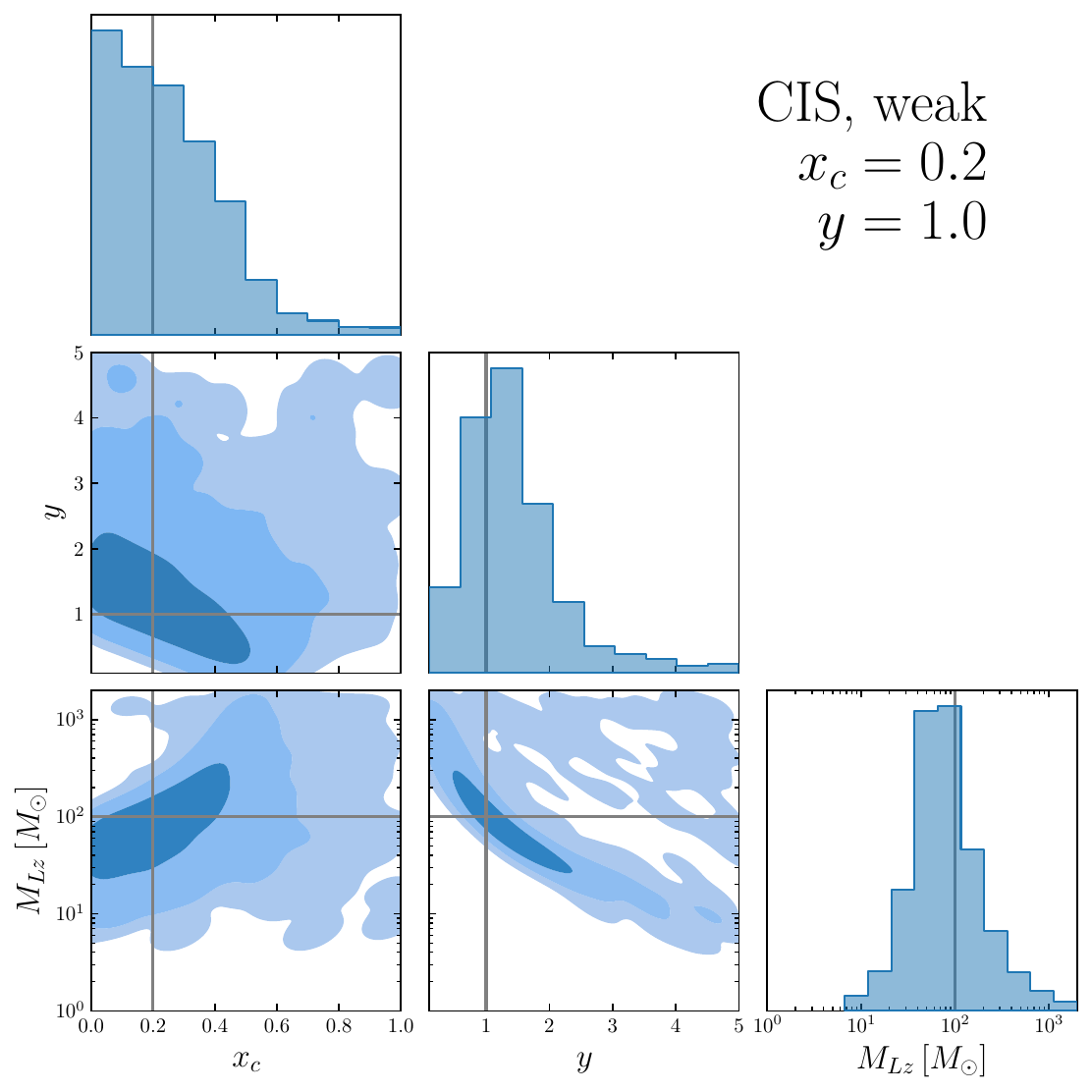}
    \includegraphics[width=0.32\textwidth]{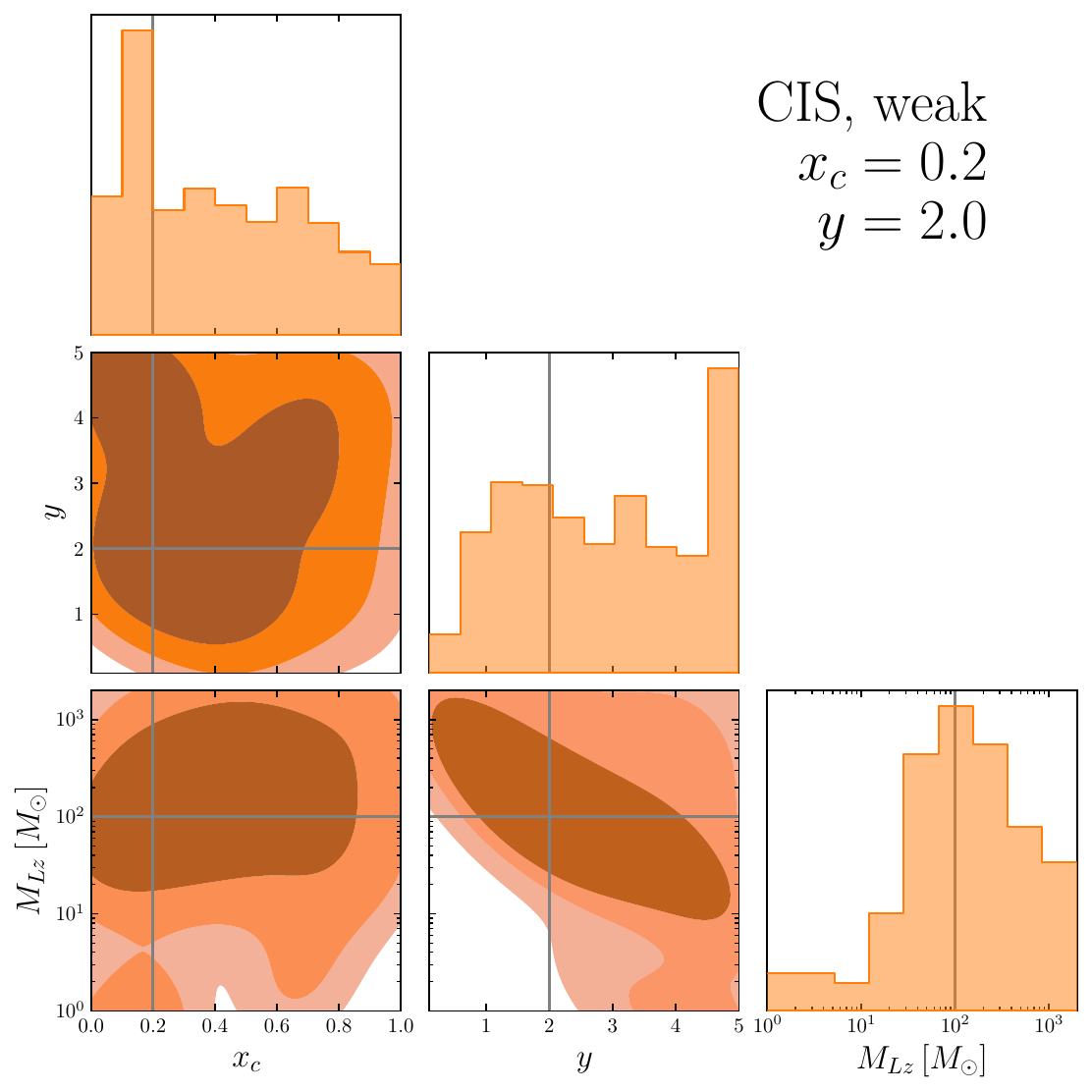}
    \includegraphics[width=0.32\textwidth]{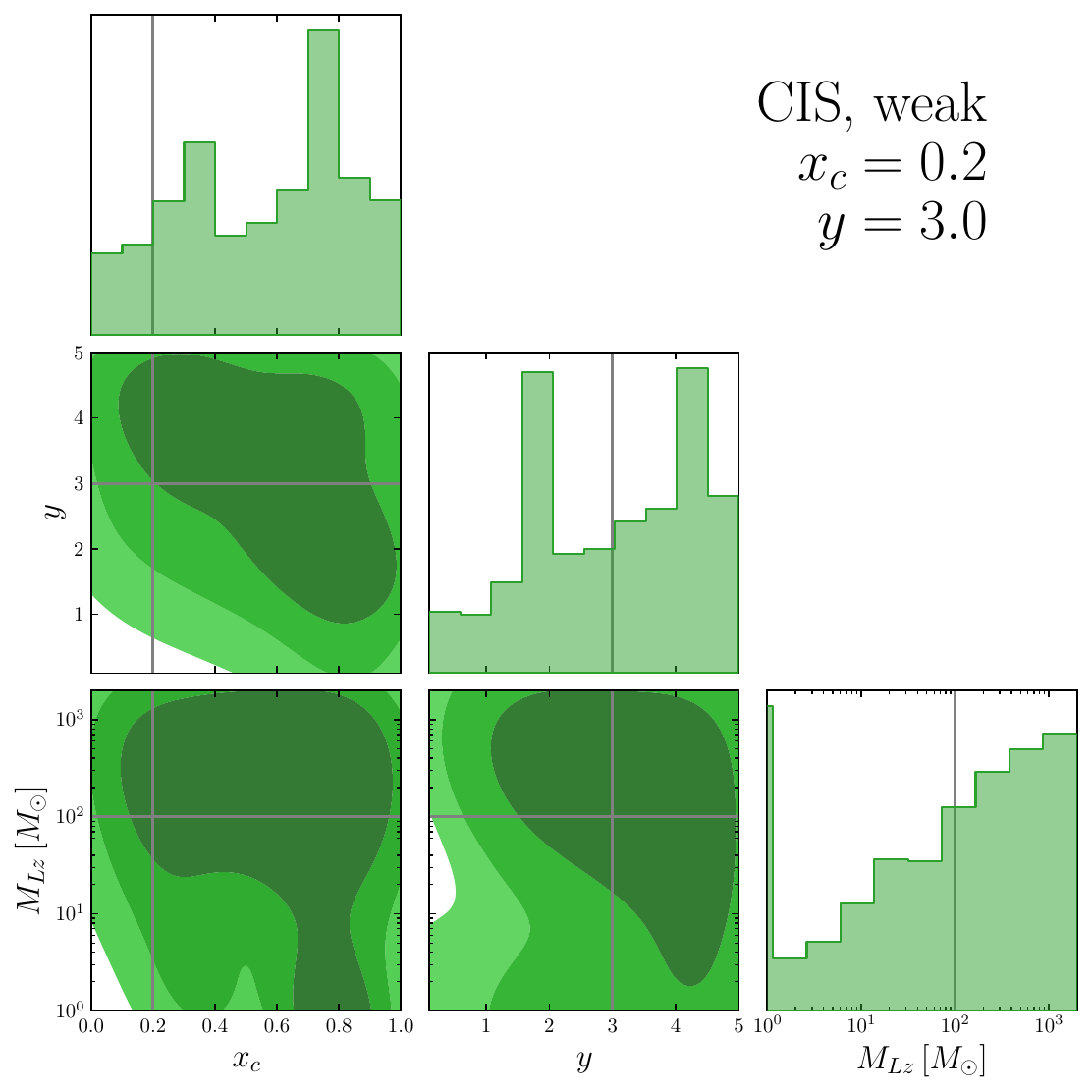}
    \caption{
        Parameter estimation results for the CIS lens for strong (top row, one panel) and weak (bottom row, three panels) lensing.
        The injected values are marked by the grey lines.
    }
    \label{fig:PE_CIS}
\end{figure*}

For the NFW and CIS lenses, we perform one strong lensing run.
For the gSIS lens, we perform multiple runs ($y \in \{1, 2, 3\}$) for the strong lensing case because the strong lensing regime extends to arbitrary $y$ as long as $k > 1$.
For the marginal $k = 1$ case corresponding to injecting an SIS lens, we skip $y = 1$ because it coincides with the caustic at $y_{\rm crit} = 1$.
We use $y \in \{1, 2, 3\}$ for the weak lensing case of all lenses to test the marginal value of $y$ for which lensing effects are detectable, even when there is only a single image.
For all runs, we use a lens mass $M_{Lz} = 100 M_\odot$, a source luminosity distance $d_L = 500 {\rm Mpc}$, and a BBH with nonspinning progenitors of equal mass $30 + 30 M_\odot$ for simplicity, injected into the LIGO Hanford and Livingston detector network at O4 design sensitivity, with an unlensed network signal-to-noise ratio (SNR) $\sim 30$.
We use the \texttt{IMRPhenomXHM} waveform model~\cite{Garcia-Quiros:2020qpx} and include all intrinsic, extrinsic and lensing parameters in our analysis.
More details of the injection run settings can be found in Appendix~\ref{app:PE_details}.

The results of all of the parameter estimation runs for the NFW lens are shown in Fig.~\ref{fig:PE_NFW}.
While parameter estimation is performed on all parameters, we only show the corner plots of the lensing-related parameters for simplicity.
The full corner plots including the posterior distribution of all other intrinsic and extrinsic parameters of the GW source are available on the \texttt{glworia} \texttt{GitHub} repository~\cite{glworia_github}.
For the strong lensing case, all of the lensing-related parameters can be constrained, although there are degeneracies between them.
In the weak lensing case, for $y = 1.0$ the parameter posteriors are informative, but for $y \geq 2.0$ the posterior for $y$ rails significantly at the upper bound of $y = 5.0$.
At a high $y$, the GWs are minimally affected by lensing and the signal approaches an unlensed signal, so a railing against the upper bound signifies non-detection of lensing.
Therefore, the posterior of $y$ should be used as a first check of whether we have measured the effects of lensing.
The results in Fig.~\ref{fig:PE_NFW} show that, at least for $\kappa = 3.0$, these effects cannot be measured if $y \gtrsim 2$.
For the cases with higher degeneracy, $y$ is often lower and $M_{Lz}$ is often higher for samples with a higher luminosity distance $d_L$ (not shown in the plots).
Such a degeneracy is expected because in the geometrical optics approximation, the reduction in the magnification has the same effect as a higher luminosity distance, and can be compensated by a lower $y$ and a higher $M_{Lz}$.

Similar trends can be observed in the parameter estimation results for the gSIS and CIS lenses shown in Figs.~\ref{fig:PE_gSIS} and \ref{fig:PE_CIS} respectively.
For the gSIS lens, in the strong lensing limit the lensing effects can be measured up to $y = 2.0$, but only for $y = 1.0$ in the weak lensing limit.
For the CIS lens, the effects can also only be measured for $y = 1.0$.
However, when interpreting these results, note that the same impact parameter $y$ could correspond to different physical scales when comparing between different lens models, or even when comparing between realizations of the same lens model with different values of $l$.

While the posterior of $y$ is useful for determining whether or not lensing is measured, the physical properties of the lens are encoded in the lens mass and in the lens parameter.
Even if the posterior of $y$ is informative, it does not necessarily mean that the lens mass or lens parameter will be well measured.
For example, the slope $k$ of the gSIS lens is often poorly constrained (i.e., there is significant railing at the lower or upper bounds) even if $y$ is well constrained.
The mass $M_{Lz}$ of the lens is relatively better constrained in all cases across lens models, although the posterior often has support over orders of magnitudes in $M_{Lz}$.

While the unlensed GW signal that we used has an SNR of $\sim 30$, the lensing effects could magnify the signal and increase the SNR.
For example, the strong lensing results could have an SNR $\gtrsim 70$.
In this regime, the Fisher Information Matrix (FIM) can be used to estimate the measurement uncertainties.
Therefore, we can compare the uncertainties in our PE results with the estimates in, e.g., Ref.~\cite{Tambalo:2022wlm}.
For example, we find that the strong lensing result for the CIS lens ($x_c = 0.02$, $y = 0.3$, top panel of Fig.~\ref{fig:PE_CIS}) agrees within a factor of a few with those in Ref.~\cite{Tambalo:2022wlm}.

Although we used an interpolation table to speed up the evaluation of the lensing amplification factor, this calculation is still the bottleneck in computational cost when compared to the evaluation of the unlensed waveform.
While performing the interpolation takes negligible time, we need to use a very fine grid in time (with $2^{16}$ points) to ensure that the FFT results are accurate in the high-frequency regime.
Broadly speaking, a single likelihood evaluation of a lensed model is a few times more expensive than an unlensed (\texttt{IMRPhenomXHM}) one.
Moreover, assuming that there is only one lens parameter $l$ (as is the case for all lens models in this work), the lensed waveform model will contain three additional parameters ($l, M_{Lz}, y$), which increases the dimensionality and hence the time it takes for the sampler to converge.
Overall, depending on whether the injected signal shows prominent lensing features, using a lensed model could introduce as high as an order of magnitude increase in the sampling time.

\section{Discussion}\label{sec:discussion}

In the previous section we have shown that, in principle, diffraction lensing effects can be measured by current ground-based GW detectors, as long as the impact parameter $y$ is small enough, even in the weak lensing limit where there are no multiple images.
The parameters $l \in \{\kappa, k, x_c\}$ characterizing the lens could also be constrained, meaning that wave-optics lensing of GWs could be a unique probe of the properties of minihalos.

Of course, whether we will observe these types of lensing events depends on the abundance of minihalos in the Universe and their properties (e.g., concentration).
The abundance of minihalos is closely related to the distribution of impact parameters $y$ of lensing events, and the distributions of the lens mass $M_{Lz}$ and parameter $l$ also affect the probability of detecting measurable lensing events.
In fact, given an abundance of minihalos, we can estimate the distribution of $y$, and from constraints on $y$ we can also constrain the abundance of minihalos.
However, note that there is no guarantee that the parameter ranges that are covered by our injection analysis or our prior distribution would fully encapsulate the distribution of the parameters in nature.

In this work, we estimate the marginally measurable impact parameter to be around $y \sim 1$ in the weak lensing regime, at least for the lens models and values of the lensing-related parameters we considered.
With this information, we can estimate the rate of detectable diffraction lensing events given a population of lenses.
To pinpoint the marginal impact parameter for different lens models spanning a larger range of impact parameters, more injection runs should be performed in the future.
Our results can also be used to calibrate Fisher information matrix estimates, which are more scalable over multiple events than full Bayesian parameter estimation.
In principle, our methods can be generalized to forecast the detectability of diffractive lensing by space-based detectors, although the accuracy of the amplification factor computation must be improved and the interpolation table must be pushed to higher impact parameters.

All of the parameter estimation results in this paper are obtained assuming the sensitivity curve of the fourth observing run (O4) of the LIGO detectors. Therefore the parameter estimation accuracy can be expected to improve as the detector sensitivity improves in the future, or when next-generation detectors will come online.

\section*{acknowledgments}
We thank Guilherme Brando, Mesut \c{C}al\i{}\c{s}kan, Jose Maria Ezquiaga, Otto Hannuksela, Jason Poon, Stefano Savastano, Giovanni Tambalo, Hector Villarrubia-Rojo and Simon Yeung for useful insights.
M.H.Y.C. is a Croucher Scholar supported by the Croucher Foundation.
K.K.Y.N. is supported by the Miller Fellowship at Johns Hopkins University and the Croucher Fellowship by the Croucher Foundation.
M.H.Y.C., K.K.Y.N. and E.B. are supported by NSF Grants No. AST-2006538, PHY-2207502, PHY-090003 and PHY-20043, by NASA Grants No. 20-LPS20-0011 and 21-ATP21-0010, by the John Templeton Foundation Grant 62840, by the Simons Foundation, and by the Italian Ministry of Foreign Affairs and International Cooperation Grant No.~PGR01167.
This work was carried out at the Advanced Research Computing at Hopkins (ARCH) core facility (\url{rockfish.jhu.edu}) which is supported by the NSF Grant No.~OAC-1920103.

\appendix

\section{Implementation of the numerical contour integration}\label{app:contour_step}

The computation of the lensing amplification factor requires performing a contour integral over constant time delay contour lines.
Given a point on one of these contour lines, we can perform the integration by going in steps tangential to this line, which is perpendicular to the gradient $\vec{\nabla} T$.
In our implementation, starting at a point $\vec{x}_{\rm init}$ on a contour at time delay $T_0$, we use a fourth-order Runge-Kutta method to trace the contour line by moving perpendicular to $\vec{\nabla} T$, as shown in Algorithm~\ref{alg:contour_integration}.
In the algorithm, ${\rm RK4}(f(\cdot), \vec{x}, h)$ is the Runge-Kutta step constructor for the function $f(\cdot)$ at $\vec{x}$ with step size $h$.
We adaptively control the step size by the factor $\rho$.
We use a smaller $\rho = \rho_c$ when the contour line curves more acutely, but we make sure that $\rho$ is never less than $\rho_r$, the distance from the current point on the contour to the origin, because the contour lines could be an arbitrarily small circle around the origin for some $T_0$.
Otherwise, we impose the bound $\rho_{\rm min} < \rho < \rho_{\rm max}$ to avoid using an overly small or large step.
After the Runge-Kutta proposal step $\vec{x}_{\rm prop}$, we perform an additional adjustment step $\vec{x}_{\rm adj}$ to reduce the deviation of the proposed point from the contour line.
We then evaluate the integrand at the mid-point of the full step.

\begin{algorithm}
    \SetKwInOut{Input}{input}
    \SetKwInOut{Output}{Output}
    \SetKwComment{Comment}{\#}{}
    \Input{Initial point $\vec{x}_0$, step size $h$, $\rho_{\rm min}$, $\rho_{\rm max}$, time delay function $T(\cdot)$}
    \Output{Numerical integration result $u$}
    $T_{\perp}(\cdot) \gets {\rm Rot}_{\theta = \pi/2} \vec{\nabla}T(\cdot)$\\
    $u \gets 0$\\
    $\vec{x} \gets \vec{x}_0$\\
    \While{\text{not yet traced whole contour}}{
    $\rho_{r} \gets |\vec{x}|$ \\
    $\rho_{c} \gets | \vec{\nabla}T(\vec{x})| / {\rm det}({\rm Hess}(T)(\vec{x}))$ \Comment*[r]{Inverse of contour curvature}
    $\rho \gets {\rm max}\left[ {\rm min}(\rho_{\rm min}, \rho_r), {\rm min}(\rho_{\rm max}, \rho_r, \rho_c)\right]$ \\
    $\Delta \vec{x}_{\rm prop} \gets {\rm RK4}(T_{\perp}(\cdot), \vec{x}, \rho h)$ \Comment*[r]{RK4 proposal}
    $\vec{x}_{\rm prop} \gets \vec{x} + \Delta \vec{x}_{\rm prop}$ \\
    $T_{\rm prop} \gets T(\vec{x}_{\rm prop})$ \\
    $\Delta \vec{x}_{\rm adj} \gets \frac{\vec{\nabla}T(\vec{x}_{\rm prop})}{|\vec{\nabla}T(\vec{x}_{\rm prop})|^2} (T_{\rm prop} - T_0)$ \Comment*[r]{Adjust back onto contour}
    $\vec{x}_{\rm new} \gets \vec{x}_{\rm prop} + \Delta \vec{x}_{\rm adj}$ \\
    $\Delta x \gets (\vec{x}_{\rm new} - \vec{x})$ \\
    $\vec{x}_{\rm mid} \gets (\vec{x}_{\rm new} + \vec{x})/2$ \\
    $\Delta u \gets |\Delta x| / |\vec{\nabla}T(\vec{x}_{\rm mid})|$ \Comment*[r]{Integrand}
    $u \gets u + \Delta u$ \\
    $\vec{x} \gets \vec{x}_{\rm new}$
    }
    \caption{contour integration.}
    \label{alg:contour_integration}
\end{algorithm}

\section{Error analysis}\label{app:error}

\begin{figure*}
    \centering
    \includegraphics[width=0.99\textwidth]{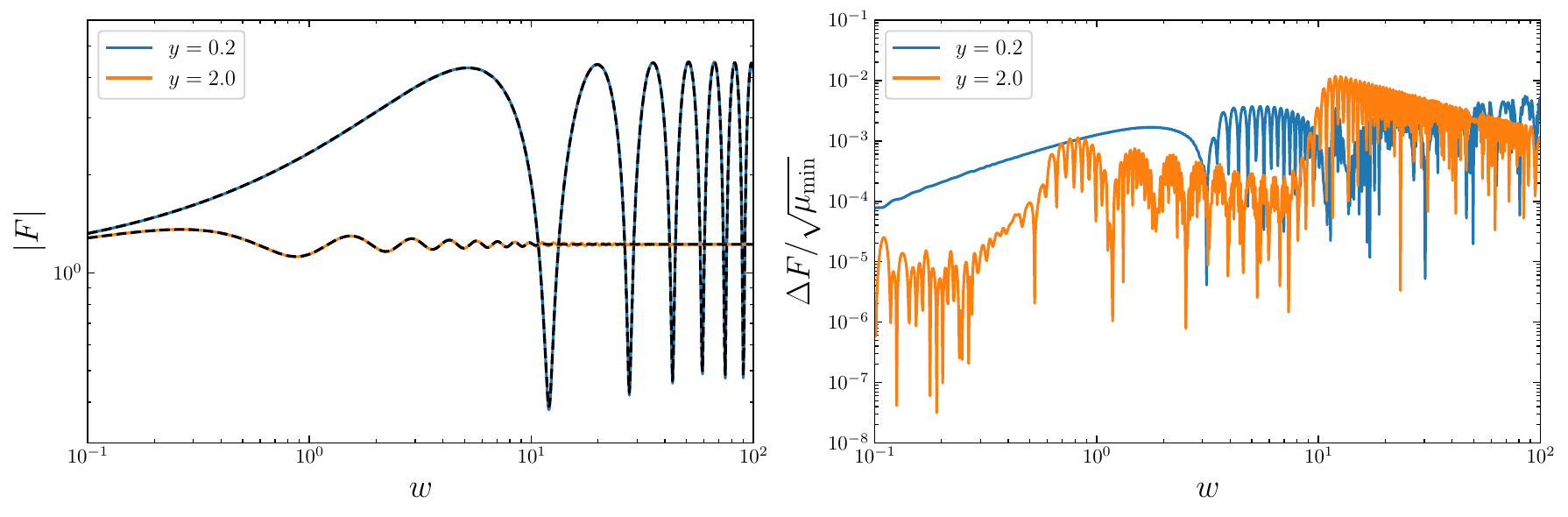}
    \caption{
        Comparison between our implementation of the contour integral method for computing the frequency domain amplification factor versus analytical series expansion results for an SIS lens.
        Left: the colored lines are results computed with our contour integration procedure, while the black dashed lines are the analytical results.
        Right: The error between the two methods, normalized by $\sqrt{|\mu_{\rm min}|}$.
    }
    \label{fig:analytic_comparison}
\end{figure*}

In this appendix, we will estimate the errors of our algorithm for computing and interpolating the amplification factor $F(w)$.
First and foremost, we test our algorithm against known analytical results for the SIS lens.
The amplification factor for an SIS lens can be analytically written as a series expansion in $w$~\cite{Matsunaga:2006uc,Tambalo:2022plm,Caliskan:2023zqm}.
In Fig.~\ref{fig:analytic_comparison} we compare our results obtained via contour integration with those from the series expansion.
We define a measure of the error by $\Delta F/\sqrt{|\mu_{\rm min}|}$, the absolute error normalized by the magnification of the minimum image.
We find that in both the strong lensing ($y = 0.2$) and weak lensing ($y = 2.0$) regimes, the error as defined above does not exceed $10^{-2}$.

\begin{figure*}
    \centering
    \includegraphics[width=0.99\textwidth]{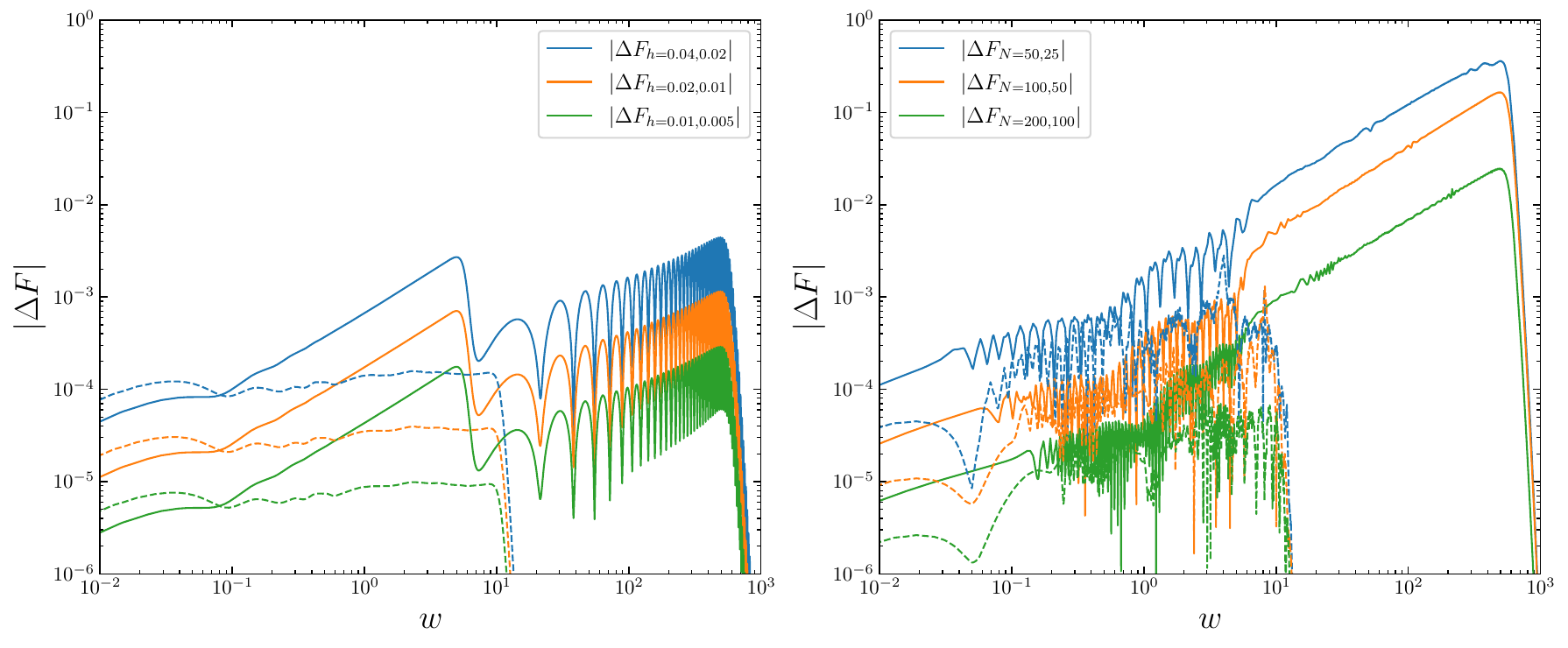}
    \caption{
        Convergence test for the Runge-Kutta integration step size $h$ (left) and the number of points $N$ in time used when computing $\tilde{I}(\tau)$ (right).
        The solid lines correspond to a strong lensing scenario and the dashed lines correspond to a weak lensing one.
        All results show approximate second-order convergence.
        The curves all drop to zero at a certain $w$ because we transition to the geometrical optics approximation there.
    }
    \label{fig:convergence}
\end{figure*}

The computation of $F(w)$ makes use of a Runge-Kutta scheme with step size $h$ for performing a contour integration over values of $\tau$ with resolution related to the number of points $N$ used in time.
The numerical error of computation will then depend on $h$ and $N$.
In Fig.~\ref{fig:convergence} we show the convergence of the numerical results when reducing $h$ and increasing $N$ for both strong lensing and weak lensing of an NFW lens.
The results exhibit approximately second-order convergence for both $h$ and $N$.

\section{Details of the injection runs} \label{app:PE_details}

For all of the injection runs we use a redshifted lens mass $M_{Lz} = 100 M_\odot$.
We use the \texttt{IMRPhenomXHM} BBH merger waveform model to generate the (unlensed) nonprecessing and noneccentric gravitational waveforms.
We use an equal mass of $m_1 = m_2 = 30 M_\odot$ for both BH progenitors, with zero spins ($|\vec{\chi_1}| = |\vec{\chi_2}| = 0$).
For the extrinsic parameters, we inject an inclination angle $\theta_{jn} = \pi/3$, polarization angle $\psi = \pi/2$, phase $\phi = \pi/2$, geocentric time $t_g = 1126259642.413$, right ascension $\alpha = 1.375$, and declination $\delta = -1.2108$.
We use a sampling rate of $2048\,{\rm Hz}$ and a minimum frequency of $f_{\rm min} = 20\,{\rm Hz}$, and inject the signals into the two LIGO detectors, both assuming design sensitivity Gaussian noise.
We use \texttt{bilby}'s implementation of the \texttt{dynesty} sampler for nested sampling.
For all of the intrinsic and extrinsic parameters of the waveform, we use the standard priors as implemented in \texttt{bilby}.
For the lensing-related parameters, we use uniform priors with $M_{Lz} \in (0.1, 2000)$, $y \in (0.1, 5.0)$, and $\kappa \in (0.1, 10)$ for the NFW lens, $k \in (0.1, 1.9)$ for the gSIS lens, and $x_c \in (0, 1)$ for the CIS lens when sampling.
When plotting the results in the main text we reweight the prior of $M_{Lz}$ into a log-uniform prior.
We use a uniform prior for all lensing-related parameters when sampling to make sure that the results will not be biased due to strong degeneracies.
All of the settings files and scripts for submitting the runs are publicly available on the \texttt{glworia} \texttt{GitHub} repository~\cite{glworia_github}.

\bibliography{Glworia}

\begin{thebibliography}{123}%
\makeatletter
\providecommand \@ifxundefined [1]{%
 \@ifx{#1\undefined}
}%
\providecommand \@ifnum [1]{%
 \ifnum #1\expandafter \@firstoftwo
 \else \expandafter \@secondoftwo
 \fi
}%
\providecommand \@ifx [1]{%
 \ifx #1\expandafter \@firstoftwo
 \else \expandafter \@secondoftwo
 \fi
}%
\providecommand \natexlab [1]{#1}%
\providecommand \enquote  [1]{``#1''}%
\providecommand \bibnamefont  [1]{#1}%
\providecommand \bibfnamefont [1]{#1}%
\providecommand \citenamefont [1]{#1}%
\providecommand \href@noop [0]{\@secondoftwo}%
\providecommand \href [0]{\begingroup \@sanitize@url \@href}%
\providecommand \@href[1]{\@@startlink{#1}\@@href}%
\providecommand \@@href[1]{\endgroup#1\@@endlink}%
\providecommand \@sanitize@url [0]{\catcode `\\12\catcode `\$12\catcode `\&12\catcode `\#12\catcode `\^12\catcode `\_12\catcode `\%12\relax}%
\providecommand \@@startlink[1]{}%
\providecommand \@@endlink[0]{}%
\providecommand \url  [0]{\begingroup\@sanitize@url \@url }%
\providecommand \@url [1]{\endgroup\@href {#1}{\urlprefix }}%
\providecommand \urlprefix  [0]{URL }%
\providecommand \Eprint [0]{\href }%
\providecommand \doibase [0]{http://dx.doi.org/}%
\providecommand \selectlanguage [0]{\@gobble}%
\providecommand \bibinfo  [0]{\@secondoftwo}%
\providecommand \bibfield  [0]{\@secondoftwo}%
\providecommand \translation [1]{[#1]}%
\providecommand \BibitemOpen [0]{}%
\providecommand \bibitemStop [0]{}%
\providecommand \bibitemNoStop [0]{.\EOS\space}%
\providecommand \EOS [0]{\spacefactor3000\relax}%
\providecommand \BibitemShut  [1]{\csname bibitem#1\endcsname}%
\let\auto@bib@innerbib\@empty
%</preamble>
\bibitem [{\citenamefont {Cheung}(2024)}]{glworia_github}%
  \BibitemOpen
  \bibfield  {author} {\bibinfo {author} {\bibfnamefont {M.~H.-Y.}\ \bibnamefont {Cheung}},\ }\href@noop {} {\enquote {\bibinfo {title} {glworia},}\ }\bibinfo {howpublished} {\url{https://github.com/mhycheung/glworia}} (\bibinfo {year} {2024})\BibitemShut {NoStop}%
\bibitem [{\citenamefont {{Schneider}}\ \emph {et~al.}(1992)\citenamefont {{Schneider}}, \citenamefont {{Ehlers}},\ and\ \citenamefont {{Falco}}}]{1992grle.book.....S}%
  \BibitemOpen
  \bibfield  {author} {\bibinfo {author} {\bibfnamefont {P.}~\bibnamefont {{Schneider}}}, \bibinfo {author} {\bibfnamefont {J.}~\bibnamefont {{Ehlers}}}, \ and\ \bibinfo {author} {\bibfnamefont {E.~E.}\ \bibnamefont {{Falco}}},\ }\href {\doibase 10.1007/978-3-662-03758-4} {\emph {\bibinfo {title} {{Gravitational Lenses}}}}\ (\bibinfo {year} {1992})\BibitemShut {NoStop}%
\bibitem [{\citenamefont {Bartelmann}(2010)}]{Bartelmann:2010fz}%
  \BibitemOpen
  \bibfield  {author} {\bibinfo {author} {\bibfnamefont {M.}~\bibnamefont {Bartelmann}},\ }\href {\doibase 10.1088/0264-9381/27/23/233001} {\bibfield  {journal} {\bibinfo  {journal} {Class. Quant. Grav.}\ }\textbf {\bibinfo {volume} {27}},\ \bibinfo {pages} {233001} (\bibinfo {year} {2010})},\ \Eprint {http://arxiv.org/abs/1010.3829} {arXiv:1010.3829 [astro-ph.CO]} \BibitemShut {NoStop}%
\bibitem [{\citenamefont {Einstein}(1916)}]{Einstein:1916vd}%
  \BibitemOpen
  \bibfield  {author} {\bibinfo {author} {\bibfnamefont {A.}~\bibnamefont {Einstein}},\ }\href {\doibase 10.1002/andp.19163540702} {\bibfield  {journal} {\bibinfo  {journal} {Annalen Phys.}\ }\textbf {\bibinfo {volume} {49}},\ \bibinfo {pages} {769} (\bibinfo {year} {1916})}\BibitemShut {NoStop}%
\bibitem [{\citenamefont {Bond}\ \emph {et~al.}(2004)\citenamefont {Bond} \emph {et~al.}}]{Bond:2004qd}%
  \BibitemOpen
  \bibfield  {author} {\bibinfo {author} {\bibfnamefont {I.~A.}\ \bibnamefont {Bond}} \emph {et~al.},\ }\href {\doibase 10.1086/420928} {\bibfield  {journal} {\bibinfo  {journal} {Astrophys. J. Lett.}\ }\textbf {\bibinfo {volume} {606}},\ \bibinfo {pages} {L155} (\bibinfo {year} {2004})},\ \Eprint {http://arxiv.org/abs/astro-ph/0404309} {arXiv:astro-ph/0404309} \BibitemShut {NoStop}%
\bibitem [{\citenamefont {Coe}\ \emph {et~al.}(2013)\citenamefont {Coe} \emph {et~al.}}]{Coe:2012wj}%
  \BibitemOpen
  \bibfield  {author} {\bibinfo {author} {\bibfnamefont {D.}~\bibnamefont {Coe}} \emph {et~al.},\ }\href {\doibase 10.1088/0004-637X/762/1/32} {\bibfield  {journal} {\bibinfo  {journal} {Astrophys. J.}\ }\textbf {\bibinfo {volume} {762}},\ \bibinfo {pages} {32} (\bibinfo {year} {2013})},\ \Eprint {http://arxiv.org/abs/1211.3663} {arXiv:1211.3663 [astro-ph.CO]} \BibitemShut {NoStop}%
\bibitem [{\citenamefont {{Welch}}\ \emph {et~al.}(2022)\citenamefont {{Welch}}, \citenamefont {{Coe}}, \citenamefont {{Diego}}, \citenamefont {{Zitrin}}, \citenamefont {{Zackrisson}}, \citenamefont {{Dimauro}}, \citenamefont {{Jim{\'e}nez-Teja}}, \citenamefont {{Kelly}}, \citenamefont {{Mahler}}, \citenamefont {{Oguri}}, \citenamefont {{Timmes}}, \citenamefont {{Windhorst}}, \citenamefont {{Florian}}, \citenamefont {{de Mink}}, \citenamefont {{Avila}}, \citenamefont {{Anderson}}, \citenamefont {{Bradley}}, \citenamefont {{Sharon}}, \citenamefont {{Vikaeus}}, \citenamefont {{McCandliss}}, \citenamefont {{Brada{\v{c}}}}, \citenamefont {{Rigby}}, \citenamefont {{Frye}}, \citenamefont {{Toft}}, \citenamefont {{Strait}}, \citenamefont {{Trenti}}, \citenamefont {{Sharma}}, \citenamefont {{Andrade-Santos}},\ and\ \citenamefont {{Broadhurst}}}]{2022Natur.603..815W}%
  \BibitemOpen
  \bibfield  {author} {\bibinfo {author} {\bibfnamefont {B.}~\bibnamefont {{Welch}}}, \bibinfo {author} {\bibfnamefont {D.}~\bibnamefont {{Coe}}}, \bibinfo {author} {\bibfnamefont {J.~M.}\ \bibnamefont {{Diego}}}, \bibinfo {author} {\bibfnamefont {A.}~\bibnamefont {{Zitrin}}}, \bibinfo {author} {\bibfnamefont {E.}~\bibnamefont {{Zackrisson}}}, \bibinfo {author} {\bibfnamefont {P.}~\bibnamefont {{Dimauro}}}, \bibinfo {author} {\bibfnamefont {Y.}~\bibnamefont {{Jim{\'e}nez-Teja}}}, \bibinfo {author} {\bibfnamefont {P.}~\bibnamefont {{Kelly}}}, \bibinfo {author} {\bibfnamefont {G.}~\bibnamefont {{Mahler}}}, \bibinfo {author} {\bibfnamefont {M.}~\bibnamefont {{Oguri}}}, \bibinfo {author} {\bibfnamefont {F.~X.}\ \bibnamefont {{Timmes}}}, \bibinfo {author} {\bibfnamefont {R.}~\bibnamefont {{Windhorst}}}, \bibinfo {author} {\bibfnamefont {M.}~\bibnamefont {{Florian}}}, \bibinfo {author} {\bibfnamefont {S.~E.}\ \bibnamefont {{de Mink}}}, \bibinfo {author} {\bibfnamefont {R.~J.}\ \bibnamefont {{Avila}}}, \bibinfo {author} {\bibfnamefont {J.}~\bibnamefont {{Anderson}}}, \bibinfo {author} {\bibfnamefont {L.}~\bibnamefont {{Bradley}}}, \bibinfo {author} {\bibfnamefont {K.}~\bibnamefont {{Sharon}}}, \bibinfo {author} {\bibfnamefont {A.}~\bibnamefont {{Vikaeus}}}, \bibinfo {author} {\bibfnamefont {S.}~\bibnamefont {{McCandliss}}}, \bibinfo {author} {\bibfnamefont {M.}~\bibnamefont {{Brada{\v{c}}}}}, \bibinfo {author} {\bibfnamefont {J.}~\bibnamefont {{Rigby}}}, \bibinfo {author} {\bibfnamefont {B.}~\bibnamefont {{Frye}}}, \bibinfo {author} {\bibfnamefont {S.}~\bibnamefont {{Toft}}}, \bibinfo {author} {\bibfnamefont {V.}~\bibnamefont {{Strait}}}, \bibinfo {author} {\bibfnamefont {M.}~\bibnamefont {{Trenti}}}, \bibinfo {author} {\bibfnamefont {S.}~\bibnamefont {{Sharma}}}, \bibinfo {author} {\bibfnamefont {F.}~\bibnamefont {{Andrade-Santos}}}, \ and\ \bibinfo {author} {\bibfnamefont {T.}~\bibnamefont {{Broadhurst}}},\ }\href {\doibase 10.1038/s41586-022-04449-y} {\bibfield  {journal} {\bibinfo  {journal} {\nat}\ }\textbf {\bibinfo {volume} {603}},\ \bibinfo {pages} {815} (\bibinfo {year} {2022})},\ \Eprint {http://arxiv.org/abs/2209.14866} {arXiv:2209.14866 [astro-ph.GA]} \BibitemShut {NoStop}%
\bibitem [{\citenamefont {Clowe}\ \emph {et~al.}(2004)\citenamefont {Clowe}, \citenamefont {Gonzalez},\ and\ \citenamefont {Markevitch}}]{Clowe:2003tk}%
  \BibitemOpen
  \bibfield  {author} {\bibinfo {author} {\bibfnamefont {D.}~\bibnamefont {Clowe}}, \bibinfo {author} {\bibfnamefont {A.}~\bibnamefont {Gonzalez}}, \ and\ \bibinfo {author} {\bibfnamefont {M.}~\bibnamefont {Markevitch}},\ }\href {\doibase 10.1086/381970} {\bibfield  {journal} {\bibinfo  {journal} {Astrophys. J.}\ }\textbf {\bibinfo {volume} {604}},\ \bibinfo {pages} {596} (\bibinfo {year} {2004})},\ \Eprint {http://arxiv.org/abs/astro-ph/0312273} {arXiv:astro-ph/0312273} \BibitemShut {NoStop}%
\bibitem [{\citenamefont {Markevitch}\ \emph {et~al.}(2004)\citenamefont {Markevitch}, \citenamefont {Gonzalez}, \citenamefont {Clowe}, \citenamefont {Vikhlinin}, \citenamefont {David}, \citenamefont {Forman}, \citenamefont {Jones}, \citenamefont {Murray},\ and\ \citenamefont {Tucker}}]{Markevitch:2003at}%
  \BibitemOpen
  \bibfield  {author} {\bibinfo {author} {\bibfnamefont {M.}~\bibnamefont {Markevitch}}, \bibinfo {author} {\bibfnamefont {A.~H.}\ \bibnamefont {Gonzalez}}, \bibinfo {author} {\bibfnamefont {D.}~\bibnamefont {Clowe}}, \bibinfo {author} {\bibfnamefont {A.}~\bibnamefont {Vikhlinin}}, \bibinfo {author} {\bibfnamefont {L.}~\bibnamefont {David}}, \bibinfo {author} {\bibfnamefont {W.}~\bibnamefont {Forman}}, \bibinfo {author} {\bibfnamefont {C.}~\bibnamefont {Jones}}, \bibinfo {author} {\bibfnamefont {S.}~\bibnamefont {Murray}}, \ and\ \bibinfo {author} {\bibfnamefont {W.}~\bibnamefont {Tucker}},\ }\href {\doibase 10.1086/383178} {\bibfield  {journal} {\bibinfo  {journal} {Astrophys. J.}\ }\textbf {\bibinfo {volume} {606}},\ \bibinfo {pages} {819} (\bibinfo {year} {2004})},\ \Eprint {http://arxiv.org/abs/astro-ph/0309303} {arXiv:astro-ph/0309303} \BibitemShut {NoStop}%
\bibitem [{\citenamefont {Aghanim}\ \emph {et~al.}(2020)\citenamefont {Aghanim} \emph {et~al.}}]{Planck:2018vyg}%
  \BibitemOpen
  \bibfield  {author} {\bibinfo {author} {\bibfnamefont {N.}~\bibnamefont {Aghanim}} \emph {et~al.} (\bibinfo {collaboration} {Planck}),\ }\href {\doibase 10.1051/0004-6361/201833910} {\bibfield  {journal} {\bibinfo  {journal} {Astron. Astrophys.}\ }\textbf {\bibinfo {volume} {641}},\ \bibinfo {pages} {A6} (\bibinfo {year} {2020})},\ \bibinfo {note} {[Erratum: Astron.Astrophys. 652, C4 (2021)]},\ \Eprint {http://arxiv.org/abs/1807.06209} {arXiv:1807.06209 [astro-ph.CO]} \BibitemShut {NoStop}%
\bibitem [{\citenamefont {Abbott}\ \emph {et~al.}(2016)\citenamefont {Abbott} \emph {et~al.}}]{LIGOScientific:2016aoc}%
  \BibitemOpen
  \bibfield  {author} {\bibinfo {author} {\bibfnamefont {B.~P.}\ \bibnamefont {Abbott}} \emph {et~al.} (\bibinfo {collaboration} {LIGO Scientific, Virgo}),\ }\href {\doibase 10.1103/PhysRevLett.116.061102} {\bibfield  {journal} {\bibinfo  {journal} {Phys. Rev. Lett.}\ }\textbf {\bibinfo {volume} {116}},\ \bibinfo {pages} {061102} (\bibinfo {year} {2016})},\ \Eprint {http://arxiv.org/abs/1602.03837} {arXiv:1602.03837 [gr-qc]} \BibitemShut {NoStop}%
\bibitem [{\citenamefont {Maggiore}(2007)}]{Maggiore:2007ulw}%
  \BibitemOpen
  \bibfield  {author} {\bibinfo {author} {\bibfnamefont {M.}~\bibnamefont {Maggiore}},\ }\href {\doibase 10.1093/acprof:oso/9780198570745.001.0001} {\emph {\bibinfo {title} {{Gravitational Waves. Vol. 1: Theory and Experiments}}}}\ (\bibinfo  {publisher} {Oxford University Press},\ \bibinfo {year} {2007})\BibitemShut {NoStop}%
\bibitem [{\citenamefont {Abbott}\ \emph {et~al.}(2019)\citenamefont {Abbott} \emph {et~al.}}]{LIGOScientific:2018mvr}%
  \BibitemOpen
  \bibfield  {author} {\bibinfo {author} {\bibfnamefont {B.~P.}\ \bibnamefont {Abbott}} \emph {et~al.} (\bibinfo {collaboration} {LIGO Scientific, Virgo}),\ }\href {\doibase 10.1103/PhysRevX.9.031040} {\bibfield  {journal} {\bibinfo  {journal} {Phys. Rev. X}\ }\textbf {\bibinfo {volume} {9}},\ \bibinfo {pages} {031040} (\bibinfo {year} {2019})},\ \Eprint {http://arxiv.org/abs/1811.12907} {arXiv:1811.12907 [astro-ph.HE]} \BibitemShut {NoStop}%
\bibitem [{\citenamefont {Abbott}\ \emph {et~al.}(2021{\natexlab{a}})\citenamefont {Abbott} \emph {et~al.}}]{LIGOScientific:2020ibl}%
  \BibitemOpen
  \bibfield  {author} {\bibinfo {author} {\bibfnamefont {R.}~\bibnamefont {Abbott}} \emph {et~al.} (\bibinfo {collaboration} {LIGO Scientific, Virgo}),\ }\href {\doibase 10.1103/PhysRevX.11.021053} {\bibfield  {journal} {\bibinfo  {journal} {Phys. Rev. X}\ }\textbf {\bibinfo {volume} {11}},\ \bibinfo {pages} {021053} (\bibinfo {year} {2021}{\natexlab{a}})},\ \Eprint {http://arxiv.org/abs/2010.14527} {arXiv:2010.14527 [gr-qc]} \BibitemShut {NoStop}%
\bibitem [{\citenamefont {Abbott}\ \emph {et~al.}(2023{\natexlab{a}})\citenamefont {Abbott} \emph {et~al.}}]{KAGRA:2021vkt}%
  \BibitemOpen
  \bibfield  {author} {\bibinfo {author} {\bibfnamefont {R.}~\bibnamefont {Abbott}} \emph {et~al.} (\bibinfo {collaboration} {KAGRA, VIRGO, LIGO Scientific}),\ }\href {\doibase 10.1103/PhysRevX.13.041039} {\bibfield  {journal} {\bibinfo  {journal} {Phys. Rev. X}\ }\textbf {\bibinfo {volume} {13}},\ \bibinfo {pages} {041039} (\bibinfo {year} {2023}{\natexlab{a}})},\ \Eprint {http://arxiv.org/abs/2111.03606} {arXiv:2111.03606 [gr-qc]} \BibitemShut {NoStop}%
\bibitem [{\citenamefont {Ohanian}(1974)}]{Ohanian:1974ys}%
  \BibitemOpen
  \bibfield  {author} {\bibinfo {author} {\bibfnamefont {H.~C.}\ \bibnamefont {Ohanian}},\ }\href {\doibase 10.1007/BF01810927} {\bibfield  {journal} {\bibinfo  {journal} {Int. J. Theor. Phys.}\ }\textbf {\bibinfo {volume} {9}},\ \bibinfo {pages} {425} (\bibinfo {year} {1974})}\BibitemShut {NoStop}%
\bibitem [{\citenamefont {Oguri}(2019)}]{Oguri:2019fix}%
  \BibitemOpen
  \bibfield  {author} {\bibinfo {author} {\bibfnamefont {M.}~\bibnamefont {Oguri}},\ }\href {\doibase 10.1088/1361-6633/ab4fc5} {\bibfield  {journal} {\bibinfo  {journal} {Rept. Prog. Phys.}\ }\textbf {\bibinfo {volume} {82}},\ \bibinfo {pages} {126901} (\bibinfo {year} {2019})},\ \Eprint {http://arxiv.org/abs/1907.06830} {arXiv:1907.06830 [astro-ph.CO]} \BibitemShut {NoStop}%
\bibitem [{\citenamefont {Liao}\ \emph {et~al.}(2022)\citenamefont {Liao}, \citenamefont {Biesiada},\ and\ \citenamefont {Zhu}}]{Liao:2022gde}%
  \BibitemOpen
  \bibfield  {author} {\bibinfo {author} {\bibfnamefont {K.}~\bibnamefont {Liao}}, \bibinfo {author} {\bibfnamefont {M.}~\bibnamefont {Biesiada}}, \ and\ \bibinfo {author} {\bibfnamefont {Z.-H.}\ \bibnamefont {Zhu}},\ }\href {\doibase 10.1088/0256-307X/39/11/119801} {\bibfield  {journal} {\bibinfo  {journal} {Chin. Phys. Lett.}\ }\textbf {\bibinfo {volume} {39}},\ \bibinfo {pages} {119801} (\bibinfo {year} {2022})},\ \Eprint {http://arxiv.org/abs/2207.13489} {arXiv:2207.13489 [astro-ph.HE]} \BibitemShut {NoStop}%
\bibitem [{\citenamefont {Sereno}\ \emph {et~al.}(2011)\citenamefont {Sereno}, \citenamefont {Jetzer}, \citenamefont {Sesana},\ and\ \citenamefont {Volonteri}}]{Sereno:2011ty}%
  \BibitemOpen
  \bibfield  {author} {\bibinfo {author} {\bibfnamefont {M.}~\bibnamefont {Sereno}}, \bibinfo {author} {\bibfnamefont {P.}~\bibnamefont {Jetzer}}, \bibinfo {author} {\bibfnamefont {A.}~\bibnamefont {Sesana}}, \ and\ \bibinfo {author} {\bibfnamefont {M.}~\bibnamefont {Volonteri}},\ }\href {\doibase 10.1111/j.1365-2966.2011.18895.x} {\bibfield  {journal} {\bibinfo  {journal} {Mon. Not. Roy. Astron. Soc.}\ }\textbf {\bibinfo {volume} {415}},\ \bibinfo {pages} {2773} (\bibinfo {year} {2011})},\ \Eprint {http://arxiv.org/abs/1104.1977} {arXiv:1104.1977 [astro-ph.CO]} \BibitemShut {NoStop}%
\bibitem [{\citenamefont {Liao}\ \emph {et~al.}(2017)\citenamefont {Liao}, \citenamefont {Fan}, \citenamefont {Ding}, \citenamefont {Biesiada},\ and\ \citenamefont {Zhu}}]{Liao:2017ioi}%
  \BibitemOpen
  \bibfield  {author} {\bibinfo {author} {\bibfnamefont {K.}~\bibnamefont {Liao}}, \bibinfo {author} {\bibfnamefont {X.-L.}\ \bibnamefont {Fan}}, \bibinfo {author} {\bibfnamefont {X.-H.}\ \bibnamefont {Ding}}, \bibinfo {author} {\bibfnamefont {M.}~\bibnamefont {Biesiada}}, \ and\ \bibinfo {author} {\bibfnamefont {Z.-H.}\ \bibnamefont {Zhu}},\ }\href {\doibase 10.1038/s41467-017-01152-9} {\bibfield  {journal} {\bibinfo  {journal} {Nature Commun.}\ }\textbf {\bibinfo {volume} {8}},\ \bibinfo {pages} {1148} (\bibinfo {year} {2017})},\ \bibinfo {note} {[Erratum: Nature Commun. 8, 2136 (2017)]},\ \Eprint {http://arxiv.org/abs/1703.04151} {arXiv:1703.04151 [astro-ph.CO]} \BibitemShut {NoStop}%
\bibitem [{\citenamefont {Cao}\ \emph {et~al.}(2019)\citenamefont {Cao}, \citenamefont {Qi}, \citenamefont {Cao}, \citenamefont {Biesiada}, \citenamefont {Li}, \citenamefont {Pan},\ and\ \citenamefont {Zhu}}]{Cao:2019kgn}%
  \BibitemOpen
  \bibfield  {author} {\bibinfo {author} {\bibfnamefont {S.}~\bibnamefont {Cao}}, \bibinfo {author} {\bibfnamefont {J.}~\bibnamefont {Qi}}, \bibinfo {author} {\bibfnamefont {Z.}~\bibnamefont {Cao}}, \bibinfo {author} {\bibfnamefont {M.}~\bibnamefont {Biesiada}}, \bibinfo {author} {\bibfnamefont {J.}~\bibnamefont {Li}}, \bibinfo {author} {\bibfnamefont {Y.}~\bibnamefont {Pan}}, \ and\ \bibinfo {author} {\bibfnamefont {Z.-H.}\ \bibnamefont {Zhu}},\ }\href {\doibase 10.1038/s41598-019-47616-4} {\bibfield  {journal} {\bibinfo  {journal} {Sci. Rep.}\ }\textbf {\bibinfo {volume} {9}},\ \bibinfo {pages} {11608} (\bibinfo {year} {2019})},\ \Eprint {http://arxiv.org/abs/1910.10365} {arXiv:1910.10365 [astro-ph.CO]} \BibitemShut {NoStop}%
\bibitem [{\citenamefont {Jana}\ \emph {et~al.}(2023)\citenamefont {Jana}, \citenamefont {Kapadia}, \citenamefont {Venumadhav},\ and\ \citenamefont {Ajith}}]{Jana:2022shb}%
  \BibitemOpen
  \bibfield  {author} {\bibinfo {author} {\bibfnamefont {S.}~\bibnamefont {Jana}}, \bibinfo {author} {\bibfnamefont {S.~J.}\ \bibnamefont {Kapadia}}, \bibinfo {author} {\bibfnamefont {T.}~\bibnamefont {Venumadhav}}, \ and\ \bibinfo {author} {\bibfnamefont {P.}~\bibnamefont {Ajith}},\ }\href {\doibase 10.1103/PhysRevLett.130.261401} {\bibfield  {journal} {\bibinfo  {journal} {Phys. Rev. Lett.}\ }\textbf {\bibinfo {volume} {130}},\ \bibinfo {pages} {261401} (\bibinfo {year} {2023})},\ \Eprint {http://arxiv.org/abs/2211.12212} {arXiv:2211.12212 [astro-ph.CO]} \BibitemShut {NoStop}%
\bibitem [{\citenamefont {Cremonese}\ \emph {et~al.}(2021)\citenamefont {Cremonese}, \citenamefont {Ezquiaga},\ and\ \citenamefont {Salzano}}]{Cremonese:2021puh}%
  \BibitemOpen
  \bibfield  {author} {\bibinfo {author} {\bibfnamefont {P.}~\bibnamefont {Cremonese}}, \bibinfo {author} {\bibfnamefont {J.~M.}\ \bibnamefont {Ezquiaga}}, \ and\ \bibinfo {author} {\bibfnamefont {V.}~\bibnamefont {Salzano}},\ }\href {\doibase 10.1103/PhysRevD.104.023503} {\bibfield  {journal} {\bibinfo  {journal} {Phys. Rev. D}\ }\textbf {\bibinfo {volume} {104}},\ \bibinfo {pages} {023503} (\bibinfo {year} {2021})},\ \Eprint {http://arxiv.org/abs/2104.07055} {arXiv:2104.07055 [astro-ph.CO]} \BibitemShut {NoStop}%
\bibitem [{\citenamefont {Ezquiaga}\ and\ \citenamefont {Zumalac\'arregui}(2020)}]{Ezquiaga:2020dao}%
  \BibitemOpen
  \bibfield  {author} {\bibinfo {author} {\bibfnamefont {J.~M.}\ \bibnamefont {Ezquiaga}}\ and\ \bibinfo {author} {\bibfnamefont {M.}~\bibnamefont {Zumalac\'arregui}},\ }\href {\doibase 10.1103/PhysRevD.102.124048} {\bibfield  {journal} {\bibinfo  {journal} {Phys. Rev. D}\ }\textbf {\bibinfo {volume} {102}},\ \bibinfo {pages} {124048} (\bibinfo {year} {2020})},\ \Eprint {http://arxiv.org/abs/2009.12187} {arXiv:2009.12187 [gr-qc]} \BibitemShut {NoStop}%
\bibitem [{\citenamefont {Wang}\ \emph {et~al.}(2022)\citenamefont {Wang}, \citenamefont {Brown}, \citenamefont {Shao},\ and\ \citenamefont {Zhao}}]{Wang:2021gqm}%
  \BibitemOpen
  \bibfield  {author} {\bibinfo {author} {\bibfnamefont {Y.-F.}\ \bibnamefont {Wang}}, \bibinfo {author} {\bibfnamefont {S.~M.}\ \bibnamefont {Brown}}, \bibinfo {author} {\bibfnamefont {L.}~\bibnamefont {Shao}}, \ and\ \bibinfo {author} {\bibfnamefont {W.}~\bibnamefont {Zhao}},\ }\href {\doibase 10.1103/PhysRevD.106.084005} {\bibfield  {journal} {\bibinfo  {journal} {Phys. Rev. D}\ }\textbf {\bibinfo {volume} {106}},\ \bibinfo {pages} {084005} (\bibinfo {year} {2022})},\ \Eprint {http://arxiv.org/abs/2109.09718} {arXiv:2109.09718 [astro-ph.HE]} \BibitemShut {NoStop}%
\bibitem [{\citenamefont {Goyal}\ \emph {et~al.}(2023)\citenamefont {Goyal}, \citenamefont {Vijaykumar}, \citenamefont {Ezquiaga},\ and\ \citenamefont {Zumalacarregui}}]{Goyal:2023uvm}%
  \BibitemOpen
  \bibfield  {author} {\bibinfo {author} {\bibfnamefont {S.}~\bibnamefont {Goyal}}, \bibinfo {author} {\bibfnamefont {A.}~\bibnamefont {Vijaykumar}}, \bibinfo {author} {\bibfnamefont {J.~M.}\ \bibnamefont {Ezquiaga}}, \ and\ \bibinfo {author} {\bibfnamefont {M.}~\bibnamefont {Zumalacarregui}},\ }\href {\doibase 10.1103/PhysRevD.108.024052} {\bibfield  {journal} {\bibinfo  {journal} {Phys. Rev. D}\ }\textbf {\bibinfo {volume} {108}},\ \bibinfo {pages} {024052} (\bibinfo {year} {2023})},\ \Eprint {http://arxiv.org/abs/2301.04826} {arXiv:2301.04826 [gr-qc]} \BibitemShut {NoStop}%
\bibitem [{\citenamefont {Baker}\ and\ \citenamefont {Trodden}(2017)}]{Baker:2016reh}%
  \BibitemOpen
  \bibfield  {author} {\bibinfo {author} {\bibfnamefont {T.}~\bibnamefont {Baker}}\ and\ \bibinfo {author} {\bibfnamefont {M.}~\bibnamefont {Trodden}},\ }\href {\doibase 10.1103/PhysRevD.95.063512} {\bibfield  {journal} {\bibinfo  {journal} {Phys. Rev. D}\ }\textbf {\bibinfo {volume} {95}},\ \bibinfo {pages} {063512} (\bibinfo {year} {2017})},\ \Eprint {http://arxiv.org/abs/1612.02004} {arXiv:1612.02004 [astro-ph.CO]} \BibitemShut {NoStop}%
\bibitem [{\citenamefont {Collett}\ and\ \citenamefont {Bacon}(2017)}]{Collett:2016dey}%
  \BibitemOpen
  \bibfield  {author} {\bibinfo {author} {\bibfnamefont {T.~E.}\ \bibnamefont {Collett}}\ and\ \bibinfo {author} {\bibfnamefont {D.}~\bibnamefont {Bacon}},\ }\href {\doibase 10.1103/PhysRevLett.118.091101} {\bibfield  {journal} {\bibinfo  {journal} {Phys. Rev. Lett.}\ }\textbf {\bibinfo {volume} {118}},\ \bibinfo {pages} {091101} (\bibinfo {year} {2017})},\ \Eprint {http://arxiv.org/abs/1602.05882} {arXiv:1602.05882 [astro-ph.HE]} \BibitemShut {NoStop}%
\bibitem [{\citenamefont {Hannuksela}\ \emph {et~al.}(2020)\citenamefont {Hannuksela}, \citenamefont {Collett}, \citenamefont {\c{C}al\i{}\c{s}kan},\ and\ \citenamefont {Li}}]{Hannuksela:2020xor}%
  \BibitemOpen
  \bibfield  {author} {\bibinfo {author} {\bibfnamefont {O.~A.}\ \bibnamefont {Hannuksela}}, \bibinfo {author} {\bibfnamefont {T.~E.}\ \bibnamefont {Collett}}, \bibinfo {author} {\bibfnamefont {M.}~\bibnamefont {\c{C}al\i{}\c{s}kan}}, \ and\ \bibinfo {author} {\bibfnamefont {T.~G.~F.}\ \bibnamefont {Li}},\ }\href {\doibase 10.1093/mnras/staa2577} {\bibfield  {journal} {\bibinfo  {journal} {Mon. Not. Roy. Astron. Soc.}\ }\textbf {\bibinfo {volume} {498}},\ \bibinfo {pages} {3395} (\bibinfo {year} {2020})},\ \Eprint {http://arxiv.org/abs/2004.13811} {arXiv:2004.13811 [astro-ph.HE]} \BibitemShut {NoStop}%
\bibitem [{\citenamefont {Yu}\ \emph {et~al.}(2020)\citenamefont {Yu}, \citenamefont {Zhang},\ and\ \citenamefont {Wang}}]{Yu:2020agu}%
  \BibitemOpen
  \bibfield  {author} {\bibinfo {author} {\bibfnamefont {H.}~\bibnamefont {Yu}}, \bibinfo {author} {\bibfnamefont {P.}~\bibnamefont {Zhang}}, \ and\ \bibinfo {author} {\bibfnamefont {F.-Y.}\ \bibnamefont {Wang}},\ }\href {\doibase 10.1093/mnras/staa1952} {\bibfield  {journal} {\bibinfo  {journal} {Mon. Not. Roy. Astron. Soc.}\ }\textbf {\bibinfo {volume} {497}},\ \bibinfo {pages} {204} (\bibinfo {year} {2020})},\ \Eprint {http://arxiv.org/abs/2007.00828} {arXiv:2007.00828 [astro-ph.CO]} \BibitemShut {NoStop}%
\bibitem [{\citenamefont {Takahashi}\ and\ \citenamefont {Nakamura}(2003)}]{Takahashi:2003ix}%
  \BibitemOpen
  \bibfield  {author} {\bibinfo {author} {\bibfnamefont {R.}~\bibnamefont {Takahashi}}\ and\ \bibinfo {author} {\bibfnamefont {T.}~\bibnamefont {Nakamura}},\ }\href {\doibase 10.1086/377430} {\bibfield  {journal} {\bibinfo  {journal} {Astrophys. J.}\ }\textbf {\bibinfo {volume} {595}},\ \bibinfo {pages} {1039} (\bibinfo {year} {2003})},\ \Eprint {http://arxiv.org/abs/astro-ph/0305055} {arXiv:astro-ph/0305055} \BibitemShut {NoStop}%
\bibitem [{\citenamefont {Ezquiaga}\ \emph {et~al.}(2021)\citenamefont {Ezquiaga}, \citenamefont {Holz}, \citenamefont {Hu}, \citenamefont {Lagos},\ and\ \citenamefont {Wald}}]{Ezquiaga:2020gdt}%
  \BibitemOpen
  \bibfield  {author} {\bibinfo {author} {\bibfnamefont {J.~M.}\ \bibnamefont {Ezquiaga}}, \bibinfo {author} {\bibfnamefont {D.~E.}\ \bibnamefont {Holz}}, \bibinfo {author} {\bibfnamefont {W.}~\bibnamefont {Hu}}, \bibinfo {author} {\bibfnamefont {M.}~\bibnamefont {Lagos}}, \ and\ \bibinfo {author} {\bibfnamefont {R.~M.}\ \bibnamefont {Wald}},\ }\href {\doibase 10.1103/PhysRevD.103.064047} {\bibfield  {journal} {\bibinfo  {journal} {Phys. Rev. D}\ }\textbf {\bibinfo {volume} {103}},\ \bibinfo {pages} {064047} (\bibinfo {year} {2021})},\ \Eprint {http://arxiv.org/abs/2008.12814} {arXiv:2008.12814 [gr-qc]} \BibitemShut {NoStop}%
\bibitem [{\citenamefont {Dai}\ and\ \citenamefont {Venumadhav}(2017)}]{Dai:2017huk}%
  \BibitemOpen
  \bibfield  {author} {\bibinfo {author} {\bibfnamefont {L.}~\bibnamefont {Dai}}\ and\ \bibinfo {author} {\bibfnamefont {T.}~\bibnamefont {Venumadhav}},\ }\href@noop {} {\  (\bibinfo {year} {2017})},\ \Eprint {http://arxiv.org/abs/1702.04724} {arXiv:1702.04724 [gr-qc]} \BibitemShut {NoStop}%
\bibitem [{\citenamefont {Lai}\ \emph {et~al.}(2018)\citenamefont {Lai}, \citenamefont {Hannuksela}, \citenamefont {Herrera-Mart\'\i{}n}, \citenamefont {Diego}, \citenamefont {Broadhurst},\ and\ \citenamefont {Li}}]{Lai:2018rto}%
  \BibitemOpen
  \bibfield  {author} {\bibinfo {author} {\bibfnamefont {K.-H.}\ \bibnamefont {Lai}}, \bibinfo {author} {\bibfnamefont {O.~A.}\ \bibnamefont {Hannuksela}}, \bibinfo {author} {\bibfnamefont {A.}~\bibnamefont {Herrera-Mart\'\i{}n}}, \bibinfo {author} {\bibfnamefont {J.~M.}\ \bibnamefont {Diego}}, \bibinfo {author} {\bibfnamefont {T.}~\bibnamefont {Broadhurst}}, \ and\ \bibinfo {author} {\bibfnamefont {T.~G.~F.}\ \bibnamefont {Li}},\ }\href {\doibase 10.1103/PhysRevD.98.083005} {\bibfield  {journal} {\bibinfo  {journal} {Phys. Rev. D}\ }\textbf {\bibinfo {volume} {98}},\ \bibinfo {pages} {083005} (\bibinfo {year} {2018})},\ \Eprint {http://arxiv.org/abs/1801.07840} {arXiv:1801.07840 [gr-qc]} \BibitemShut {NoStop}%
\bibitem [{\citenamefont {Gais}\ \emph {et~al.}(2022)\citenamefont {Gais}, \citenamefont {Ng}, \citenamefont {Seo}, \citenamefont {Wong},\ and\ \citenamefont {Li}}]{Gais:2022xir}%
  \BibitemOpen
  \bibfield  {author} {\bibinfo {author} {\bibfnamefont {J.}~\bibnamefont {Gais}}, \bibinfo {author} {\bibfnamefont {K.~K.~Y.}\ \bibnamefont {Ng}}, \bibinfo {author} {\bibfnamefont {E.}~\bibnamefont {Seo}}, \bibinfo {author} {\bibfnamefont {K.~W.~K.}\ \bibnamefont {Wong}}, \ and\ \bibinfo {author} {\bibfnamefont {T.~G.~F.}\ \bibnamefont {Li}},\ }\href {\doibase 10.3847/2041-8213/ac7052} {\bibfield  {journal} {\bibinfo  {journal} {Astrophys. J. Lett.}\ }\textbf {\bibinfo {volume} {932}},\ \bibinfo {pages} {L4} (\bibinfo {year} {2022})},\ \Eprint {http://arxiv.org/abs/2201.01817} {arXiv:2201.01817 [gr-qc]} \BibitemShut {NoStop}%
\bibitem [{\citenamefont {Liu}\ \emph {et~al.}(2023)\citenamefont {Liu}, \citenamefont {Wong}, \citenamefont {Leong}, \citenamefont {More}, \citenamefont {Hannuksela},\ and\ \citenamefont {Li}}]{Liu:2023ikc}%
  \BibitemOpen
  \bibfield  {author} {\bibinfo {author} {\bibfnamefont {A.}~\bibnamefont {Liu}}, \bibinfo {author} {\bibfnamefont {I.~C.~F.}\ \bibnamefont {Wong}}, \bibinfo {author} {\bibfnamefont {S.~H.~W.}\ \bibnamefont {Leong}}, \bibinfo {author} {\bibfnamefont {A.}~\bibnamefont {More}}, \bibinfo {author} {\bibfnamefont {O.~A.}\ \bibnamefont {Hannuksela}}, \ and\ \bibinfo {author} {\bibfnamefont {T.~G.~F.}\ \bibnamefont {Li}},\ }\href {\doibase 10.1093/mnras/stad1302} {\bibfield  {journal} {\bibinfo  {journal} {Mon. Not. Roy. Astron. Soc.}\ }\textbf {\bibinfo {volume} {525}},\ \bibinfo {pages} {4149} (\bibinfo {year} {2023})},\ \Eprint {http://arxiv.org/abs/2302.09870} {arXiv:2302.09870 [gr-qc]} \BibitemShut {NoStop}%
\bibitem [{\citenamefont {Basak}\ \emph {et~al.}(2022)\citenamefont {Basak}, \citenamefont {Ganguly}, \citenamefont {Haris}, \citenamefont {Kapadia}, \citenamefont {Mehta},\ and\ \citenamefont {Ajith}}]{Basak:2021ten}%
  \BibitemOpen
  \bibfield  {author} {\bibinfo {author} {\bibfnamefont {S.}~\bibnamefont {Basak}}, \bibinfo {author} {\bibfnamefont {A.}~\bibnamefont {Ganguly}}, \bibinfo {author} {\bibfnamefont {K.}~\bibnamefont {Haris}}, \bibinfo {author} {\bibfnamefont {S.}~\bibnamefont {Kapadia}}, \bibinfo {author} {\bibfnamefont {A.~K.}\ \bibnamefont {Mehta}}, \ and\ \bibinfo {author} {\bibfnamefont {P.}~\bibnamefont {Ajith}},\ }\href {\doibase 10.3847/2041-8213/ac4dfa} {\bibfield  {journal} {\bibinfo  {journal} {Astrophys. J. Lett.}\ }\textbf {\bibinfo {volume} {926}},\ \bibinfo {pages} {L28} (\bibinfo {year} {2022})},\ \Eprint {http://arxiv.org/abs/2109.06456} {arXiv:2109.06456 [gr-qc]} \BibitemShut {NoStop}%
\bibitem [{\citenamefont {Meena}\ and\ \citenamefont {Bagla}(2020)}]{Meena:2019ate}%
  \BibitemOpen
  \bibfield  {author} {\bibinfo {author} {\bibfnamefont {A.~K.}\ \bibnamefont {Meena}}\ and\ \bibinfo {author} {\bibfnamefont {J.~S.}\ \bibnamefont {Bagla}},\ }\href {\doibase 10.1093/mnras/stz3509} {\bibfield  {journal} {\bibinfo  {journal} {Mon. Not. Roy. Astron. Soc.}\ }\textbf {\bibinfo {volume} {492}},\ \bibinfo {pages} {1127} (\bibinfo {year} {2020})},\ \Eprint {http://arxiv.org/abs/1903.11809} {arXiv:1903.11809 [astro-ph.CO]} \BibitemShut {NoStop}%
\bibitem [{\citenamefont {Jung}\ and\ \citenamefont {Shin}(2019)}]{Jung:2017flg}%
  \BibitemOpen
  \bibfield  {author} {\bibinfo {author} {\bibfnamefont {S.}~\bibnamefont {Jung}}\ and\ \bibinfo {author} {\bibfnamefont {C.~S.}\ \bibnamefont {Shin}},\ }\href {\doibase 10.1103/PhysRevLett.122.041103} {\bibfield  {journal} {\bibinfo  {journal} {Phys. Rev. Lett.}\ }\textbf {\bibinfo {volume} {122}},\ \bibinfo {pages} {041103} (\bibinfo {year} {2019})},\ \Eprint {http://arxiv.org/abs/1712.01396} {arXiv:1712.01396 [astro-ph.CO]} \BibitemShut {NoStop}%
\bibitem [{\citenamefont {Diego}(2020)}]{Diego:2019rzc}%
  \BibitemOpen
  \bibfield  {author} {\bibinfo {author} {\bibfnamefont {J.~M.}\ \bibnamefont {Diego}},\ }\href {\doibase 10.1103/PhysRevD.101.123512} {\bibfield  {journal} {\bibinfo  {journal} {Phys. Rev. D}\ }\textbf {\bibinfo {volume} {101}},\ \bibinfo {pages} {123512} (\bibinfo {year} {2020})},\ \Eprint {http://arxiv.org/abs/1911.05736} {arXiv:1911.05736 [astro-ph.CO]} \BibitemShut {NoStop}%
\bibitem [{\citenamefont {Urrutia}\ and\ \citenamefont {Vaskonen}(2021)}]{Urrutia:2021qak}%
  \BibitemOpen
  \bibfield  {author} {\bibinfo {author} {\bibfnamefont {J.}~\bibnamefont {Urrutia}}\ and\ \bibinfo {author} {\bibfnamefont {V.}~\bibnamefont {Vaskonen}},\ }\href {\doibase 10.1093/mnras/stab3118} {\bibfield  {journal} {\bibinfo  {journal} {Mon. Not. Roy. Astron. Soc.}\ }\textbf {\bibinfo {volume} {509}},\ \bibinfo {pages} {1358} (\bibinfo {year} {2021})},\ \Eprint {http://arxiv.org/abs/2109.03213} {arXiv:2109.03213 [astro-ph.CO]} \BibitemShut {NoStop}%
\bibitem [{\citenamefont {Zhou}\ \emph {et~al.}(2022)\citenamefont {Zhou}, \citenamefont {Li}, \citenamefont {Liao},\ and\ \citenamefont {Huang}}]{Zhou:2022yeo}%
  \BibitemOpen
  \bibfield  {author} {\bibinfo {author} {\bibfnamefont {H.}~\bibnamefont {Zhou}}, \bibinfo {author} {\bibfnamefont {Z.}~\bibnamefont {Li}}, \bibinfo {author} {\bibfnamefont {K.}~\bibnamefont {Liao}}, \ and\ \bibinfo {author} {\bibfnamefont {Z.}~\bibnamefont {Huang}},\ }\href {\doibase 10.1093/mnras/stac2944} {\bibfield  {journal} {\bibinfo  {journal} {Mon. Not. Roy. Astron. Soc.}\ }\textbf {\bibinfo {volume} {518}},\ \bibinfo {pages} {149} (\bibinfo {year} {2022})},\ \Eprint {http://arxiv.org/abs/2206.13128} {arXiv:2206.13128 [astro-ph.CO]} \BibitemShut {NoStop}%
\bibitem [{\citenamefont {Tambalo}\ \emph {et~al.}(2023{\natexlab{a}})\citenamefont {Tambalo}, \citenamefont {Zumalac\'arregui}, \citenamefont {Dai},\ and\ \citenamefont {Cheung}}]{Tambalo:2022wlm}%
  \BibitemOpen
  \bibfield  {author} {\bibinfo {author} {\bibfnamefont {G.}~\bibnamefont {Tambalo}}, \bibinfo {author} {\bibfnamefont {M.}~\bibnamefont {Zumalac\'arregui}}, \bibinfo {author} {\bibfnamefont {L.}~\bibnamefont {Dai}}, \ and\ \bibinfo {author} {\bibfnamefont {M.~H.-Y.}\ \bibnamefont {Cheung}},\ }\href {\doibase 10.1103/PhysRevD.108.103529} {\bibfield  {journal} {\bibinfo  {journal} {Phys. Rev. D}\ }\textbf {\bibinfo {volume} {108}},\ \bibinfo {pages} {103529} (\bibinfo {year} {2023}{\natexlab{a}})},\ \Eprint {http://arxiv.org/abs/2212.11960} {arXiv:2212.11960 [astro-ph.CO]} \BibitemShut {NoStop}%
\bibitem [{\citenamefont {Dai}\ \emph {et~al.}(2018)\citenamefont {Dai}, \citenamefont {Li}, \citenamefont {Zackay}, \citenamefont {Mao},\ and\ \citenamefont {Lu}}]{Dai:2018enj}%
  \BibitemOpen
  \bibfield  {author} {\bibinfo {author} {\bibfnamefont {L.}~\bibnamefont {Dai}}, \bibinfo {author} {\bibfnamefont {S.-S.}\ \bibnamefont {Li}}, \bibinfo {author} {\bibfnamefont {B.}~\bibnamefont {Zackay}}, \bibinfo {author} {\bibfnamefont {S.}~\bibnamefont {Mao}}, \ and\ \bibinfo {author} {\bibfnamefont {Y.}~\bibnamefont {Lu}},\ }\href {\doibase 10.1103/PhysRevD.98.104029} {\bibfield  {journal} {\bibinfo  {journal} {Phys. Rev. D}\ }\textbf {\bibinfo {volume} {98}},\ \bibinfo {pages} {104029} (\bibinfo {year} {2018})},\ \Eprint {http://arxiv.org/abs/1810.00003} {arXiv:1810.00003 [gr-qc]} \BibitemShut {NoStop}%
\bibitem [{\citenamefont {Oguri}\ and\ \citenamefont {Takahashi}(2020)}]{Oguri:2020ldf}%
  \BibitemOpen
  \bibfield  {author} {\bibinfo {author} {\bibfnamefont {M.}~\bibnamefont {Oguri}}\ and\ \bibinfo {author} {\bibfnamefont {R.}~\bibnamefont {Takahashi}},\ }\href {\doibase 10.3847/1538-4357/abafab} {\bibfield  {journal} {\bibinfo  {journal} {Astrophys. J.}\ }\textbf {\bibinfo {volume} {901}},\ \bibinfo {pages} {58} (\bibinfo {year} {2020})},\ \Eprint {http://arxiv.org/abs/2007.01936} {arXiv:2007.01936 [astro-ph.CO]} \BibitemShut {NoStop}%
\bibitem [{\citenamefont {Guo}\ and\ \citenamefont {Lu}(2022)}]{Guo:2022dre}%
  \BibitemOpen
  \bibfield  {author} {\bibinfo {author} {\bibfnamefont {X.}~\bibnamefont {Guo}}\ and\ \bibinfo {author} {\bibfnamefont {Y.}~\bibnamefont {Lu}},\ }\href {\doibase 10.1103/PhysRevD.106.023018} {\bibfield  {journal} {\bibinfo  {journal} {Phys. Rev. D}\ }\textbf {\bibinfo {volume} {106}},\ \bibinfo {pages} {023018} (\bibinfo {year} {2022})},\ \Eprint {http://arxiv.org/abs/2207.00325} {arXiv:2207.00325 [astro-ph.CO]} \BibitemShut {NoStop}%
\bibitem [{\citenamefont {\c{C}al\i{}\c{s}kan}\ \emph {et~al.}(2023{\natexlab{a}})\citenamefont {\c{C}al\i{}\c{s}kan}, \citenamefont {Ji}, \citenamefont {Cotesta}, \citenamefont {Berti}, \citenamefont {Kamionkowski},\ and\ \citenamefont {Marsat}}]{Caliskan:2022hbu}%
  \BibitemOpen
  \bibfield  {author} {\bibinfo {author} {\bibfnamefont {M.}~\bibnamefont {\c{C}al\i{}\c{s}kan}}, \bibinfo {author} {\bibfnamefont {L.}~\bibnamefont {Ji}}, \bibinfo {author} {\bibfnamefont {R.}~\bibnamefont {Cotesta}}, \bibinfo {author} {\bibfnamefont {E.}~\bibnamefont {Berti}}, \bibinfo {author} {\bibfnamefont {M.}~\bibnamefont {Kamionkowski}}, \ and\ \bibinfo {author} {\bibfnamefont {S.}~\bibnamefont {Marsat}},\ }\href {\doibase 10.1103/PhysRevD.107.043029} {\bibfield  {journal} {\bibinfo  {journal} {Phys. Rev. D}\ }\textbf {\bibinfo {volume} {107}},\ \bibinfo {pages} {043029} (\bibinfo {year} {2023}{\natexlab{a}})},\ \Eprint {http://arxiv.org/abs/2206.02803} {arXiv:2206.02803 [astro-ph.CO]} \BibitemShut {NoStop}%
\bibitem [{\citenamefont {Savastano}\ \emph {et~al.}(2023)\citenamefont {Savastano}, \citenamefont {Tambalo}, \citenamefont {Villarrubia-Rojo},\ and\ \citenamefont {Zumalacarregui}}]{Savastano:2023spl}%
  \BibitemOpen
  \bibfield  {author} {\bibinfo {author} {\bibfnamefont {S.}~\bibnamefont {Savastano}}, \bibinfo {author} {\bibfnamefont {G.}~\bibnamefont {Tambalo}}, \bibinfo {author} {\bibfnamefont {H.}~\bibnamefont {Villarrubia-Rojo}}, \ and\ \bibinfo {author} {\bibfnamefont {M.}~\bibnamefont {Zumalacarregui}},\ }\href {\doibase 10.1103/PhysRevD.108.103532} {\bibfield  {journal} {\bibinfo  {journal} {Phys. Rev. D}\ }\textbf {\bibinfo {volume} {108}},\ \bibinfo {pages} {103532} (\bibinfo {year} {2023})},\ \Eprint {http://arxiv.org/abs/2306.05282} {arXiv:2306.05282 [gr-qc]} \BibitemShut {NoStop}%
\bibitem [{\citenamefont {\c{C}al\i{}\c{s}kan}\ \emph {et~al.}(2023{\natexlab{b}})\citenamefont {\c{C}al\i{}\c{s}kan}, \citenamefont {Anil~Kumar}, \citenamefont {Ji}, \citenamefont {Ezquiaga}, \citenamefont {Cotesta}, \citenamefont {Berti},\ and\ \citenamefont {Kamionkowski}}]{Caliskan:2023zqm}%
  \BibitemOpen
  \bibfield  {author} {\bibinfo {author} {\bibfnamefont {M.}~\bibnamefont {\c{C}al\i{}\c{s}kan}}, \bibinfo {author} {\bibfnamefont {N.}~\bibnamefont {Anil~Kumar}}, \bibinfo {author} {\bibfnamefont {L.}~\bibnamefont {Ji}}, \bibinfo {author} {\bibfnamefont {J.~M.}\ \bibnamefont {Ezquiaga}}, \bibinfo {author} {\bibfnamefont {R.}~\bibnamefont {Cotesta}}, \bibinfo {author} {\bibfnamefont {E.}~\bibnamefont {Berti}}, \ and\ \bibinfo {author} {\bibfnamefont {M.}~\bibnamefont {Kamionkowski}},\ }\href {\doibase 10.1103/PhysRevD.108.123543} {\bibfield  {journal} {\bibinfo  {journal} {Phys. Rev. D}\ }\textbf {\bibinfo {volume} {108}},\ \bibinfo {pages} {123543} (\bibinfo {year} {2023}{\natexlab{b}})},\ \Eprint {http://arxiv.org/abs/2307.06990} {arXiv:2307.06990 [astro-ph.CO]} \BibitemShut {NoStop}%
\bibitem [{\citenamefont {Sereno}\ \emph {et~al.}(2010)\citenamefont {Sereno}, \citenamefont {Sesana}, \citenamefont {Bleuler}, \citenamefont {Jetzer}, \citenamefont {Volonteri},\ and\ \citenamefont {Begelman}}]{Sereno:2010dr}%
  \BibitemOpen
  \bibfield  {author} {\bibinfo {author} {\bibfnamefont {M.}~\bibnamefont {Sereno}}, \bibinfo {author} {\bibfnamefont {A.}~\bibnamefont {Sesana}}, \bibinfo {author} {\bibfnamefont {A.}~\bibnamefont {Bleuler}}, \bibinfo {author} {\bibfnamefont {P.}~\bibnamefont {Jetzer}}, \bibinfo {author} {\bibfnamefont {M.}~\bibnamefont {Volonteri}}, \ and\ \bibinfo {author} {\bibfnamefont {M.~C.}\ \bibnamefont {Begelman}},\ }\href {\doibase 10.1103/PhysRevLett.105.251101} {\bibfield  {journal} {\bibinfo  {journal} {Phys. Rev. Lett.}\ }\textbf {\bibinfo {volume} {105}},\ \bibinfo {pages} {251101} (\bibinfo {year} {2010})},\ \Eprint {http://arxiv.org/abs/1011.5238} {arXiv:1011.5238 [astro-ph.CO]} \BibitemShut {NoStop}%
\bibitem [{\citenamefont {Gao}\ \emph {et~al.}(2022)\citenamefont {Gao}, \citenamefont {Chen}, \citenamefont {Hu}, \citenamefont {Zhang},\ and\ \citenamefont {Huang}}]{Gao:2021sxw}%
  \BibitemOpen
  \bibfield  {author} {\bibinfo {author} {\bibfnamefont {Z.}~\bibnamefont {Gao}}, \bibinfo {author} {\bibfnamefont {X.}~\bibnamefont {Chen}}, \bibinfo {author} {\bibfnamefont {Y.-M.}\ \bibnamefont {Hu}}, \bibinfo {author} {\bibfnamefont {J.-D.}\ \bibnamefont {Zhang}}, \ and\ \bibinfo {author} {\bibfnamefont {S.-J.}\ \bibnamefont {Huang}},\ }\href {\doibase 10.1093/mnras/stac365} {\bibfield  {journal} {\bibinfo  {journal} {Mon. Not. Roy. Astron. Soc.}\ }\textbf {\bibinfo {volume} {512}},\ \bibinfo {pages} {1} (\bibinfo {year} {2022})},\ \Eprint {http://arxiv.org/abs/2102.10295} {arXiv:2102.10295 [astro-ph.CO]} \BibitemShut {NoStop}%
\bibitem [{\citenamefont {Choi}\ \emph {et~al.}(2021)\citenamefont {Choi}, \citenamefont {Park},\ and\ \citenamefont {Jung}}]{Choi:2021bkx}%
  \BibitemOpen
  \bibfield  {author} {\bibinfo {author} {\bibfnamefont {H.~G.}\ \bibnamefont {Choi}}, \bibinfo {author} {\bibfnamefont {C.}~\bibnamefont {Park}}, \ and\ \bibinfo {author} {\bibfnamefont {S.}~\bibnamefont {Jung}},\ }\href {\doibase 10.1103/PhysRevD.104.063001} {\bibfield  {journal} {\bibinfo  {journal} {Phys. Rev. D}\ }\textbf {\bibinfo {volume} {104}},\ \bibinfo {pages} {063001} (\bibinfo {year} {2021})},\ \Eprint {http://arxiv.org/abs/2103.08618} {arXiv:2103.08618 [astro-ph.CO]} \BibitemShut {NoStop}%
\bibitem [{\citenamefont {Cremonese}\ \emph {et~al.}(2023)\citenamefont {Cremonese}, \citenamefont {Mota},\ and\ \citenamefont {Salzano}}]{Cremonese:2021ahz}%
  \BibitemOpen
  \bibfield  {author} {\bibinfo {author} {\bibfnamefont {P.}~\bibnamefont {Cremonese}}, \bibinfo {author} {\bibfnamefont {D.~F.}\ \bibnamefont {Mota}}, \ and\ \bibinfo {author} {\bibfnamefont {V.}~\bibnamefont {Salzano}},\ }\href {\doibase 10.1002/andp.202300040} {\bibfield  {journal} {\bibinfo  {journal} {Annalen Phys.}\ }\textbf {\bibinfo {volume} {535}},\ \bibinfo {pages} {2300040} (\bibinfo {year} {2023})},\ \Eprint {http://arxiv.org/abs/2111.01163} {arXiv:2111.01163 [astro-ph.CO]} \BibitemShut {NoStop}%
\bibitem [{\citenamefont {Fairbairn}\ \emph {et~al.}(2023)\citenamefont {Fairbairn}, \citenamefont {Urrutia},\ and\ \citenamefont {Vaskonen}}]{Fairbairn:2022xln}%
  \BibitemOpen
  \bibfield  {author} {\bibinfo {author} {\bibfnamefont {M.}~\bibnamefont {Fairbairn}}, \bibinfo {author} {\bibfnamefont {J.}~\bibnamefont {Urrutia}}, \ and\ \bibinfo {author} {\bibfnamefont {V.}~\bibnamefont {Vaskonen}},\ }\href {\doibase 10.1088/1475-7516/2023/07/007} {\bibfield  {journal} {\bibinfo  {journal} {JCAP}\ }\textbf {\bibinfo {volume} {07}},\ \bibinfo {pages} {007} (\bibinfo {year} {2023})},\ \Eprint {http://arxiv.org/abs/2210.13436} {arXiv:2210.13436 [astro-ph.CO]} \BibitemShut {NoStop}%
\bibitem [{\citenamefont {Christian}\ \emph {et~al.}(2018)\citenamefont {Christian}, \citenamefont {Vitale},\ and\ \citenamefont {Loeb}}]{Christian:2018vsi}%
  \BibitemOpen
  \bibfield  {author} {\bibinfo {author} {\bibfnamefont {P.}~\bibnamefont {Christian}}, \bibinfo {author} {\bibfnamefont {S.}~\bibnamefont {Vitale}}, \ and\ \bibinfo {author} {\bibfnamefont {A.}~\bibnamefont {Loeb}},\ }\href {\doibase 10.1103/PhysRevD.98.103022} {\bibfield  {journal} {\bibinfo  {journal} {Phys. Rev. D}\ }\textbf {\bibinfo {volume} {98}},\ \bibinfo {pages} {103022} (\bibinfo {year} {2018})},\ \Eprint {http://arxiv.org/abs/1802.02586} {arXiv:1802.02586 [astro-ph.HE]} \BibitemShut {NoStop}%
\bibitem [{\citenamefont {Diego}\ \emph {et~al.}(2019)\citenamefont {Diego}, \citenamefont {Hannuksela}, \citenamefont {Kelly}, \citenamefont {Broadhurst}, \citenamefont {Kim}, \citenamefont {Li}, \citenamefont {Smoot},\ and\ \citenamefont {Pagano}}]{Diego:2019lcd}%
  \BibitemOpen
  \bibfield  {author} {\bibinfo {author} {\bibfnamefont {J.~M.}\ \bibnamefont {Diego}}, \bibinfo {author} {\bibfnamefont {O.~A.}\ \bibnamefont {Hannuksela}}, \bibinfo {author} {\bibfnamefont {P.~L.}\ \bibnamefont {Kelly}}, \bibinfo {author} {\bibfnamefont {T.}~\bibnamefont {Broadhurst}}, \bibinfo {author} {\bibfnamefont {K.}~\bibnamefont {Kim}}, \bibinfo {author} {\bibfnamefont {T.~G.~F.}\ \bibnamefont {Li}}, \bibinfo {author} {\bibfnamefont {G.~F.}\ \bibnamefont {Smoot}}, \ and\ \bibinfo {author} {\bibfnamefont {G.}~\bibnamefont {Pagano}},\ }\href {\doibase 10.1051/0004-6361/201935490} {\bibfield  {journal} {\bibinfo  {journal} {Astron. Astrophys.}\ }\textbf {\bibinfo {volume} {627}},\ \bibinfo {pages} {A130} (\bibinfo {year} {2019})},\ \Eprint {http://arxiv.org/abs/1903.04513} {arXiv:1903.04513 [astro-ph.CO]} \BibitemShut {NoStop}%
\bibitem [{\citenamefont {Cheung}\ \emph {et~al.}(2021)\citenamefont {Cheung}, \citenamefont {Gais}, \citenamefont {Hannuksela},\ and\ \citenamefont {Li}}]{Cheung:2020okf}%
  \BibitemOpen
  \bibfield  {author} {\bibinfo {author} {\bibfnamefont {M.~H.~Y.}\ \bibnamefont {Cheung}}, \bibinfo {author} {\bibfnamefont {J.}~\bibnamefont {Gais}}, \bibinfo {author} {\bibfnamefont {O.~A.}\ \bibnamefont {Hannuksela}}, \ and\ \bibinfo {author} {\bibfnamefont {T.~G.~F.}\ \bibnamefont {Li}},\ }\href {\doibase 10.1093/mnras/stab579} {\bibfield  {journal} {\bibinfo  {journal} {Mon. Not. Roy. Astron. Soc.}\ }\textbf {\bibinfo {volume} {503}},\ \bibinfo {pages} {3326} (\bibinfo {year} {2021})},\ \Eprint {http://arxiv.org/abs/2012.07800} {arXiv:2012.07800 [astro-ph.HE]} \BibitemShut {NoStop}%
\bibitem [{\citenamefont {Mishra}\ \emph {et~al.}(2021)\citenamefont {Mishra}, \citenamefont {Meena}, \citenamefont {More}, \citenamefont {Bose},\ and\ \citenamefont {Bagla}}]{Mishra:2021xzz}%
  \BibitemOpen
  \bibfield  {author} {\bibinfo {author} {\bibfnamefont {A.}~\bibnamefont {Mishra}}, \bibinfo {author} {\bibfnamefont {A.~K.}\ \bibnamefont {Meena}}, \bibinfo {author} {\bibfnamefont {A.}~\bibnamefont {More}}, \bibinfo {author} {\bibfnamefont {S.}~\bibnamefont {Bose}}, \ and\ \bibinfo {author} {\bibfnamefont {J.~S.}\ \bibnamefont {Bagla}},\ }\href {\doibase 10.1093/mnras/stab2875} {\bibfield  {journal} {\bibinfo  {journal} {Mon. Not. Roy. Astron. Soc.}\ }\textbf {\bibinfo {volume} {508}},\ \bibinfo {pages} {4869} (\bibinfo {year} {2021})},\ \Eprint {http://arxiv.org/abs/2102.03946} {arXiv:2102.03946 [astro-ph.CO]} \BibitemShut {NoStop}%
\bibitem [{\citenamefont {Suvorov}(2022)}]{Suvorov:2021uvd}%
  \BibitemOpen
  \bibfield  {author} {\bibinfo {author} {\bibfnamefont {A.~G.}\ \bibnamefont {Suvorov}},\ }\href {\doibase 10.3847/1538-4357/ac5f45} {\bibfield  {journal} {\bibinfo  {journal} {Astrophys. J.}\ }\textbf {\bibinfo {volume} {930}},\ \bibinfo {pages} {13} (\bibinfo {year} {2022})},\ \Eprint {http://arxiv.org/abs/2112.01670} {arXiv:2112.01670 [astro-ph.HE]} \BibitemShut {NoStop}%
\bibitem [{\citenamefont {Meena}\ \emph {et~al.}(2022)\citenamefont {Meena}, \citenamefont {Mishra}, \citenamefont {More}, \citenamefont {Bose},\ and\ \citenamefont {Bagla}}]{Meena:2022unp}%
  \BibitemOpen
  \bibfield  {author} {\bibinfo {author} {\bibfnamefont {A.~K.}\ \bibnamefont {Meena}}, \bibinfo {author} {\bibfnamefont {A.}~\bibnamefont {Mishra}}, \bibinfo {author} {\bibfnamefont {A.}~\bibnamefont {More}}, \bibinfo {author} {\bibfnamefont {S.}~\bibnamefont {Bose}}, \ and\ \bibinfo {author} {\bibfnamefont {J.~S.}\ \bibnamefont {Bagla}},\ }\href {\doibase 10.1093/mnras/stac2721} {\bibfield  {journal} {\bibinfo  {journal} {Mon. Not. Roy. Astron. Soc.}\ }\textbf {\bibinfo {volume} {517}},\ \bibinfo {pages} {872} (\bibinfo {year} {2022})},\ \Eprint {http://arxiv.org/abs/2205.05409} {arXiv:2205.05409 [astro-ph.GA]} \BibitemShut {NoStop}%
\bibitem [{\citenamefont {Yeung}\ \emph {et~al.}(2023)\citenamefont {Yeung}, \citenamefont {Cheung}, \citenamefont {Seo}, \citenamefont {Gais}, \citenamefont {Hannuksela},\ and\ \citenamefont {Li}}]{Yeung:2023mbs}%
  \BibitemOpen
  \bibfield  {author} {\bibinfo {author} {\bibfnamefont {S.~M.~C.}\ \bibnamefont {Yeung}}, \bibinfo {author} {\bibfnamefont {M.~H.~Y.}\ \bibnamefont {Cheung}}, \bibinfo {author} {\bibfnamefont {E.}~\bibnamefont {Seo}}, \bibinfo {author} {\bibfnamefont {J.~A.~J.}\ \bibnamefont {Gais}}, \bibinfo {author} {\bibfnamefont {O.~A.}\ \bibnamefont {Hannuksela}}, \ and\ \bibinfo {author} {\bibfnamefont {T.~G.~F.}\ \bibnamefont {Li}},\ }\href {\doibase 10.1093/mnras/stad2772} {\bibfield  {journal} {\bibinfo  {journal} {Mon. Not. Roy. Astron. Soc.}\ }\textbf {\bibinfo {volume} {526}},\ \bibinfo {pages} {2230} (\bibinfo {year} {2023})}\BibitemShut {NoStop}%
\bibitem [{\citenamefont {Shan}\ \emph {et~al.}(2023{\natexlab{a}})\citenamefont {Shan}, \citenamefont {Chen}, \citenamefont {Hu},\ and\ \citenamefont {Cai}}]{Shan:2023ngi}%
  \BibitemOpen
  \bibfield  {author} {\bibinfo {author} {\bibfnamefont {X.}~\bibnamefont {Shan}}, \bibinfo {author} {\bibfnamefont {X.}~\bibnamefont {Chen}}, \bibinfo {author} {\bibfnamefont {B.}~\bibnamefont {Hu}}, \ and\ \bibinfo {author} {\bibfnamefont {R.-G.}\ \bibnamefont {Cai}},\ }\href@noop {} {\  (\bibinfo {year} {2023}{\natexlab{a}})},\ \Eprint {http://arxiv.org/abs/2301.06117} {arXiv:2301.06117 [astro-ph.IM]} \BibitemShut {NoStop}%
\bibitem [{\citenamefont {Shan}\ \emph {et~al.}(2023{\natexlab{b}})\citenamefont {Shan}, \citenamefont {Chen}, \citenamefont {Hu},\ and\ \citenamefont {Li}}]{Shan:2023qvd}%
  \BibitemOpen
  \bibfield  {author} {\bibinfo {author} {\bibfnamefont {X.}~\bibnamefont {Shan}}, \bibinfo {author} {\bibfnamefont {X.}~\bibnamefont {Chen}}, \bibinfo {author} {\bibfnamefont {B.}~\bibnamefont {Hu}}, \ and\ \bibinfo {author} {\bibfnamefont {G.}~\bibnamefont {Li}},\ }\href@noop {} {\  (\bibinfo {year} {2023}{\natexlab{b}})},\ \Eprint {http://arxiv.org/abs/2306.14796} {arXiv:2306.14796 [astro-ph.CO]} \BibitemShut {NoStop}%
\bibitem [{\citenamefont {Hui}\ \emph {et~al.}(2017)\citenamefont {Hui}, \citenamefont {Ostriker}, \citenamefont {Tremaine},\ and\ \citenamefont {Witten}}]{Hui:2016ltb}%
  \BibitemOpen
  \bibfield  {author} {\bibinfo {author} {\bibfnamefont {L.}~\bibnamefont {Hui}}, \bibinfo {author} {\bibfnamefont {J.~P.}\ \bibnamefont {Ostriker}}, \bibinfo {author} {\bibfnamefont {S.}~\bibnamefont {Tremaine}}, \ and\ \bibinfo {author} {\bibfnamefont {E.}~\bibnamefont {Witten}},\ }\href {\doibase 10.1103/PhysRevD.95.043541} {\bibfield  {journal} {\bibinfo  {journal} {Phys. Rev. D}\ }\textbf {\bibinfo {volume} {95}},\ \bibinfo {pages} {043541} (\bibinfo {year} {2017})},\ \Eprint {http://arxiv.org/abs/1610.08297} {arXiv:1610.08297 [astro-ph.CO]} \BibitemShut {NoStop}%
\bibitem [{\citenamefont {Arvanitaki}\ \emph {et~al.}(2020)\citenamefont {Arvanitaki}, \citenamefont {Dimopoulos}, \citenamefont {Galanis}, \citenamefont {Lehner}, \citenamefont {Thompson},\ and\ \citenamefont {Van~Tilburg}}]{Arvanitaki:2019rax}%
  \BibitemOpen
  \bibfield  {author} {\bibinfo {author} {\bibfnamefont {A.}~\bibnamefont {Arvanitaki}}, \bibinfo {author} {\bibfnamefont {S.}~\bibnamefont {Dimopoulos}}, \bibinfo {author} {\bibfnamefont {M.}~\bibnamefont {Galanis}}, \bibinfo {author} {\bibfnamefont {L.}~\bibnamefont {Lehner}}, \bibinfo {author} {\bibfnamefont {J.~O.}\ \bibnamefont {Thompson}}, \ and\ \bibinfo {author} {\bibfnamefont {K.}~\bibnamefont {Van~Tilburg}},\ }\href {\doibase 10.1103/PhysRevD.101.083014} {\bibfield  {journal} {\bibinfo  {journal} {Phys. Rev. D}\ }\textbf {\bibinfo {volume} {101}},\ \bibinfo {pages} {083014} (\bibinfo {year} {2020})},\ \Eprint {http://arxiv.org/abs/1909.11665} {arXiv:1909.11665 [astro-ph.CO]} \BibitemShut {NoStop}%
\bibitem [{\citenamefont {Gilman}\ \emph {et~al.}(2021)\citenamefont {Gilman}, \citenamefont {Bovy}, \citenamefont {Treu}, \citenamefont {Nierenberg}, \citenamefont {Birrer}, \citenamefont {Benson},\ and\ \citenamefont {Sameie}}]{Gilman:2021sdr}%
  \BibitemOpen
  \bibfield  {author} {\bibinfo {author} {\bibfnamefont {D.}~\bibnamefont {Gilman}}, \bibinfo {author} {\bibfnamefont {J.}~\bibnamefont {Bovy}}, \bibinfo {author} {\bibfnamefont {T.}~\bibnamefont {Treu}}, \bibinfo {author} {\bibfnamefont {A.}~\bibnamefont {Nierenberg}}, \bibinfo {author} {\bibfnamefont {S.}~\bibnamefont {Birrer}}, \bibinfo {author} {\bibfnamefont {A.}~\bibnamefont {Benson}}, \ and\ \bibinfo {author} {\bibfnamefont {O.}~\bibnamefont {Sameie}},\ }\href {\doibase 10.1093/mnras/stab2335} {\bibfield  {journal} {\bibinfo  {journal} {Mon. Not. Roy. Astron. Soc.}\ }\textbf {\bibinfo {volume} {507}},\ \bibinfo {pages} {2432} (\bibinfo {year} {2021})},\ \Eprint {http://arxiv.org/abs/2105.05259} {arXiv:2105.05259 [astro-ph.CO]} \BibitemShut {NoStop}%
\bibitem [{\citenamefont {Adhikari}\ \emph {et~al.}(2022)\citenamefont {Adhikari} \emph {et~al.}}]{Adhikari:2022sbh}%
  \BibitemOpen
  \bibfield  {author} {\bibinfo {author} {\bibfnamefont {S.}~\bibnamefont {Adhikari}} \emph {et~al.},\ }\href@noop {} {\  (\bibinfo {year} {2022})},\ \Eprint {http://arxiv.org/abs/2207.10638} {arXiv:2207.10638 [astro-ph.CO]} \BibitemShut {NoStop}%
\bibitem [{\citenamefont {Delos}\ and\ \citenamefont {White}(2022)}]{Delos:2022yhn}%
  \BibitemOpen
  \bibfield  {author} {\bibinfo {author} {\bibfnamefont {M.~S.}\ \bibnamefont {Delos}}\ and\ \bibinfo {author} {\bibfnamefont {S.~D.~M.}\ \bibnamefont {White}},\ }\href {\doibase 10.1093/mnras/stac3373} {\bibfield  {journal} {\bibinfo  {journal} {Mon. Not. Roy. Astron. Soc.}\ }\textbf {\bibinfo {volume} {518}},\ \bibinfo {pages} {3509} (\bibinfo {year} {2022})},\ \Eprint {http://arxiv.org/abs/2207.05082} {arXiv:2207.05082 [astro-ph.CO]} \BibitemShut {NoStop}%
\bibitem [{\citenamefont {Delos}(2023)}]{Delos:2023exh}%
  \BibitemOpen
  \bibfield  {author} {\bibinfo {author} {\bibfnamefont {M.~S.}\ \bibnamefont {Delos}},\ }\href {\doibase 10.1093/mnrasl/slad043} {\bibfield  {journal} {\bibinfo  {journal} {Mon. Not. Roy. Astron. Soc.}\ }\textbf {\bibinfo {volume} {522}},\ \bibinfo {pages} {L78} (\bibinfo {year} {2023})},\ \Eprint {http://arxiv.org/abs/2302.03040} {arXiv:2302.03040 [astro-ph.CO]} \BibitemShut {NoStop}%
\bibitem [{\citenamefont {Serpico}\ \emph {et~al.}(2020)\citenamefont {Serpico}, \citenamefont {Poulin}, \citenamefont {Inman},\ and\ \citenamefont {Kohri}}]{Serpico:2020ehh}%
  \BibitemOpen
  \bibfield  {author} {\bibinfo {author} {\bibfnamefont {P.~D.}\ \bibnamefont {Serpico}}, \bibinfo {author} {\bibfnamefont {V.}~\bibnamefont {Poulin}}, \bibinfo {author} {\bibfnamefont {D.}~\bibnamefont {Inman}}, \ and\ \bibinfo {author} {\bibfnamefont {K.}~\bibnamefont {Kohri}},\ }\href {\doibase 10.1103/PhysRevResearch.2.023204} {\bibfield  {journal} {\bibinfo  {journal} {Phys. Rev. Res.}\ }\textbf {\bibinfo {volume} {2}},\ \bibinfo {pages} {023204} (\bibinfo {year} {2020})},\ \Eprint {http://arxiv.org/abs/2002.10771} {arXiv:2002.10771 [astro-ph.CO]} \BibitemShut {NoStop}%
\bibitem [{\citenamefont {Esteban-Guti\'errez}\ \emph {et~al.}(2023)\citenamefont {Esteban-Guti\'errez}, \citenamefont {Mediavilla}, \citenamefont {Jim\'enez-Vicente},\ and\ \citenamefont {Mu\~noz}}]{Esteban-Gutierrez:2023qcz}%
  \BibitemOpen
  \bibfield  {author} {\bibinfo {author} {\bibfnamefont {A.}~\bibnamefont {Esteban-Guti\'errez}}, \bibinfo {author} {\bibfnamefont {E.}~\bibnamefont {Mediavilla}}, \bibinfo {author} {\bibfnamefont {J.}~\bibnamefont {Jim\'enez-Vicente}}, \ and\ \bibinfo {author} {\bibfnamefont {J.~A.}\ \bibnamefont {Mu\~noz}},\ }\href {\doibase 10.3847/1538-4357/ace62f} {\bibfield  {journal} {\bibinfo  {journal} {Astrophys. J.}\ }\textbf {\bibinfo {volume} {954}},\ \bibinfo {pages} {172} (\bibinfo {year} {2023})},\ \Eprint {http://arxiv.org/abs/2307.07473} {arXiv:2307.07473 [astro-ph.CO]} \BibitemShut {NoStop}%
\bibitem [{\citenamefont {Blaineau}\ \emph {et~al.}(2022)\citenamefont {Blaineau} \emph {et~al.}}]{Blaineau:2022nhy}%
  \BibitemOpen
  \bibfield  {author} {\bibinfo {author} {\bibfnamefont {T.}~\bibnamefont {Blaineau}} \emph {et~al.},\ }\href {\doibase 10.1051/0004-6361/202243430} {\bibfield  {journal} {\bibinfo  {journal} {Astron. Astrophys.}\ }\textbf {\bibinfo {volume} {664}},\ \bibinfo {pages} {A106} (\bibinfo {year} {2022})},\ \Eprint {http://arxiv.org/abs/2202.13819} {arXiv:2202.13819 [astro-ph.GA]} \BibitemShut {NoStop}%
\bibitem [{\citenamefont {Oguri}\ \emph {et~al.}(2018)\citenamefont {Oguri}, \citenamefont {Diego}, \citenamefont {Kaiser}, \citenamefont {Kelly},\ and\ \citenamefont {Broadhurst}}]{Oguri:2017ock}%
  \BibitemOpen
  \bibfield  {author} {\bibinfo {author} {\bibfnamefont {M.}~\bibnamefont {Oguri}}, \bibinfo {author} {\bibfnamefont {J.~M.}\ \bibnamefont {Diego}}, \bibinfo {author} {\bibfnamefont {N.}~\bibnamefont {Kaiser}}, \bibinfo {author} {\bibfnamefont {P.~L.}\ \bibnamefont {Kelly}}, \ and\ \bibinfo {author} {\bibfnamefont {T.}~\bibnamefont {Broadhurst}},\ }\href {\doibase 10.1103/PhysRevD.97.023518} {\bibfield  {journal} {\bibinfo  {journal} {Phys. Rev. D}\ }\textbf {\bibinfo {volume} {97}},\ \bibinfo {pages} {023518} (\bibinfo {year} {2018})},\ \Eprint {http://arxiv.org/abs/1710.00148} {arXiv:1710.00148 [astro-ph.CO]} \BibitemShut {NoStop}%
\bibitem [{\citenamefont {Zumalacarregui}\ and\ \citenamefont {Seljak}(2018)}]{Zumalacarregui:2017qqd}%
  \BibitemOpen
  \bibfield  {author} {\bibinfo {author} {\bibfnamefont {M.}~\bibnamefont {Zumalacarregui}}\ and\ \bibinfo {author} {\bibfnamefont {U.}~\bibnamefont {Seljak}},\ }\href {\doibase 10.1103/PhysRevLett.121.141101} {\bibfield  {journal} {\bibinfo  {journal} {Phys. Rev. Lett.}\ }\textbf {\bibinfo {volume} {121}},\ \bibinfo {pages} {141101} (\bibinfo {year} {2018})},\ \Eprint {http://arxiv.org/abs/1712.02240} {arXiv:1712.02240 [astro-ph.CO]} \BibitemShut {NoStop}%
\bibitem [{\citenamefont {Mu\~noz}\ \emph {et~al.}(2016)\citenamefont {Mu\~noz}, \citenamefont {Kovetz}, \citenamefont {Dai},\ and\ \citenamefont {Kamionkowski}}]{Munoz:2016tmg}%
  \BibitemOpen
  \bibfield  {author} {\bibinfo {author} {\bibfnamefont {J.~B.}\ \bibnamefont {Mu\~noz}}, \bibinfo {author} {\bibfnamefont {E.~D.}\ \bibnamefont {Kovetz}}, \bibinfo {author} {\bibfnamefont {L.}~\bibnamefont {Dai}}, \ and\ \bibinfo {author} {\bibfnamefont {M.}~\bibnamefont {Kamionkowski}},\ }\href {\doibase 10.1103/PhysRevLett.117.091301} {\bibfield  {journal} {\bibinfo  {journal} {Phys. Rev. Lett.}\ }\textbf {\bibinfo {volume} {117}},\ \bibinfo {pages} {091301} (\bibinfo {year} {2016})},\ \Eprint {http://arxiv.org/abs/1605.00008} {arXiv:1605.00008 [astro-ph.CO]} \BibitemShut {NoStop}%
\bibitem [{\citenamefont {Gil~Choi}\ \emph {et~al.}(2023)\citenamefont {Gil~Choi}, \citenamefont {Jung}, \citenamefont {Lu},\ and\ \citenamefont {Takhistov}}]{GilChoi:2023qrz}%
  \BibitemOpen
  \bibfield  {author} {\bibinfo {author} {\bibfnamefont {H.}~\bibnamefont {Gil~Choi}}, \bibinfo {author} {\bibfnamefont {S.}~\bibnamefont {Jung}}, \bibinfo {author} {\bibfnamefont {P.}~\bibnamefont {Lu}}, \ and\ \bibinfo {author} {\bibfnamefont {V.}~\bibnamefont {Takhistov}},\ }\href@noop {} {\  (\bibinfo {year} {2023})},\ \Eprint {http://arxiv.org/abs/2311.17829} {arXiv:2311.17829 [astro-ph.CO]} \BibitemShut {NoStop}%
\bibitem [{\citenamefont {Hannuksela}\ \emph {et~al.}(2019)\citenamefont {Hannuksela}, \citenamefont {Haris}, \citenamefont {Ng}, \citenamefont {Kumar}, \citenamefont {Mehta}, \citenamefont {Keitel}, \citenamefont {Li},\ and\ \citenamefont {Ajith}}]{Hannuksela:2019kle}%
  \BibitemOpen
  \bibfield  {author} {\bibinfo {author} {\bibfnamefont {O.~A.}\ \bibnamefont {Hannuksela}}, \bibinfo {author} {\bibfnamefont {K.}~\bibnamefont {Haris}}, \bibinfo {author} {\bibfnamefont {K.~K.~Y.}\ \bibnamefont {Ng}}, \bibinfo {author} {\bibfnamefont {S.}~\bibnamefont {Kumar}}, \bibinfo {author} {\bibfnamefont {A.~K.}\ \bibnamefont {Mehta}}, \bibinfo {author} {\bibfnamefont {D.}~\bibnamefont {Keitel}}, \bibinfo {author} {\bibfnamefont {T.~G.~F.}\ \bibnamefont {Li}}, \ and\ \bibinfo {author} {\bibfnamefont {P.}~\bibnamefont {Ajith}},\ }\href {\doibase 10.3847/2041-8213/ab0c0f} {\bibfield  {journal} {\bibinfo  {journal} {Astrophys. J. Lett.}\ }\textbf {\bibinfo {volume} {874}},\ \bibinfo {pages} {L2} (\bibinfo {year} {2019})},\ \Eprint {http://arxiv.org/abs/1901.02674} {arXiv:1901.02674 [gr-qc]} \BibitemShut {NoStop}%
\bibitem [{\citenamefont {Abbott}\ \emph {et~al.}(2021{\natexlab{b}})\citenamefont {Abbott} \emph {et~al.}}]{LIGOScientific:2021izm}%
  \BibitemOpen
  \bibfield  {author} {\bibinfo {author} {\bibfnamefont {R.}~\bibnamefont {Abbott}} \emph {et~al.} (\bibinfo {collaboration} {LIGO Scientific, VIRGO}),\ }\href {\doibase 10.3847/1538-4357/ac23db} {\bibfield  {journal} {\bibinfo  {journal} {Astrophys. J.}\ }\textbf {\bibinfo {volume} {923}},\ \bibinfo {pages} {14} (\bibinfo {year} {2021}{\natexlab{b}})},\ \Eprint {http://arxiv.org/abs/2105.06384} {arXiv:2105.06384 [gr-qc]} \BibitemShut {NoStop}%
\bibitem [{\citenamefont {Abbott}\ \emph {et~al.}(2023{\natexlab{b}})\citenamefont {Abbott} \emph {et~al.}}]{LIGOScientific:2023bwz}%
  \BibitemOpen
  \bibfield  {author} {\bibinfo {author} {\bibfnamefont {R.}~\bibnamefont {Abbott}} \emph {et~al.} (\bibinfo {collaboration} {LIGO Scientific, VIRGO, KAGRA}),\ }\href@noop {} {\  (\bibinfo {year} {2023}{\natexlab{b}})},\ \Eprint {http://arxiv.org/abs/2304.08393} {arXiv:2304.08393 [gr-qc]} \BibitemShut {NoStop}%
\bibitem [{\citenamefont {Liu}\ \emph {et~al.}(2021)\citenamefont {Liu}, \citenamefont {Magana~Hernandez},\ and\ \citenamefont {Creighton}}]{Liu:2020par}%
  \BibitemOpen
  \bibfield  {author} {\bibinfo {author} {\bibfnamefont {X.}~\bibnamefont {Liu}}, \bibinfo {author} {\bibfnamefont {I.}~\bibnamefont {Magana~Hernandez}}, \ and\ \bibinfo {author} {\bibfnamefont {J.}~\bibnamefont {Creighton}},\ }\href {\doibase 10.3847/1538-4357/abd7eb} {\bibfield  {journal} {\bibinfo  {journal} {Astrophys. J.}\ }\textbf {\bibinfo {volume} {908}},\ \bibinfo {pages} {97} (\bibinfo {year} {2021})},\ \Eprint {http://arxiv.org/abs/2009.06539} {arXiv:2009.06539 [astro-ph.HE]} \BibitemShut {NoStop}%
\bibitem [{\citenamefont {Dai}\ \emph {et~al.}(2020)\citenamefont {Dai}, \citenamefont {Zackay}, \citenamefont {Venumadhav}, \citenamefont {Roulet},\ and\ \citenamefont {Zaldarriaga}}]{Dai:2020tpj}%
  \BibitemOpen
  \bibfield  {author} {\bibinfo {author} {\bibfnamefont {L.}~\bibnamefont {Dai}}, \bibinfo {author} {\bibfnamefont {B.}~\bibnamefont {Zackay}}, \bibinfo {author} {\bibfnamefont {T.}~\bibnamefont {Venumadhav}}, \bibinfo {author} {\bibfnamefont {J.}~\bibnamefont {Roulet}}, \ and\ \bibinfo {author} {\bibfnamefont {M.}~\bibnamefont {Zaldarriaga}},\ }\href@noop {} {\  (\bibinfo {year} {2020})},\ \Eprint {http://arxiv.org/abs/2007.12709} {arXiv:2007.12709 [astro-ph.HE]} \BibitemShut {NoStop}%
\bibitem [{\citenamefont {Haris}\ \emph {et~al.}(2018)\citenamefont {Haris}, \citenamefont {Mehta}, \citenamefont {Kumar}, \citenamefont {Venumadhav},\ and\ \citenamefont {Ajith}}]{Haris:2018vmn}%
  \BibitemOpen
  \bibfield  {author} {\bibinfo {author} {\bibfnamefont {K.}~\bibnamefont {Haris}}, \bibinfo {author} {\bibfnamefont {A.~K.}\ \bibnamefont {Mehta}}, \bibinfo {author} {\bibfnamefont {S.}~\bibnamefont {Kumar}}, \bibinfo {author} {\bibfnamefont {T.}~\bibnamefont {Venumadhav}}, \ and\ \bibinfo {author} {\bibfnamefont {P.}~\bibnamefont {Ajith}},\ }\href@noop {} {\  (\bibinfo {year} {2018})},\ \Eprint {http://arxiv.org/abs/1807.07062} {arXiv:1807.07062 [gr-qc]} \BibitemShut {NoStop}%
\bibitem [{\citenamefont {Lo}\ and\ \citenamefont {Magana~Hernandez}(2023)}]{Lo:2021nae}%
  \BibitemOpen
  \bibfield  {author} {\bibinfo {author} {\bibfnamefont {R.~K.~L.}\ \bibnamefont {Lo}}\ and\ \bibinfo {author} {\bibfnamefont {I.}~\bibnamefont {Magana~Hernandez}},\ }\href {\doibase 10.1103/PhysRevD.107.123015} {\bibfield  {journal} {\bibinfo  {journal} {Phys. Rev. D}\ }\textbf {\bibinfo {volume} {107}},\ \bibinfo {pages} {123015} (\bibinfo {year} {2023})},\ \Eprint {http://arxiv.org/abs/2104.09339} {arXiv:2104.09339 [gr-qc]} \BibitemShut {NoStop}%
\bibitem [{\citenamefont {Janquart}\ \emph {et~al.}(2022)\citenamefont {Janquart}, \citenamefont {Hannuksela}, \citenamefont {Haris},\ and\ \citenamefont {Van~den Broeck}}]{Janquart:2022wxc}%
  \BibitemOpen
  \bibfield  {author} {\bibinfo {author} {\bibfnamefont {J.}~\bibnamefont {Janquart}}, \bibinfo {author} {\bibfnamefont {O.~A.}\ \bibnamefont {Hannuksela}}, \bibinfo {author} {\bibfnamefont {K.}~\bibnamefont {Haris}}, \ and\ \bibinfo {author} {\bibfnamefont {C.}~\bibnamefont {Van~den Broeck}},\ }in\ \href@noop {} {\emph {\bibinfo {booktitle} {{56th Rencontres de Moriond on Gravitation}}}}\ (\bibinfo {year} {2022})\ \Eprint {http://arxiv.org/abs/2203.06444} {arXiv:2203.06444 [gr-qc]} \BibitemShut {NoStop}%
\bibitem [{\citenamefont {Ali}\ \emph {et~al.}(2023)\citenamefont {Ali}, \citenamefont {Stoikos}, \citenamefont {Meade}, \citenamefont {Kesden},\ and\ \citenamefont {King}}]{Ali:2022guz}%
  \BibitemOpen
  \bibfield  {author} {\bibinfo {author} {\bibfnamefont {S.}~\bibnamefont {Ali}}, \bibinfo {author} {\bibfnamefont {E.}~\bibnamefont {Stoikos}}, \bibinfo {author} {\bibfnamefont {E.}~\bibnamefont {Meade}}, \bibinfo {author} {\bibfnamefont {M.}~\bibnamefont {Kesden}}, \ and\ \bibinfo {author} {\bibfnamefont {L.}~\bibnamefont {King}},\ }\href {\doibase 10.1103/PhysRevD.107.103023} {\bibfield  {journal} {\bibinfo  {journal} {Phys. Rev. D}\ }\textbf {\bibinfo {volume} {107}},\ \bibinfo {pages} {103023} (\bibinfo {year} {2023})},\ \Eprint {http://arxiv.org/abs/2210.01873} {arXiv:2210.01873 [gr-qc]} \BibitemShut {NoStop}%
\bibitem [{\citenamefont {Ezquiaga}\ \emph {et~al.}(2023)\citenamefont {Ezquiaga}, \citenamefont {Hu},\ and\ \citenamefont {Lo}}]{Ezquiaga:2023xfe}%
  \BibitemOpen
  \bibfield  {author} {\bibinfo {author} {\bibfnamefont {J.~M.}\ \bibnamefont {Ezquiaga}}, \bibinfo {author} {\bibfnamefont {W.}~\bibnamefont {Hu}}, \ and\ \bibinfo {author} {\bibfnamefont {R.~K.~L.}\ \bibnamefont {Lo}},\ }\href {\doibase 10.1103/PhysRevD.108.103520} {\bibfield  {journal} {\bibinfo  {journal} {Phys. Rev. D}\ }\textbf {\bibinfo {volume} {108}},\ \bibinfo {pages} {103520} (\bibinfo {year} {2023})},\ \Eprint {http://arxiv.org/abs/2308.06616} {arXiv:2308.06616 [astro-ph.CO]} \BibitemShut {NoStop}%
\bibitem [{\citenamefont {Li}\ \emph {et~al.}(2023{\natexlab{a}})\citenamefont {Li}, \citenamefont {Lo}, \citenamefont {Sachdev}, \citenamefont {Chan}, \citenamefont {Lin}, \citenamefont {Li},\ and\ \citenamefont {Weinstein}}]{Li:2019osa}%
  \BibitemOpen
  \bibfield  {author} {\bibinfo {author} {\bibfnamefont {A.~K.~Y.}\ \bibnamefont {Li}}, \bibinfo {author} {\bibfnamefont {R.~K.~L.}\ \bibnamefont {Lo}}, \bibinfo {author} {\bibfnamefont {S.}~\bibnamefont {Sachdev}}, \bibinfo {author} {\bibfnamefont {J.~C.~L.}\ \bibnamefont {Chan}}, \bibinfo {author} {\bibfnamefont {E.~T.}\ \bibnamefont {Lin}}, \bibinfo {author} {\bibfnamefont {T.~G.~F.}\ \bibnamefont {Li}}, \ and\ \bibinfo {author} {\bibfnamefont {A.~J.}\ \bibnamefont {Weinstein}} (\bibinfo {collaboration} {LIGO Scientific, Virgo}),\ }\href {\doibase 10.1103/PhysRevD.107.123014} {\bibfield  {journal} {\bibinfo  {journal} {Phys. Rev. D}\ }\textbf {\bibinfo {volume} {107}},\ \bibinfo {pages} {123014} (\bibinfo {year} {2023}{\natexlab{a}})},\ \Eprint {http://arxiv.org/abs/1904.06020} {arXiv:1904.06020 [gr-qc]} \BibitemShut {NoStop}%
\bibitem [{\citenamefont {Li}\ \emph {et~al.}(2023{\natexlab{b}})\citenamefont {Li}, \citenamefont {Chan}, \citenamefont {Fong}, \citenamefont {Chong}, \citenamefont {Weinstein},\ and\ \citenamefont {Ezquiaga}}]{Li:2023zdl}%
  \BibitemOpen
  \bibfield  {author} {\bibinfo {author} {\bibfnamefont {A.~K.~Y.}\ \bibnamefont {Li}}, \bibinfo {author} {\bibfnamefont {J.~C.~L.}\ \bibnamefont {Chan}}, \bibinfo {author} {\bibfnamefont {H.}~\bibnamefont {Fong}}, \bibinfo {author} {\bibfnamefont {A.~H.~Y.}\ \bibnamefont {Chong}}, \bibinfo {author} {\bibfnamefont {A.~J.}\ \bibnamefont {Weinstein}}, \ and\ \bibinfo {author} {\bibfnamefont {J.~M.}\ \bibnamefont {Ezquiaga}},\ }\href@noop {} {\  (\bibinfo {year} {2023}{\natexlab{b}})},\ \Eprint {http://arxiv.org/abs/2311.06416} {arXiv:2311.06416 [gr-qc]} \BibitemShut {NoStop}%
\bibitem [{\citenamefont {McIsaac}\ \emph {et~al.}(2020)\citenamefont {McIsaac}, \citenamefont {Keitel}, \citenamefont {Collett}, \citenamefont {Harry}, \citenamefont {Mozzon}, \citenamefont {Edy},\ and\ \citenamefont {Bacon}}]{McIsaac:2019use}%
  \BibitemOpen
  \bibfield  {author} {\bibinfo {author} {\bibfnamefont {C.}~\bibnamefont {McIsaac}}, \bibinfo {author} {\bibfnamefont {D.}~\bibnamefont {Keitel}}, \bibinfo {author} {\bibfnamefont {T.}~\bibnamefont {Collett}}, \bibinfo {author} {\bibfnamefont {I.}~\bibnamefont {Harry}}, \bibinfo {author} {\bibfnamefont {S.}~\bibnamefont {Mozzon}}, \bibinfo {author} {\bibfnamefont {O.}~\bibnamefont {Edy}}, \ and\ \bibinfo {author} {\bibfnamefont {D.}~\bibnamefont {Bacon}},\ }\href {\doibase 10.1103/PhysRevD.102.084031} {\bibfield  {journal} {\bibinfo  {journal} {Phys. Rev. D}\ }\textbf {\bibinfo {volume} {102}},\ \bibinfo {pages} {084031} (\bibinfo {year} {2020})},\ \Eprint {http://arxiv.org/abs/1912.05389} {arXiv:1912.05389 [gr-qc]} \BibitemShut {NoStop}%
\bibitem [{\citenamefont {Ng}\ \emph {et~al.}(2018)\citenamefont {Ng}, \citenamefont {Wong}, \citenamefont {Broadhurst},\ and\ \citenamefont {Li}}]{Ng:2017yiu}%
  \BibitemOpen
  \bibfield  {author} {\bibinfo {author} {\bibfnamefont {K.~K.~Y.}\ \bibnamefont {Ng}}, \bibinfo {author} {\bibfnamefont {K.~W.~K.}\ \bibnamefont {Wong}}, \bibinfo {author} {\bibfnamefont {T.}~\bibnamefont {Broadhurst}}, \ and\ \bibinfo {author} {\bibfnamefont {T.~G.~F.}\ \bibnamefont {Li}},\ }\href {\doibase 10.1103/PhysRevD.97.023012} {\bibfield  {journal} {\bibinfo  {journal} {Phys. Rev. D}\ }\textbf {\bibinfo {volume} {97}},\ \bibinfo {pages} {023012} (\bibinfo {year} {2018})},\ \Eprint {http://arxiv.org/abs/1703.06319} {arXiv:1703.06319 [astro-ph.CO]} \BibitemShut {NoStop}%
\bibitem [{\citenamefont {Li}\ \emph {et~al.}(2018)\citenamefont {Li}, \citenamefont {Mao}, \citenamefont {Zhao},\ and\ \citenamefont {Lu}}]{Li:2018prc}%
  \BibitemOpen
  \bibfield  {author} {\bibinfo {author} {\bibfnamefont {S.-S.}\ \bibnamefont {Li}}, \bibinfo {author} {\bibfnamefont {S.}~\bibnamefont {Mao}}, \bibinfo {author} {\bibfnamefont {Y.}~\bibnamefont {Zhao}}, \ and\ \bibinfo {author} {\bibfnamefont {Y.}~\bibnamefont {Lu}},\ }\href {\doibase 10.1093/mnras/sty411} {\bibfield  {journal} {\bibinfo  {journal} {Mon. Not. Roy. Astron. Soc.}\ }\textbf {\bibinfo {volume} {476}},\ \bibinfo {pages} {2220} (\bibinfo {year} {2018})},\ \Eprint {http://arxiv.org/abs/1802.05089} {arXiv:1802.05089 [astro-ph.CO]} \BibitemShut {NoStop}%
\bibitem [{\citenamefont {Xu}\ \emph {et~al.}(2022)\citenamefont {Xu}, \citenamefont {Ezquiaga},\ and\ \citenamefont {Holz}}]{Xu:2021bfn}%
  \BibitemOpen
  \bibfield  {author} {\bibinfo {author} {\bibfnamefont {F.}~\bibnamefont {Xu}}, \bibinfo {author} {\bibfnamefont {J.~M.}\ \bibnamefont {Ezquiaga}}, \ and\ \bibinfo {author} {\bibfnamefont {D.~E.}\ \bibnamefont {Holz}},\ }\href {\doibase 10.3847/1538-4357/ac58f8} {\bibfield  {journal} {\bibinfo  {journal} {Astrophys. J.}\ }\textbf {\bibinfo {volume} {929}},\ \bibinfo {pages} {9} (\bibinfo {year} {2022})},\ \Eprint {http://arxiv.org/abs/2105.14390} {arXiv:2105.14390 [astro-ph.CO]} \BibitemShut {NoStop}%
\bibitem [{\citenamefont {Wierda}\ \emph {et~al.}(2021)\citenamefont {Wierda}, \citenamefont {Wempe}, \citenamefont {Hannuksela}, \citenamefont {Koopmans},\ and\ \citenamefont {Van Den~Broeck}}]{Wierda:2021upe}%
  \BibitemOpen
  \bibfield  {author} {\bibinfo {author} {\bibfnamefont {A.~R. A.~C.}\ \bibnamefont {Wierda}}, \bibinfo {author} {\bibfnamefont {E.}~\bibnamefont {Wempe}}, \bibinfo {author} {\bibfnamefont {O.~A.}\ \bibnamefont {Hannuksela}}, \bibinfo {author} {\bibfnamefont {L.~e. V.~E.}\ \bibnamefont {Koopmans}}, \ and\ \bibinfo {author} {\bibfnamefont {C.}~\bibnamefont {Van Den~Broeck}},\ }\href {\doibase 10.3847/1538-4357/ac1bb4} {\bibfield  {journal} {\bibinfo  {journal} {Astrophys. J.}\ }\textbf {\bibinfo {volume} {921}},\ \bibinfo {pages} {154} (\bibinfo {year} {2021})},\ \Eprint {http://arxiv.org/abs/2106.06303} {arXiv:2106.06303 [astro-ph.HE]} \BibitemShut {NoStop}%
\bibitem [{\citenamefont {Smith}\ \emph {et~al.}(2018)\citenamefont {Smith}, \citenamefont {Jauzac}, \citenamefont {Veitch}, \citenamefont {Farr}, \citenamefont {Massey},\ and\ \citenamefont {Richard}}]{Smith:2017mqu}%
  \BibitemOpen
  \bibfield  {author} {\bibinfo {author} {\bibfnamefont {G.~P.}\ \bibnamefont {Smith}}, \bibinfo {author} {\bibfnamefont {M.}~\bibnamefont {Jauzac}}, \bibinfo {author} {\bibfnamefont {J.}~\bibnamefont {Veitch}}, \bibinfo {author} {\bibfnamefont {W.~M.}\ \bibnamefont {Farr}}, \bibinfo {author} {\bibfnamefont {R.}~\bibnamefont {Massey}}, \ and\ \bibinfo {author} {\bibfnamefont {J.}~\bibnamefont {Richard}},\ }\href {\doibase 10.1093/mnras/sty031} {\bibfield  {journal} {\bibinfo  {journal} {Mon. Not. Roy. Astron. Soc.}\ }\textbf {\bibinfo {volume} {475}},\ \bibinfo {pages} {3823} (\bibinfo {year} {2018})},\ \Eprint {http://arxiv.org/abs/1707.03412} {arXiv:1707.03412 [astro-ph.HE]} \BibitemShut {NoStop}%
\bibitem [{\citenamefont {Oguri}(2018)}]{Oguri:2018muv}%
  \BibitemOpen
  \bibfield  {author} {\bibinfo {author} {\bibfnamefont {M.}~\bibnamefont {Oguri}},\ }\href {\doibase 10.1093/mnras/sty2145} {\bibfield  {journal} {\bibinfo  {journal} {Mon. Not. Roy. Astron. Soc.}\ }\textbf {\bibinfo {volume} {480}},\ \bibinfo {pages} {3842} (\bibinfo {year} {2018})},\ \Eprint {http://arxiv.org/abs/1807.02584} {arXiv:1807.02584 [astro-ph.CO]} \BibitemShut {NoStop}%
\bibitem [{\citenamefont {Cao}\ \emph {et~al.}(2014)\citenamefont {Cao}, \citenamefont {Li},\ and\ \citenamefont {Wang}}]{Cao:2014oaa}%
  \BibitemOpen
  \bibfield  {author} {\bibinfo {author} {\bibfnamefont {Z.}~\bibnamefont {Cao}}, \bibinfo {author} {\bibfnamefont {L.-F.}\ \bibnamefont {Li}}, \ and\ \bibinfo {author} {\bibfnamefont {Y.}~\bibnamefont {Wang}},\ }\href {\doibase 10.1103/PhysRevD.90.062003} {\bibfield  {journal} {\bibinfo  {journal} {Phys. Rev. D}\ }\textbf {\bibinfo {volume} {90}},\ \bibinfo {pages} {062003} (\bibinfo {year} {2014})}\BibitemShut {NoStop}%
\bibitem [{\citenamefont {Dai}\ \emph {et~al.}(2017)\citenamefont {Dai}, \citenamefont {Venumadhav},\ and\ \citenamefont {Sigurdson}}]{Dai:2016igl}%
  \BibitemOpen
  \bibfield  {author} {\bibinfo {author} {\bibfnamefont {L.}~\bibnamefont {Dai}}, \bibinfo {author} {\bibfnamefont {T.}~\bibnamefont {Venumadhav}}, \ and\ \bibinfo {author} {\bibfnamefont {K.}~\bibnamefont {Sigurdson}},\ }\href {\doibase 10.1103/PhysRevD.95.044011} {\bibfield  {journal} {\bibinfo  {journal} {Phys. Rev. D}\ }\textbf {\bibinfo {volume} {95}},\ \bibinfo {pages} {044011} (\bibinfo {year} {2017})},\ \Eprint {http://arxiv.org/abs/1605.09398} {arXiv:1605.09398 [astro-ph.CO]} \BibitemShut {NoStop}%
\bibitem [{\citenamefont {Cusin}\ and\ \citenamefont {Tamanini}(2021)}]{Cusin:2020ezb}%
  \BibitemOpen
  \bibfield  {author} {\bibinfo {author} {\bibfnamefont {G.}~\bibnamefont {Cusin}}\ and\ \bibinfo {author} {\bibfnamefont {N.}~\bibnamefont {Tamanini}},\ }\href {\doibase 10.1093/mnras/stab1130} {\bibfield  {journal} {\bibinfo  {journal} {Mon. Not. Roy. Astron. Soc.}\ }\textbf {\bibinfo {volume} {504}},\ \bibinfo {pages} {3610} (\bibinfo {year} {2021})},\ \Eprint {http://arxiv.org/abs/2011.15109} {arXiv:2011.15109 [astro-ph.CO]} \BibitemShut {NoStop}%
\bibitem [{\citenamefont {Mukherjee}\ \emph {et~al.}(2021)\citenamefont {Mukherjee}, \citenamefont {Broadhurst}, \citenamefont {Diego}, \citenamefont {Silk},\ and\ \citenamefont {Smoot}}]{Mukherjee:2021qam}%
  \BibitemOpen
  \bibfield  {author} {\bibinfo {author} {\bibfnamefont {S.}~\bibnamefont {Mukherjee}}, \bibinfo {author} {\bibfnamefont {T.}~\bibnamefont {Broadhurst}}, \bibinfo {author} {\bibfnamefont {J.~M.}\ \bibnamefont {Diego}}, \bibinfo {author} {\bibfnamefont {J.}~\bibnamefont {Silk}}, \ and\ \bibinfo {author} {\bibfnamefont {G.~F.}\ \bibnamefont {Smoot}},\ }\href {\doibase 10.1093/mnras/stab1980} {\bibfield  {journal} {\bibinfo  {journal} {Mon. Not. Roy. Astron. Soc.}\ }\textbf {\bibinfo {volume} {506}},\ \bibinfo {pages} {3751} (\bibinfo {year} {2021})},\ \Eprint {http://arxiv.org/abs/2106.00392} {arXiv:2106.00392 [gr-qc]} \BibitemShut {NoStop}%
\bibitem [{\citenamefont {Gould}(1991)}]{Gould:1991td}%
  \BibitemOpen
  \bibfield  {author} {\bibinfo {author} {\bibfnamefont {A.}~\bibnamefont {Gould}},\ }\href@noop {} {\  (\bibinfo {year} {1991})}\BibitemShut {NoStop}%
\bibitem [{\citenamefont {Barnacka}\ \emph {et~al.}(2012)\citenamefont {Barnacka}, \citenamefont {Glicenstein},\ and\ \citenamefont {Moderski}}]{Barnacka:2012bm}%
  \BibitemOpen
  \bibfield  {author} {\bibinfo {author} {\bibfnamefont {A.}~\bibnamefont {Barnacka}}, \bibinfo {author} {\bibfnamefont {J.~F.}\ \bibnamefont {Glicenstein}}, \ and\ \bibinfo {author} {\bibfnamefont {R.}~\bibnamefont {Moderski}},\ }\href {\doibase 10.1103/PhysRevD.86.043001} {\bibfield  {journal} {\bibinfo  {journal} {Phys. Rev. D}\ }\textbf {\bibinfo {volume} {86}},\ \bibinfo {pages} {043001} (\bibinfo {year} {2012})},\ \Eprint {http://arxiv.org/abs/1204.2056} {arXiv:1204.2056 [astro-ph.CO]} \BibitemShut {NoStop}%
\bibitem [{\citenamefont {Matsunaga}\ and\ \citenamefont {Yamamoto}(2006)}]{Matsunaga:2006uc}%
  \BibitemOpen
  \bibfield  {author} {\bibinfo {author} {\bibfnamefont {N.}~\bibnamefont {Matsunaga}}\ and\ \bibinfo {author} {\bibfnamefont {K.}~\bibnamefont {Yamamoto}},\ }\href {\doibase 10.1088/1475-7516/2006/01/023} {\bibfield  {journal} {\bibinfo  {journal} {JCAP}\ }\textbf {\bibinfo {volume} {01}},\ \bibinfo {pages} {023} (\bibinfo {year} {2006})},\ \Eprint {http://arxiv.org/abs/astro-ph/0601701} {arXiv:astro-ph/0601701} \BibitemShut {NoStop}%
\bibitem [{\citenamefont {Katz}\ \emph {et~al.}(2018)\citenamefont {Katz}, \citenamefont {Kopp}, \citenamefont {Sibiryakov},\ and\ \citenamefont {Xue}}]{Katz:2018zrn}%
  \BibitemOpen
  \bibfield  {author} {\bibinfo {author} {\bibfnamefont {A.}~\bibnamefont {Katz}}, \bibinfo {author} {\bibfnamefont {J.}~\bibnamefont {Kopp}}, \bibinfo {author} {\bibfnamefont {S.}~\bibnamefont {Sibiryakov}}, \ and\ \bibinfo {author} {\bibfnamefont {W.}~\bibnamefont {Xue}},\ }\href {\doibase 10.1088/1475-7516/2018/12/005} {\bibfield  {journal} {\bibinfo  {journal} {JCAP}\ }\textbf {\bibinfo {volume} {12}},\ \bibinfo {pages} {005} (\bibinfo {year} {2018})},\ \Eprint {http://arxiv.org/abs/1807.11495} {arXiv:1807.11495 [astro-ph.CO]} \BibitemShut {NoStop}%
\bibitem [{\citenamefont {Dai}\ and\ \citenamefont {Lu}(2017)}]{Dai:2017twh}%
  \BibitemOpen
  \bibfield  {author} {\bibinfo {author} {\bibfnamefont {L.}~\bibnamefont {Dai}}\ and\ \bibinfo {author} {\bibfnamefont {W.}~\bibnamefont {Lu}},\ }\href {\doibase 10.3847/1538-4357/aa8873} {\bibfield  {journal} {\bibinfo  {journal} {Astrophys. J.}\ }\textbf {\bibinfo {volume} {847}},\ \bibinfo {pages} {19} (\bibinfo {year} {2017})},\ \Eprint {http://arxiv.org/abs/1706.06103} {arXiv:1706.06103 [astro-ph.HE]} \BibitemShut {NoStop}%
\bibitem [{\citenamefont {Laha}(2020)}]{Laha:2018zav}%
  \BibitemOpen
  \bibfield  {author} {\bibinfo {author} {\bibfnamefont {R.}~\bibnamefont {Laha}},\ }\href {\doibase 10.1103/PhysRevD.102.023016} {\bibfield  {journal} {\bibinfo  {journal} {Phys. Rev. D}\ }\textbf {\bibinfo {volume} {102}},\ \bibinfo {pages} {023016} (\bibinfo {year} {2020})},\ \Eprint {http://arxiv.org/abs/1812.11810} {arXiv:1812.11810 [astro-ph.CO]} \BibitemShut {NoStop}%
\bibitem [{\citenamefont {Zheng}\ \emph {et~al.}(2014)\citenamefont {Zheng}, \citenamefont {Ofek}, \citenamefont {Kulkarni}, \citenamefont {Neill},\ and\ \citenamefont {Juric}}]{Zheng:2014rpa}%
  \BibitemOpen
  \bibfield  {author} {\bibinfo {author} {\bibfnamefont {Z.}~\bibnamefont {Zheng}}, \bibinfo {author} {\bibfnamefont {E.~O.}\ \bibnamefont {Ofek}}, \bibinfo {author} {\bibfnamefont {S.~R.}\ \bibnamefont {Kulkarni}}, \bibinfo {author} {\bibfnamefont {J.~D.}\ \bibnamefont {Neill}}, \ and\ \bibinfo {author} {\bibfnamefont {M.}~\bibnamefont {Juric}},\ }\href {\doibase 10.1088/0004-637X/797/1/71} {\bibfield  {journal} {\bibinfo  {journal} {Astrophys. J.}\ }\textbf {\bibinfo {volume} {797}},\ \bibinfo {pages} {71} (\bibinfo {year} {2014})},\ \Eprint {http://arxiv.org/abs/1409.3244} {arXiv:1409.3244 [astro-ph.HE]} \BibitemShut {NoStop}%
\bibitem [{\citenamefont {Eichler}(2017)}]{Eichler:2017eid}%
  \BibitemOpen
  \bibfield  {author} {\bibinfo {author} {\bibfnamefont {D.}~\bibnamefont {Eichler}},\ }\href {\doibase 10.3847/1538-4357/aa8b70} {\bibfield  {journal} {\bibinfo  {journal} {Astrophys. J.}\ }\textbf {\bibinfo {volume} {850}},\ \bibinfo {pages} {159} (\bibinfo {year} {2017})},\ \Eprint {http://arxiv.org/abs/1711.04764} {arXiv:1711.04764 [astro-ph.HE]} \BibitemShut {NoStop}%
\bibitem [{\citenamefont {Katz}\ \emph {et~al.}(2020)\citenamefont {Katz}, \citenamefont {Kopp}, \citenamefont {Sibiryakov},\ and\ \citenamefont {Xue}}]{Katz:2019qug}%
  \BibitemOpen
  \bibfield  {author} {\bibinfo {author} {\bibfnamefont {A.}~\bibnamefont {Katz}}, \bibinfo {author} {\bibfnamefont {J.}~\bibnamefont {Kopp}}, \bibinfo {author} {\bibfnamefont {S.}~\bibnamefont {Sibiryakov}}, \ and\ \bibinfo {author} {\bibfnamefont {W.}~\bibnamefont {Xue}},\ }\href {\doibase 10.1093/mnras/staa1497} {\bibfield  {journal} {\bibinfo  {journal} {Mon. Not. Roy. Astron. Soc.}\ }\textbf {\bibinfo {volume} {496}},\ \bibinfo {pages} {564} (\bibinfo {year} {2020})},\ \Eprint {http://arxiv.org/abs/1912.07620} {arXiv:1912.07620 [astro-ph.CO]} \BibitemShut {NoStop}%
\bibitem [{\citenamefont {Leung}\ \emph {et~al.}(2022)\citenamefont {Leung} \emph {et~al.}}]{Leung:2022vcx}%
  \BibitemOpen
  \bibfield  {author} {\bibinfo {author} {\bibfnamefont {C.}~\bibnamefont {Leung}} \emph {et~al.},\ }\href {\doibase 10.1103/PhysRevD.106.043017} {\bibfield  {journal} {\bibinfo  {journal} {Phys. Rev. D}\ }\textbf {\bibinfo {volume} {106}},\ \bibinfo {pages} {043017} (\bibinfo {year} {2022})},\ \Eprint {http://arxiv.org/abs/2204.06001} {arXiv:2204.06001 [astro-ph.HE]} \BibitemShut {NoStop}%
\bibitem [{\citenamefont {Jow}\ \emph {et~al.}(2020)\citenamefont {Jow}, \citenamefont {Foreman}, \citenamefont {Pen},\ and\ \citenamefont {Zhu}}]{Jow:2020rcy}%
  \BibitemOpen
  \bibfield  {author} {\bibinfo {author} {\bibfnamefont {D.~L.}\ \bibnamefont {Jow}}, \bibinfo {author} {\bibfnamefont {S.}~\bibnamefont {Foreman}}, \bibinfo {author} {\bibfnamefont {U.-L.}\ \bibnamefont {Pen}}, \ and\ \bibinfo {author} {\bibfnamefont {W.}~\bibnamefont {Zhu}},\ }\href {\doibase 10.1093/mnras/staa2230} {\bibfield  {journal} {\bibinfo  {journal} {Mon. Not. Roy. Astron. Soc.}\ }\textbf {\bibinfo {volume} {497}},\ \bibinfo {pages} {4956} (\bibinfo {year} {2020})},\ \Eprint {http://arxiv.org/abs/2002.01570} {arXiv:2002.01570 [astro-ph.HE]} \BibitemShut {NoStop}%
\bibitem [{\citenamefont {Takahashi}(2004)}]{Takahashi:2004mc}%
  \BibitemOpen
  \bibfield  {author} {\bibinfo {author} {\bibfnamefont {R.}~\bibnamefont {Takahashi}},\ }\href {\doibase 10.1051/0004-6361:20040212} {\bibfield  {journal} {\bibinfo  {journal} {Astron. Astrophys.}\ }\textbf {\bibinfo {volume} {423}},\ \bibinfo {pages} {787} (\bibinfo {year} {2004})},\ \Eprint {http://arxiv.org/abs/astro-ph/0402165} {arXiv:astro-ph/0402165} \BibitemShut {NoStop}%
\bibitem [{\citenamefont {Kim}\ \emph {et~al.}(2022)\citenamefont {Kim}, \citenamefont {Lee}, \citenamefont {Hannuksela},\ and\ \citenamefont {Li}}]{Kim:2022lex}%
  \BibitemOpen
  \bibfield  {author} {\bibinfo {author} {\bibfnamefont {K.}~\bibnamefont {Kim}}, \bibinfo {author} {\bibfnamefont {J.}~\bibnamefont {Lee}}, \bibinfo {author} {\bibfnamefont {O.~A.}\ \bibnamefont {Hannuksela}}, \ and\ \bibinfo {author} {\bibfnamefont {T.~G.~F.}\ \bibnamefont {Li}},\ }\href {\doibase 10.3847/1538-4357/ac92f3} {\bibfield  {journal} {\bibinfo  {journal} {Astrophys. J.}\ }\textbf {\bibinfo {volume} {938}},\ \bibinfo {pages} {157} (\bibinfo {year} {2022})},\ \Eprint {http://arxiv.org/abs/2206.08234} {arXiv:2206.08234 [gr-qc]} \BibitemShut {NoStop}%
\bibitem [{\citenamefont {Ulmer}\ and\ \citenamefont {Goodman}(1995)}]{Ulmer:1994ij}%
  \BibitemOpen
  \bibfield  {author} {\bibinfo {author} {\bibfnamefont {A.}~\bibnamefont {Ulmer}}\ and\ \bibinfo {author} {\bibfnamefont {J.}~\bibnamefont {Goodman}},\ }\href {\doibase 10.1086/175422} {\bibfield  {journal} {\bibinfo  {journal} {Astrophys. J.}\ }\textbf {\bibinfo {volume} {442}},\ \bibinfo {pages} {67} (\bibinfo {year} {1995})},\ \Eprint {http://arxiv.org/abs/astro-ph/9406042} {arXiv:astro-ph/9406042} \BibitemShut {NoStop}%
\bibitem [{\citenamefont {Feldbrugge}\ \emph {et~al.}(2023)\citenamefont {Feldbrugge}, \citenamefont {Pen},\ and\ \citenamefont {Turok}}]{Feldbrugge:2019fjs}%
  \BibitemOpen
  \bibfield  {author} {\bibinfo {author} {\bibfnamefont {J.}~\bibnamefont {Feldbrugge}}, \bibinfo {author} {\bibfnamefont {U.-L.}\ \bibnamefont {Pen}}, \ and\ \bibinfo {author} {\bibfnamefont {N.}~\bibnamefont {Turok}},\ }\href {\doibase 10.1016/j.aop.2023.169255} {\bibfield  {journal} {\bibinfo  {journal} {Annals Phys.}\ }\textbf {\bibinfo {volume} {451}},\ \bibinfo {pages} {169255} (\bibinfo {year} {2023})},\ \Eprint {http://arxiv.org/abs/1909.04632} {arXiv:1909.04632 [astro-ph.HE]} \BibitemShut {NoStop}%
\bibitem [{\citenamefont {Jow}\ \emph {et~al.}(2023)\citenamefont {Jow}, \citenamefont {Pen},\ and\ \citenamefont {Feldbrugge}}]{Jow:2022pux}%
  \BibitemOpen
  \bibfield  {author} {\bibinfo {author} {\bibfnamefont {D.~L.}\ \bibnamefont {Jow}}, \bibinfo {author} {\bibfnamefont {U.-L.}\ \bibnamefont {Pen}}, \ and\ \bibinfo {author} {\bibfnamefont {J.}~\bibnamefont {Feldbrugge}},\ }\href {\doibase 10.1093/mnras/stad2332} {\bibfield  {journal} {\bibinfo  {journal} {Mon. Not. Roy. Astron. Soc.}\ }\textbf {\bibinfo {volume} {525}},\ \bibinfo {pages} {2107} (\bibinfo {year} {2023})},\ \Eprint {http://arxiv.org/abs/2204.12004} {arXiv:2204.12004 [astro-ph.HE]} \BibitemShut {NoStop}%
\bibitem [{\citenamefont {Tambalo}\ \emph {et~al.}(2023{\natexlab{b}})\citenamefont {Tambalo}, \citenamefont {Zumalac\'arregui}, \citenamefont {Dai},\ and\ \citenamefont {Cheung}}]{Tambalo:2022plm}%
  \BibitemOpen
  \bibfield  {author} {\bibinfo {author} {\bibfnamefont {G.}~\bibnamefont {Tambalo}}, \bibinfo {author} {\bibfnamefont {M.}~\bibnamefont {Zumalac\'arregui}}, \bibinfo {author} {\bibfnamefont {L.}~\bibnamefont {Dai}}, \ and\ \bibinfo {author} {\bibfnamefont {M.~H.-Y.}\ \bibnamefont {Cheung}},\ }\href {\doibase 10.1103/PhysRevD.108.043527} {\bibfield  {journal} {\bibinfo  {journal} {Phys. Rev. D}\ }\textbf {\bibinfo {volume} {108}},\ \bibinfo {pages} {043527} (\bibinfo {year} {2023}{\natexlab{b}})},\ \Eprint {http://arxiv.org/abs/2210.05658} {arXiv:2210.05658 [gr-qc]} \BibitemShut {NoStop}%
\bibitem [{\citenamefont {Seo}\ \emph {et~al.}(2022)\citenamefont {Seo}, \citenamefont {Hannuksela},\ and\ \citenamefont {Li}}]{Seo:2021psp}%
  \BibitemOpen
  \bibfield  {author} {\bibinfo {author} {\bibfnamefont {E.}~\bibnamefont {Seo}}, \bibinfo {author} {\bibfnamefont {O.~A.}\ \bibnamefont {Hannuksela}}, \ and\ \bibinfo {author} {\bibfnamefont {T.~G.~F.}\ \bibnamefont {Li}},\ }\href {\doibase 10.3847/1538-4357/ac6dea} {\bibfield  {journal} {\bibinfo  {journal} {Astrophys. J.}\ }\textbf {\bibinfo {volume} {932}},\ \bibinfo {pages} {50} (\bibinfo {year} {2022})},\ \Eprint {http://arxiv.org/abs/2110.03308} {arXiv:2110.03308 [astro-ph.HE]} \BibitemShut {NoStop}%
\bibitem [{\citenamefont {Wright}\ and\ \citenamefont {Hendry}(2021)}]{Wright:2021cbn}%
  \BibitemOpen
  \bibfield  {author} {\bibinfo {author} {\bibfnamefont {M.}~\bibnamefont {Wright}}\ and\ \bibinfo {author} {\bibfnamefont {M.}~\bibnamefont {Hendry}},\ }\href {\doibase 10.3847/1538-4357/ac7ec2} {\  (\bibinfo {year} {2021}),\ 10.3847/1538-4357/ac7ec2},\ \Eprint {http://arxiv.org/abs/2112.07012} {arXiv:2112.07012 [astro-ph.HE]} \BibitemShut {NoStop}%
\bibitem [{\citenamefont {Navarro}\ \emph {et~al.}(1996)\citenamefont {Navarro}, \citenamefont {Frenk},\ and\ \citenamefont {White}}]{Navarro:1995iw}%
  \BibitemOpen
  \bibfield  {author} {\bibinfo {author} {\bibfnamefont {J.~F.}\ \bibnamefont {Navarro}}, \bibinfo {author} {\bibfnamefont {C.~S.}\ \bibnamefont {Frenk}}, \ and\ \bibinfo {author} {\bibfnamefont {S.~D.~M.}\ \bibnamefont {White}},\ }\href {\doibase 10.1086/177173} {\bibfield  {journal} {\bibinfo  {journal} {Astrophys. J.}\ }\textbf {\bibinfo {volume} {462}},\ \bibinfo {pages} {563} (\bibinfo {year} {1996})},\ \Eprint {http://arxiv.org/abs/astro-ph/9508025} {arXiv:astro-ph/9508025} \BibitemShut {NoStop}%
\bibitem [{\citenamefont {Bartelmann}(1996)}]{Bartelmann:1996hq}%
  \BibitemOpen
  \bibfield  {author} {\bibinfo {author} {\bibfnamefont {M.}~\bibnamefont {Bartelmann}},\ }\href@noop {} {\bibfield  {journal} {\bibinfo  {journal} {Astron. Astrophys.}\ }\textbf {\bibinfo {volume} {313}},\ \bibinfo {pages} {697} (\bibinfo {year} {1996})},\ \Eprint {http://arxiv.org/abs/astro-ph/9602053} {arXiv:astro-ph/9602053} \BibitemShut {NoStop}%
\bibitem [{\citenamefont {{Hinshaw}}\ and\ \citenamefont {{Krauss}}(1987)}]{1987ApJ...320..468H}%
  \BibitemOpen
  \bibfield  {author} {\bibinfo {author} {\bibfnamefont {G.}~\bibnamefont {{Hinshaw}}}\ and\ \bibinfo {author} {\bibfnamefont {L.~M.}\ \bibnamefont {{Krauss}}},\ }\href {\doibase 10.1086/165564} {\bibfield  {journal} {\bibinfo  {journal} {\apj}\ }\textbf {\bibinfo {volume} {320}},\ \bibinfo {pages} {468} (\bibinfo {year} {1987})}\BibitemShut {NoStop}%
\bibitem [{\citenamefont {Shan}\ \emph {et~al.}(2023{\natexlab{c}})\citenamefont {Shan}, \citenamefont {Li}, \citenamefont {Chen}, \citenamefont {Zheng},\ and\ \citenamefont {Zhao}}]{Shan:2022xfx}%
  \BibitemOpen
  \bibfield  {author} {\bibinfo {author} {\bibfnamefont {X.}~\bibnamefont {Shan}}, \bibinfo {author} {\bibfnamefont {G.}~\bibnamefont {Li}}, \bibinfo {author} {\bibfnamefont {X.}~\bibnamefont {Chen}}, \bibinfo {author} {\bibfnamefont {W.}~\bibnamefont {Zheng}}, \ and\ \bibinfo {author} {\bibfnamefont {W.}~\bibnamefont {Zhao}},\ }\href {\doibase 10.1007/s11433-022-1985-3} {\bibfield  {journal} {\bibinfo  {journal} {Sci. China Phys. Mech. Astron.}\ }\textbf {\bibinfo {volume} {66}},\ \bibinfo {pages} {239511} (\bibinfo {year} {2023}{\natexlab{c}})},\ \Eprint {http://arxiv.org/abs/2208.13566} {arXiv:2208.13566 [astro-ph.CO]} \BibitemShut {NoStop}%
\bibitem [{\citenamefont {Garc\'\i{}a-Quir\'os}\ \emph {et~al.}(2020)\citenamefont {Garc\'\i{}a-Quir\'os}, \citenamefont {Colleoni}, \citenamefont {Husa}, \citenamefont {Estell\'es}, \citenamefont {Pratten}, \citenamefont {Ramos-Buades}, \citenamefont {Mateu-Lucena},\ and\ \citenamefont {Jaume}}]{Garcia-Quiros:2020qpx}%
  \BibitemOpen
  \bibfield  {author} {\bibinfo {author} {\bibfnamefont {C.}~\bibnamefont {Garc\'\i{}a-Quir\'os}}, \bibinfo {author} {\bibfnamefont {M.}~\bibnamefont {Colleoni}}, \bibinfo {author} {\bibfnamefont {S.}~\bibnamefont {Husa}}, \bibinfo {author} {\bibfnamefont {H.}~\bibnamefont {Estell\'es}}, \bibinfo {author} {\bibfnamefont {G.}~\bibnamefont {Pratten}}, \bibinfo {author} {\bibfnamefont {A.}~\bibnamefont {Ramos-Buades}}, \bibinfo {author} {\bibfnamefont {M.}~\bibnamefont {Mateu-Lucena}}, \ and\ \bibinfo {author} {\bibfnamefont {R.}~\bibnamefont {Jaume}},\ }\href {\doibase 10.1103/PhysRevD.102.064002} {\bibfield  {journal} {\bibinfo  {journal} {Phys. Rev. D}\ }\textbf {\bibinfo {volume} {102}},\ \bibinfo {pages} {064002} (\bibinfo {year} {2020})},\ \Eprint {http://arxiv.org/abs/2001.10914} {arXiv:2001.10914 [gr-qc]} \BibitemShut {NoStop}%
\end{thebibliography}%

\end{document}